\documentclass[preprintnumbers,superscriptaddress,nofootinbib,aps,prd,floatfix]{revtex4-2}

\pdfoutput=1
\usepackage{graphicx}
\usepackage{latexsym}
\usepackage{amsfonts}
\usepackage{amssymb}
\usepackage{amsmath}
\usepackage{booktabs}
\usepackage{slashed}
\usepackage{feynmp}
\usepackage{url}
\usepackage{color}
\usepackage{cancel}
\usepackage{multirow,array}
\usepackage{float}
\usepackage{subfigure}
\usepackage[dvipsnames]{xcolor}
\usepackage{xspace}
\usepackage{svg}
\usepackage{colortbl}
\usepackage{fontawesome5}

\usepackage{amsmath}
\DeclareMathOperator*{\argmax}{arg\,max}
\DeclareMathOperator*{\argmin}{arg\,min}

\colorlet{linkcolor}{BrickRed}
\usepackage[
   colorlinks=true,
    urlcolor=linkcolor,
   anchorcolor=linkcolor,
    citecolor=linkcolor,
    filecolor=linkcolor,
    linkcolor=linkcolor,
    menucolor=linkcolor,
    linktocpage=true,
    pdfproducer=medialab,
    pdfa=true
]{hyperref}

\definecolor{overleaf}{rgb}{0.0, 0.6, 0.0}

 \newcommand{\Fermi}{\emph{Fermi}\xspace}

 \newcommand{\myhref}[3][black]{\href{#2}{\color{#1}{#3}}}%

\begin{document}

\preprint{\hfill MIT-CTP/5796}

\title{Inferring the Morphology of the Galactic Center Excess with Gaussian Processes}

\author{Edward D.~Ramirez}
\affiliation{Department of Physics and Astronomy, Rutgers University, Piscataway, NJ 08854, USA}

\author{Yitian Sun}
\affiliation{Department of Physics \& Trottier Space Institute,
McGill University, Montr\'eal, Qu\'ebec H3A 2T8, Canada}
\affiliation{The NSF AI Institute for Artificial Intelligence and Fundamental Interactions}
\affiliation{Center for Theoretical Physics, Massachusetts Institute of Technology, Cambridge, MA 02139, USA}

\author{Matthew R. Buckley}
\affiliation{Department of Physics and Astronomy, Rutgers University, Piscataway, NJ 08854, USA}

\author{Siddharth Mishra-Sharma}
\affiliation{The NSF AI Institute for Artificial Intelligence and Fundamental Interactions}
\affiliation{Center for Theoretical Physics, Massachusetts Institute of Technology, Cambridge, MA 02139, USA}
\affiliation{Department of Physics, Harvard University, Cambridge, MA 02138, USA}
\thanks{Currently at Anthropic; work performed while at MIT/IAIFI.}

\author{Tracy R. Slatyer}
\affiliation{The NSF AI Institute for Artificial Intelligence and Fundamental Interactions}
\affiliation{Center for Theoretical Physics, Massachusetts Institute of Technology, Cambridge, MA 02139, USA}
\affiliation{Department of Physics, Harvard University, Cambridge, MA 02138, USA}
\affiliation{Radcliffe Institute for Advanced Study at Harvard University, Cambridge,  MA 02138, USA}

\begin{abstract}
Descriptions of the Galactic Center using \Fermi gamma-ray data have so far modeled the Galactic Center Excess (GCE) as a template with fixed spatial morphology or as a linear combination of such templates. Although these templates are informed by various physical expectations, the morphology of the excess is \textit{a priori} unknown. For the first time, we describe the GCE using a flexible, non-parametric machine learning model -- the Gaussian process (GP). We assess our model's performance on synthetic data, demonstrating that the model can recover the templates used to generate the data. We then fit the \Fermi data with our model in a single energy bin from $2-20$~GeV (leaving a spectral GP analysis of the GCE for future work) using a variety of template models of diffuse gamma-ray emission to quantify our fits' systematic uncertainties associated with diffuse emission modeling. We interpret our best-fit GP in terms of GCE templates consisting of an NFW squared template and a bulge component to determine which bulge models can best describe the fitted GP and to what extent the best-fit GP is described better by an NFW squared template versus a bulge template. The best-fit GP contains morphological features that are typically not associated with traditional GCE studies. These include a localized bright source at around $(\ell,b) = (20^{\circ}, 0^{\circ})$ and a diagonal arm extending Northwest from the Galactic Center. In spite of these novel features, the fitted GP is explained best by a template-based model consisting of the bulge presented in Coleman~{\it et al.} (2020) and a squared NFW component. Our results suggest that the physical interpretation of the GCE in terms of stellar bulge and NFW-like components is highly sensitive to the assumed morphologies, background models, and the region of the sky used for inference. \myhref{https://github.com/edwarddramirez/gce-gp}{\raisebox{0.0\height}{\faGithub}}
\end{abstract}
\maketitle

\tableofcontents

\section{Introduction}
\label{sec:intro}

Since its discovery in \Fermi{} data in 2009 \cite{2009arXiv0910.2998G}, the Galactic Center Excess (GCE) has been a puzzle of interest to particle physicists and astrophysicists alike. Distinguished from other components of the gamma-ray sky by its spectrum and morphology, the characteristics of the GCE suggest a dark matter origin~\cite{2009arXiv0910.2998G, Murgia_2020, Leane_2019, Hooper_2011_a, hooper_2011_b, Abazajian_2012, Abazajian_2014, Ackermann_2017, Karwin_2017, Agrawal_2015}, but could instead be the result of a previously unknown source in the baryonic sector. Of such baryonic explanations, an unresolved population of  millisecond pulsars (MSPs) has been a favored solution~\cite{Eckner_2018, Lee_2016, Bartels_2016, Brandt_2015}, though robust debate continues about the viability of this proposed mechanism~\cite{Dinsmore_2022, Mishra_Sharma_2022, List_2021, List_2020, Leane_2020_2, Leane_2020_1, Zhong_2020}.

A number of studies have investigated the spectrum, morphology, and statistical properties of the GCE~\cite{hooper_2013, Gordon_2013, Daylan_2016, Ajello_2016, Calore_2015, Huang_2016, Linden_2016,Di_Mauro_2021, Zhou_2015}. The GCE gamma-rays peak around a GeV in energy, and compose roughly 10\% of the non-disk GeV gamma-ray flux within $10^{\circ} - 15^{\circ}$ from the Galactic Center~\cite{Cholis_2022}. Many analyses have found that the intensity of the GCE drops off with distance from the center in a manner consistent with a Navarro-Frenk-White (NFW) profile squared~\cite{Daylan_2016, Ajello_2016, Calore:2014xka, Di_Mauro_2021, Cholis_2022, McDermott:2022zmq, navarro1997}, suggestive of a dark matter interpretation. Also suggestive are the spectrum and intensity, which are generally consistent with the annihilation of a ${\cal O}(100~{\rm GeV})$ particle with a thermal relic cross section~\cite{Hooper_2011_a, hooper_2011_b, Abazajian_2012, Abazajian_2014}. 

However, all of these pieces of evidences must be extracted from data, with different approaches yielding different results for the spatial morphology.
The GCE sits on top of a large and complex gamma-ray background, emitted by the interactions of cosmic rays with the dust and interstellar radiation field near the Galactic Center. Modeling this emission is a challenge, and variations in the resulting fits have led to significant differences in the properties of the derived GCE signal  and the interpretation of the emission mechanism. One approach to simultaneously fitting the GCE and the other components of the gamma-ray sky is to use fixed spatial templates, derived in part from measurements of dust and gas at other wavelengths, and in part from modeling of cosmic-ray propagation through the Galaxy. 

Other methods do not assume these fixed spatial templates. For example, \textsc{SkyFACT}~\cite{Storm:2017arh} utilized such an approach for the foreground model, giving substantial per-pixel freedom to the gamma-ray foreground model template and using a regularized approach to impose inductive biases like smoothness, continuity, and spectral consistency. A \textsc{SkyFACT}-derived diffuse template was used to analyze the GCE in Ref.~\cite{Bartels_2018}, finding preference for a GCE morphology tracing their model of the stellar bulge. A more recent study was conducted by Ref.~\cite{manconi24}, finding preference for a GCE made up of a bulge along with sub-threshold point sources. \textsc{D\textsuperscript{3}PO}~\cite{Selig_2015} is another template-agnostic method which decomposes the gamma-ray sky into diffuse and point-like components. An analysis of the GCE using \textsc{D\textsuperscript{3}PO} is found in Ref.~\cite{Huang_2016_d3po}.

Comparing the results of analyses using different variations in templates and fitting procedures, it is notable that some analyses find that the GCE better traces the stellar bulge component around the Galactic Center, rather than an NFW-like morphology \cite{Macias:2019omb}. The gamma-ray emission of the bulge itself is uncertain, and multiple models for this component have been constructed tracing different stellar populations.

In this work, we investigate the morphology of the GCE using a flexible model to fit the data. This model -- based on a two-dimensional spatial Gaussian process (GP) \cite{rasmussen} -- does not require a specific choice of GCE morphology. Rather, we extract the distribution of the GCE in a data-driven manner from a fit to data using a GP combined with templates for background components of the gamma-ray sky. In this approach, any spatially extended gamma-ray emission which is distinct from the distributions of the background templates will be absorbed by the GP. No fixed morphology is assumed for this component; however, a prior on the smoothness properties of the putative signal is encoded through our choice of GP hyperparameters.

We model \Fermi{} gamma-ray data with energies between 2 and 20 GeV in a $20^\circ$ circle around the Galactic Center using a combination of the GP and fixed background templates. In this work we do not fit the energy spectra of the gamma-rays. This results in a degeneracy between some of the templates and the GP. In order to break this degeneracy, we additionally use the annulus of $30^\circ-40^\circ$ around the Galactic Center -- this ring is not expected to contain signal from the GCE, and within it we fit the data only using templates and without the GP. 

In addition to fitting the GP model to the data, the uncertainty in our model's predictions of the GCE morphology can be naturally quantified using Bayesian inference. Due to the complexity of the GP model (which has a parameter for each of the spatial bins of the gamma-ray sky), traditional sampling approaches of inference (\textit{e.g.,} Markov Chain Monte Carlo -- MCMC) \cite{sharma2017markov} are infeasible for an analysis  of the GCE data. Instead, we perform inference using stochastic variational inference (SVI) \cite{jordan1999introduction, hoffman2013stochastic, kingma2016improved, blei2017variational}, which -- by casting inference as an optimization problem -- scales better with the dimensions of the model parameter space. Variational inference has been used to tackle complex inference problems in astrophysics and particle physics \cite{Gabbard:2019rde,Hortua:2020qjr,Gu:2022xhk,Yang:2024cni,Campagne:2023ter,Vallisneri:2024xfk,Uzsoy:2024zho,Srinivasan:2024uax}, including a study of the small-scale structure of the GCE with a 42-parameter model~\cite{Sun2024}. 

Scaling the computation to the $\mathcal O(10^3)$ pixels in the GCE signal region is made possible through the use of differentiable forward models realized in a probabilistic programming framework~\cite{phan2019composable, bingham2019pyro, jax2018github}. This allows for the use of sparse and variational \cite{hensman2014scalable, leibfried2022} approaches for fitting Gaussian processes as one component of a larger probabilistic model describing the gamma-ray data in the Galactic Center region. A proof-of-principle application of GPs to analyze the GCE was presented in Refs.~\cite{Mishra_Sharma_2020, Mishra_Sharma_2022}.

Beyond the flexibility, scalability, and speed of our technique, there are other advantages for our particular GP model. First, while our GP is flexible, the nature of its flexibility is set by specifying the spatial behavior of the modeled process via a kernel function which directly describes how points of the process are correlated with one another (in contrast to Bayesian neural networks whose flexibility is determined by an abstract set of transformations on the inputs). This makes the GP easier to train than other flexible models built for inference, more interpretable, and more suitable to fitting smaller datasets such as is the case here.  
Second, it is straightforward to enforce that the GCE as modeled by the GP be non-negative, unlike background subtraction techniques for inferring the GCE. Third, though we enforce that the GP is zero in the outer annulus, having both the inner disk and outer annulus contribute to the fit simultaneously allows the model to trade-off between the two regions in the optimization. This is in contrast to using the outer annulus to update our priors for the background sources of gamma-ray emission, and then using the updated priors when performing a fit to the inner disk (\textit{i.e.,} taking the outer annulus as a sideband region and the inner disk as a signal region).

Briefly summarizing our results here, we find that the GP used to model the GCE identifies a number of features near the Galactic Center that are not typically associated with the excess. These include a relatively bright source along the plane near $\ell \sim 20^\circ$ and an ``arm'' extending northwest from the Center. In addition to these features, we find that -- among the existing templates for the GCE -- the GP is most compatible with the \texttt{coleman2019} bulge combined with an NFW squared profile. Perhaps surprisingly, the GP prefers relatively flat NFW profiles, which we interpret as a lack of flexibility in the existing background templates resulting in extended emission in the residuals. This extended emission is absorbed by the GP and can only be fit by a flat NFW component, as opposed to the existing bulge templates. It is possible the features of the GP unaccounted for by former GCE template models are the result of pervasive mismodeling in the diffuse background templates. Analyzing the energy spectrum of these features could help determine if this is indeed the case --  we leave this for future work. Fixing the NFW profile to have a steeper slope results in better fits to the GP at low latitudes (though at the price of a worse overall fit). This analysis of the GP fit suggests that separation between NFW and bulge components of the GCE are very sensitive to the template models for these components, in addition to the assumed background model and the spatial region over which the model is fit.

The rest of this paper is organized as follows. We describe the dataset and templates used in this work in Section~\ref{sec:data}. We introduce the Gaussian process and its implementation as a model for the GCE in Section~\ref{sec:gp}. We discuss the fitting methodology and validate a model on synthetic data in Section~\ref{sec:methodology}. We fit to \Fermi{} data and discuss results in Section~\ref{sec:dataresults}. We state our conclusions in Section~\ref{sec:conclusions}. Additionally, we have included several appendices to supplement the text. Appendix~\ref{app:choices} discusses some of our modeling choices in more detail, Appendix~\ref{app:addendum} provides some important derivations relevant for understanding the theoretical foundations of the fitting methodology, and Appendix~\ref{app:performance} contains various studies of the model's performance.

\begin{figure}[t!] 
    \centering
    \includegraphics[width=\columnwidth]{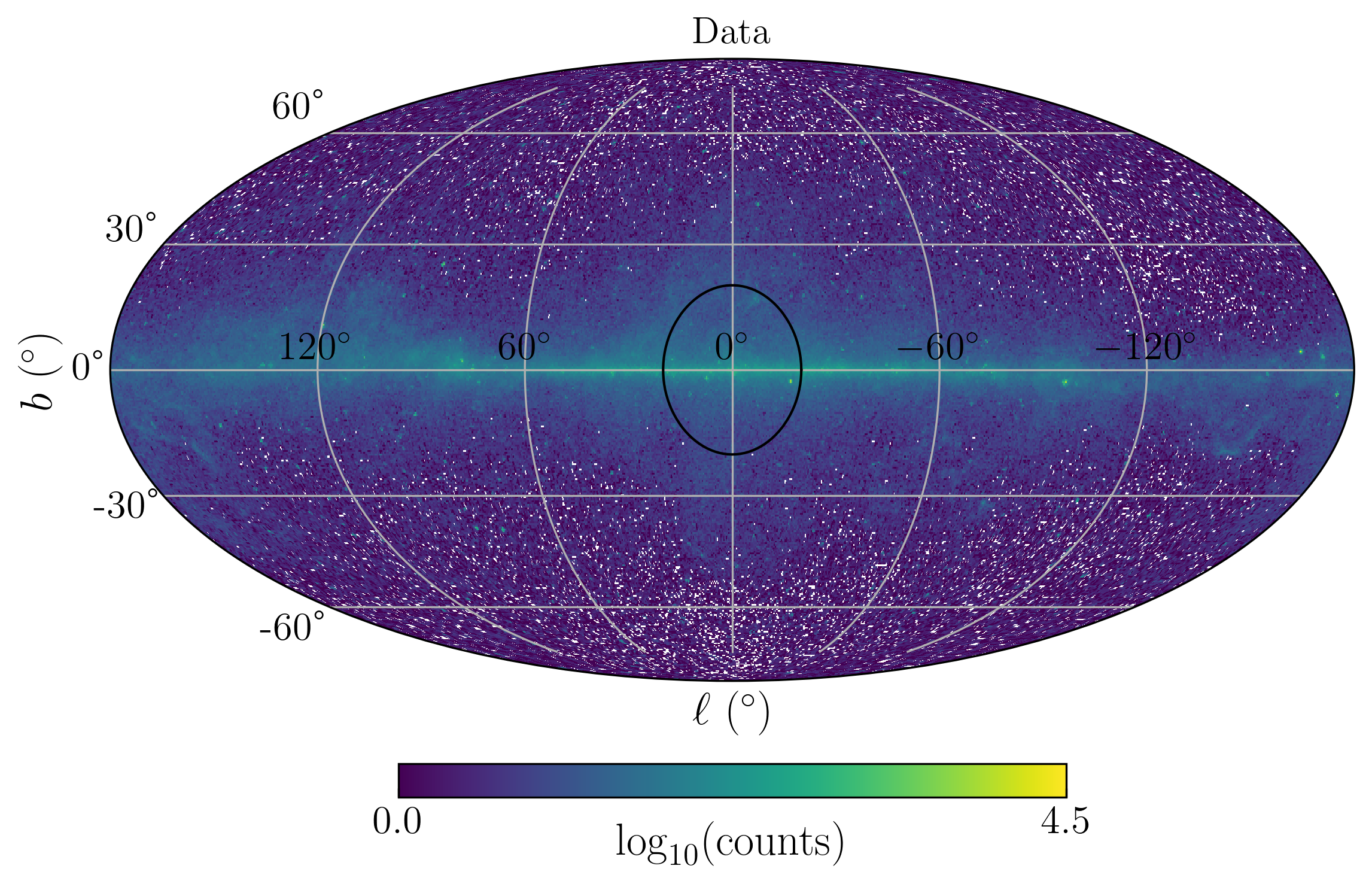}
    \includegraphics[width=\columnwidth]{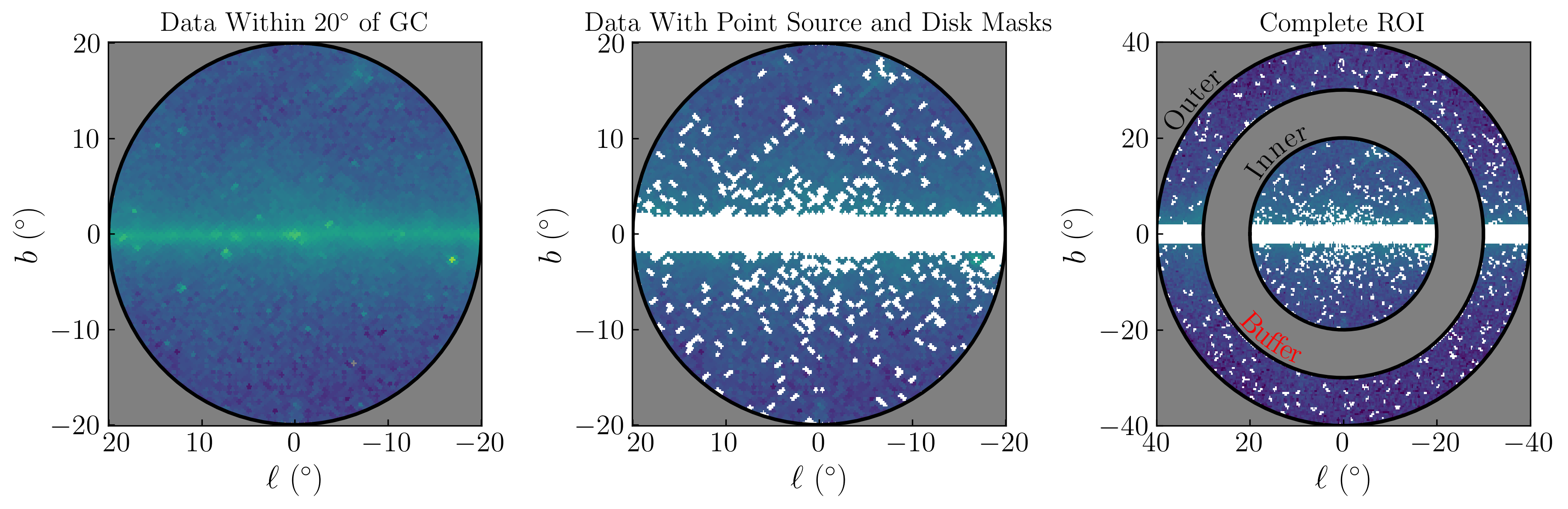}
    \caption{\textit{(Top)} Binned \textit{Fermi}-LAT data across the entire sky. \textit{(Bottom Left)} Binned data within $20^\circ$ of the GC. \textit{(Bottom Middle)} The Inner ROI consisting of pixels within $20^\circ$ of the GC, masking pixels within $2^\circ$ of the Galactic plane and within the 90\% containment radius of every 4FGL-DR2 point source. \textit{(Bottom Right)} The complete ROI consisting of the Inner ROI and the Outer ROI separated by a buffer region. The disk and point source masking applied to the Outer ROI (the annulus between $30^\circ-40^\circ$ of the GC) is constructed using the same criteria and point source catalogue as in the Inner ROI. 
    } \label{fig:data}
\end{figure}

\section{Data and Gamma-ray Source Templates}  \label{sec:data}

In this section we describe the \Fermi{} gamma-ray data used in our analysis, the regions of interest in which we perform our fits, and the fixed-morphology templates that we use to characterize the various backgrounds. As discussed in the Introduction, there are two regions of interest: an inner disk where we fit the data with both templates and a GP that models the GCE, and an outer annulus where the GCE is assumed to be absent and so the GP is zero. The GP itself is defined and discussed in Section~\ref{sec:gp} and the fitting procedure is discussed in Section~\ref{sec:methodology}.

\subsection{Data}
We use the \texttt{Pass 8} gamma-ray data collected by \textit{Fermi}-LAT up to and including May 28, 2019 (mission week 573), and employ the P8R3 instrument response functions to compute the exposure map and to process emission templates. We restrict our analysis to photons with energies between $2 - 20$ GeV. We furthermore require that photons are in the top quartile of events graded by angular reconstruction (PSF3; denoted \textsc{evtype=32}), and pass the event class cut \textsc{ULTRACLEANVETO} (denoted \textsc{evclass}=1024), corresponding to the most stringent cosmic ray rejection. We impose the quality cuts \texttt{DATA\_QUAL==1 \&\& LAT\_CONFIG==1}, and impose a maximum zenith angle of $90^\circ$.\footnote{Details of the cuts and instrument response functions may be found at \url{https://fermi.gsfc.nasa.gov/ssc/data/analysis/documentation/Cicerone/Cicerone_Data/LAT_DP.html}.} 

We process the data by binning the raw counts in angular position using \textsc{HEALPix}  \cite{gorski1999healpix} with resolution parameter $\textsc{Nside}= 128$. This sets the angular scale 
of each bin pixel to roughly $0.458^\circ$. Beyond the $2-20$~GeV selection criterion, we do not further separate the events into energy bins. The dataset across the entire sky is shown in the upper region of Fig.~\ref{fig:data}.

To mitigate the effects of mismodeling various sources of gamma-ray emission, we will fit our model (described in Section~\ref{sec:gp}) to a subset of the available pixels. To prevent overfitting on data far from the GCE, we first restrict our region-of-interest (ROI) to an inner circle around the Galactic Center. To this we add an outer ring (separated from the inner region by a buffer), where we assume the GCE is not present. As we will describe, we find this signal-free region is necessary to break degeneracies between the GP and background templates.

The Inner ROI -- containing the GCE -- is constructed from pixels inside of a $20^\circ$ circle around the Galactic Center ($(\ell,b) = (0^\circ,0^\circ)$ in Galactic coordinates).  A flat projection of the complete data within the ROI is shown to the bottom left of Fig.~\ref{fig:data}. Within this circle, we mask two bright sources of gamma-ray emission to prevent mismodeling of these sources from affecting our fits. First, we remove the Galactic disk by masking pixels whose centers are within $2^\circ$ of the Galactic plane $(|b| \leq 2^\circ)$. Next, we remove known point sources by masking pixels within the 90\% containment radius of point sources in the \textit{Fermi} 4FGL-DR2 point source catalog~\cite{4fgl-dr2}. 
We choose this point source catalog over newer releases as it uses \textit{Fermi}-LAT data collected over a time period that is closer to that of our own analysis dataset than the alternatives. We study the variation of our results with respect to the point source masking in Appendix~\ref{app:ps}. These two additional masks are shown in the bottom center of Fig.~\ref{fig:data}. From here onward, we refer to this masked collection of pixels in the inner $20^\circ$ as the Inner ROI. The Inner ROI is made up of 4,525 pixels.

We will define the GP (which models the GCE signal) only in the Inner ROI. 
As discussed in Appendix~\ref{app:degeneracy}, we find that if we fit both the GP and fixed gamma-ray source template models to the Inner ROI only, the GP's flexibility leads to degeneracies in the model that cause the fit to converge incorrectly. That is, when fitting to synthetic data generated from known templates, there exist multiple configurations of the GP and the background templates that differ significantly from the actual configuration of the synthetic gamma-ray sources, but otherwise fit the data well. In these cases, the incorrect fit returns comparable log-likelihoods as the configuration used to generate the pseudodata. 

While this degeneracy might be broken by including energy information, we leave such an analysis for future work. Instead, in this paper we include in our dataset an Outer ROI of pixels located in regions of the sky where the GCE is expected to be insignificant. In this Outer ROI, pixels will be fit only by the background templates and not the GP. For this Outer ROI, we choose an annulus extending between $30^{\circ}- 40^{\circ}$ around the Galactic Center. 

Larger ROIs have more data and can better constrain the background templates, but at the cost of greater sensitivity to mismodeling. Our choices of Inner and Outer ROI balances these competing effects, and we study the impact on our fits from varying the Outer ROI in Appendix~\ref{app:roi}. 
We apply the same disk and point source masking criteria to the Outer ROI as to the Inner ROI. After masking, the Outer ROI has roughly $1.75 \times$ as many pixels as the Inner ROI, which will transfer over to the relative weights of the template fits to the Inner and Outer ROIs in the log-likelihood. 
The complete (Outer+Inner) ROI is shown in the bottom right panel of Fig.~\ref{fig:data}. 

Neither the templates or GP will be fit to data in the buffer region: the annulus extending from $20^{\circ}-30^{\circ}$ around the Galactic Center. Inside the Buffer, the GCE is still expected to contribute a non-negligible  fraction of the gamma-ray flux (see Appendix~\ref{app:roi}). To successfully  fit the data in this region would therefore require both the background templates and the GP. Due to limits in GPU memory, we could not extend our Inner ROI to the edge of the Outer ROI, necessitating our adoption of a buffer.

\begin{figure}[t!]  
    \centering   
    \includegraphics[width = \columnwidth]{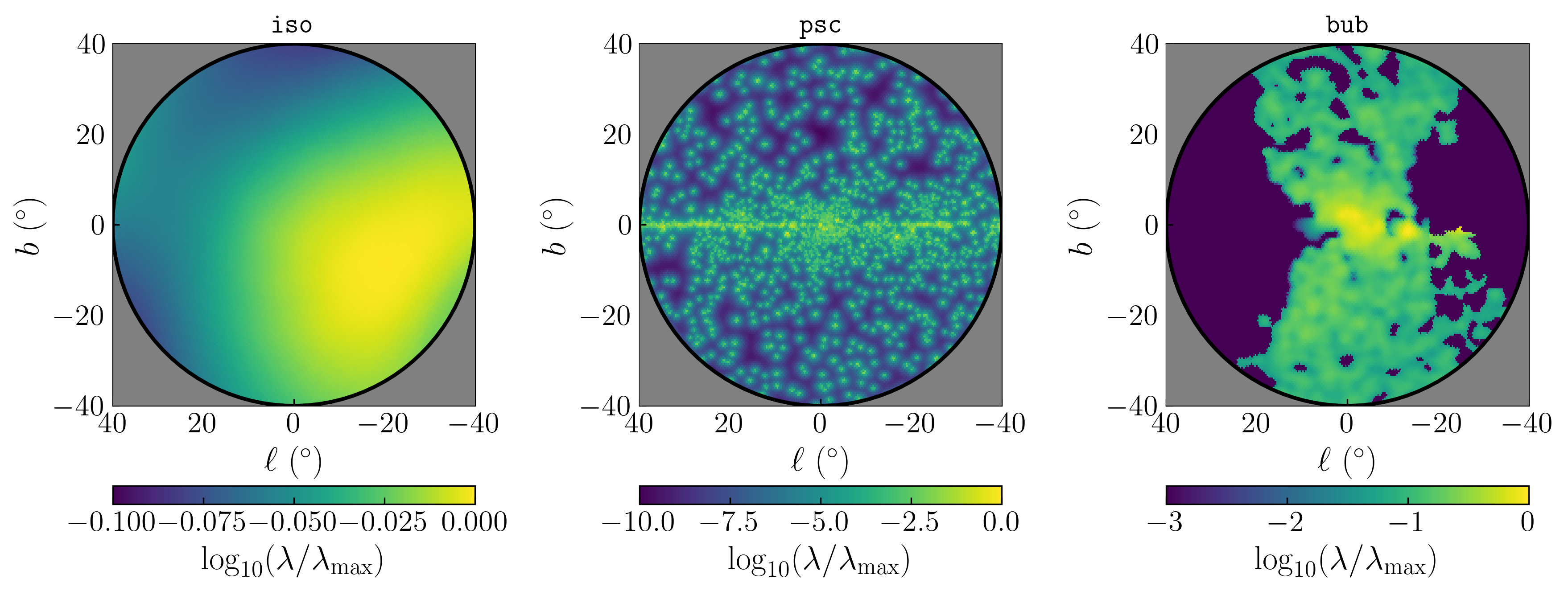} 
    \includegraphics[width = 0.66\columnwidth]{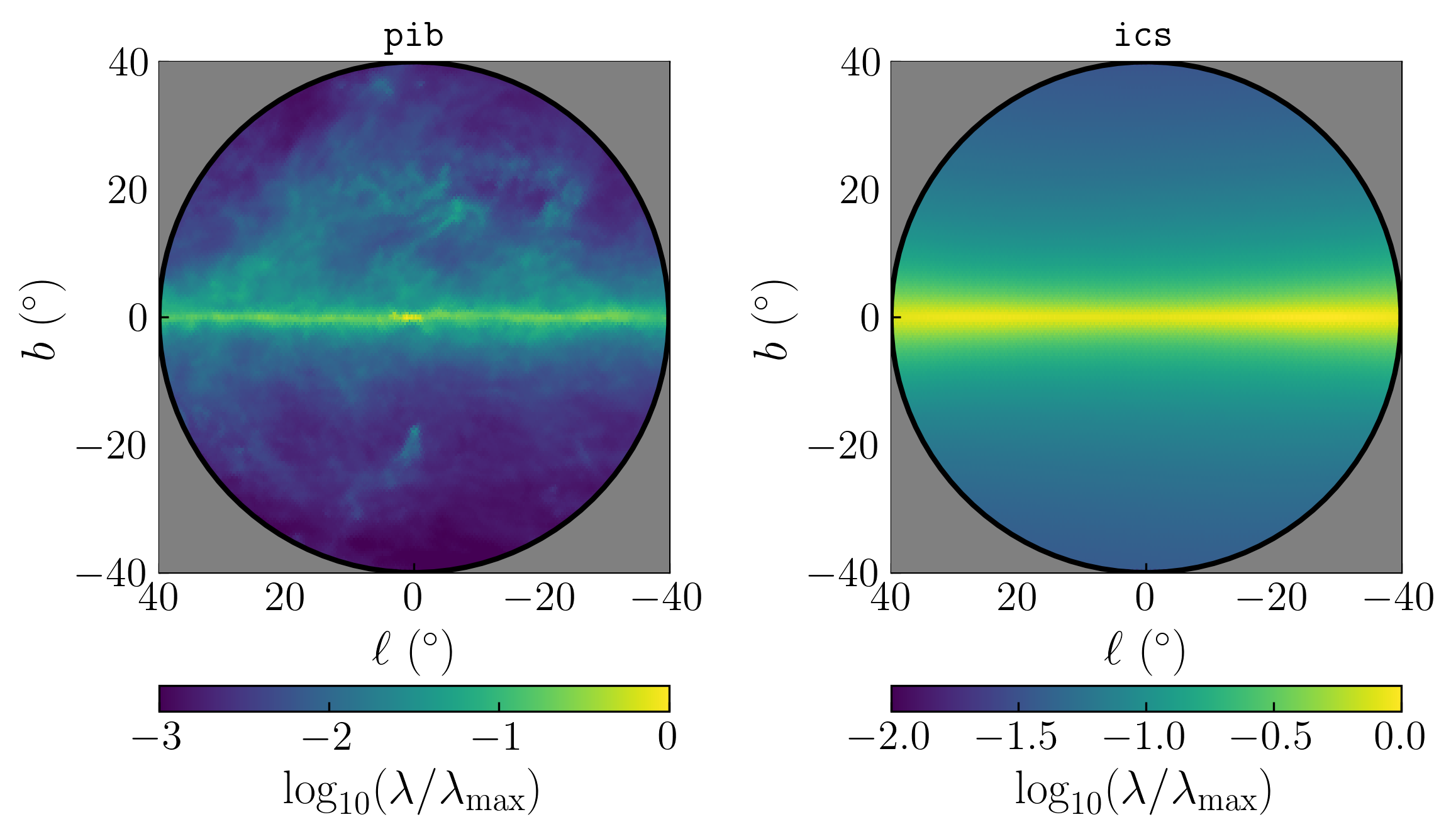}   
    \caption{Example of templates used for modeling known mechanisms of gamma-ray emission that contributes to the measured signal. \textit{(Top Left)}: The isotropic template (\texttt{iso}). \textit{(Top Center)}: The template corresponding to emission from 4FGL-DR2 point sources (\texttt{psc}). \textit{(Top Right)}: The Fermi Bubbles template (\texttt{bub}). \textit{(Bottom Left)}: The pion plus bremsstrahlung template (\texttt{pib}). \textit{(Bottom Right)}: The inverse compton scattering template (\texttt{ics}). The \texttt{pib} and \texttt{ics} templates shown here correspond to the \textsc{Galprop}-based Model X from \cite{Cholis_2022}. Templates are rates $\lambda_{j}$ (\textit{i.e.,} expected number of counts) normalized to their maximum signal. Note that the dynamic range is different for each template.} \label{fig:bkgd-templates}
\end{figure}

\subsection{Gamma-ray Source Templates} \label{sec:templates}

Having defined the dataset for our analysis, we now turn to the known sources of gamma-rays that (in addition to the GCE) contribute to the events detected by \textit{Fermi}-LAT. These are the isotropic contribution, the Fermi Bubbles, the detected point sources, inverse Compton scattering (ICS), and pion+bremsstrahlung emission. 
These appear in our gamma-ray emission model as templates with fixed morphology and floating normalizations that are fit to the data. 
Most of the templates account for the instrument response (\textit{i.e.}, the expected emission is convolved with the instrument exposure and point spread function -- PSF). 
Exposure and PSF corrections were applied using \texttt{gtsrcmaps}.\footnote{\url{https://www.slac.stanford.edu/exp/glast/wb/prod/pages/sciTools_gtsrcmaps/sciTools_gtsrcmaps_help.htm}} This tool takes into account the difference in the PSF when photons enter the instrument at different angles. 
 
We do not completely account for the instrument response for two templates. First, as we discuss below, our Fermi Bubbles template was extracted from data by Ref.~\cite{Ackermann_2017} and hence it effectively includes the exposure and PSF appropriate for the dataset used in that work. We correct for the differences in our exposures, but do not correct for the differences in our PSFs; we expect these differences to be subdominant compared to other uncertainties in the template.\footnote{The Bubbles template from Ref.~\cite{Ackermann_2017} was also smoothed with a $1^{\circ}$ Gaussian kernel, which will wash out the effects of the mismatched PSF smoothing.} Second, we do not convolve any generalized NFW squared component (see Eq.~\eqref{eq:gNFW}) to the GCE with the PSF, as its shape is allowed to float during our fits, and so it is not feasible to correct for instrument response at each iteration. We do not expect this to induce large errors in the inferred morphology, as we mask the region with $|b|<2^\circ$, and outside the central peak the NFW template is fairly smooth on the scale of the PSF. 

We start with the simplest sources. The isotropic contribution is a component accounting for the combined emission of unresolved extragalactic point sources (\textit{e.g.,} distant galaxies and ultra-high-energy cosmic rays whose sources are unknown), unresolved Galactic point sources at high latitudes, and misidentified cosmic rays \cite{Cholis_2022}. Its signal is roughly modeled as uniform across the sky, so our template traces the exposure map of \textit{Fermi}-LAT. The \texttt{iso} template is shown in Fig.~\ref{fig:bkgd-templates}.
    
The point source catalog template describes emission from point sources identified from independent fits to the \textit{Fermi}-LAT data (4FGL-DR2). Though we mask the point sources, some extended emission from each source will remain in the ROI. This residual emission is modeled by the \texttt{psc} template, shown in Fig.~\ref{fig:bkgd-templates}. We do not model unresolved point sources.

The Fermi Bubbles are a set of two roughly bilateral and latitudinal elongated lobes of gamma-ray emission emanating out from the GC and extending out to high latitudes of about $\pm50^{\circ}$. As our Bubbles template (\texttt{bub}), we combine two preexisting templates. One template (obtained from Ref.~\cite{Ackermann_2017}) is defined across the entire sky, but suffers from missing signal in locations associated with the point source masks used in their analysis. A newer template (obtained from Ref.~\cite{Macias_2019}) is an update of the previous template, where the missing signal is recovered using a Laplacian inpainting interpolation technique. However, this template is only defined for $-20^{\circ} \leq \ell,b \leq 20^{\circ}$, and so it is not compatible with our analysis requiring a larger ROI. As the newer template is a corrected version of the former, we combine these templates by using the older template in the regions either where the newer template is undefined. See Appendix~\ref{app:bub} for details of the combination method. The newer template primarily describes the Bubbles in the inner region containing the GC and the older template describes the Bubbles over the remaining area in our ROI.  The combined template is shown in Fig.~\ref{fig:bkgd-templates}.

The final components we consider are the emission mechanisms resulting from cosmic-ray interactions with the interstellar medium (ISM). These mechanisms are responsible for a majority of the gamma-rays in our energy range. There are three dominant mechanisms: $\pi^{0}$ production (followed by decay to gamma rays) from inelastic collisions of cosmic rays and heavier nuclei with ${\rm H}_{2}$, ${\rm H}_{\rm I}$, and ${\rm H}_{\rm II}$ gases in the ISM; bremsstrahlung from cosmic-ray electrons interacting with the various ISM gases; and inverse Compton scattering (ICS) of cosmic-ray electrons with low-energy photons traveling through the ISM -- the interstellar radiation field (ISRF). 

 Broadly speaking, recent studies of the GCE morphology have taken two approaches to generating fixed spatial (and spectral) templates for these sources of emission.
 One approach uses models of cosmic-ray propagation in the ISM and their subsequent interactions with interstellar gas and magnetic fields to generate a map of gamma-rays, typically performed with the \textsc{Galprop} numerical code \cite{Strong:1998pw,Moskalenko_1998,Porter:2021tlr}. The primary inputs are parameters for the source distribution and propagation for cosmic rays, a model for the interstellar radiation field, and density maps of the ISM gases. From this, a model for the gamma-ray emission over the full sky is generated for the ICS, bremsstrahlung, and $\pi^0$ components independently. To account for uncertainties in the underlying astrophysics,  the inputs are varied and the resulting models are compared to the data. The result is an ensemble of physically-motivated models that provide comparable fits to the data (as done in Ref.~\cite{Cholis:2021rpp}). We refer to these models of gamma-ray morphology as ``\textsc{Galprop}-based template models.''

In practice, we combine $\pi^{0}$ and bremsstrahlung emission from the \textsc{Galprop} output into a single template \texttt{pib} (that is, we do not float their relative normalizations) because their gamma-ray emission exhibits similar morphology, and as a result they are highly degenerate~\cite{Ackermann_2017}. We allow the normalization of the ICS component (\texttt{ics}) to vary independently. 
An example of these templates using a \textsc{Galprop}-based model is shown in Fig.~\ref{fig:bkgd-templates}.

An alternate approach is to use \textsc{Galprop} only to generate the ICS component, while modeling the other components as linearly proportional to the gas density. In this method, the Galaxy is divided into volumes within which the proportionality constant is fixed (physically, this corresponds to assuming a fixed cosmic-ray spectrum and density within that region); these constants are then fitted from the data. Typically the volumes chosen are Galactocentric cylinders (``rings''), and so these models are often referred to as ``ring-based'' ({\it e.g.}, Ref.~\cite{McDermott:2022zmq}). Depending on the approach, the normalization of the ICS template may also be simultaneously fit in each of these regions. These models may also include additional correction templates, and may contain separately fitted contributions for different components of the total gas density ({\it e.g.}, atomic and molecular hydrogen). The gas density maps may be informed by multiwavelength data and/or by simulations, see Refs.~\cite{Macias:2016nev, Pohl:2022nnd}. In general, these methods can improve the fit to data relative to the default gas maps included in \textsc{Galprop}.

Unfortunately, recent ring-based models presented in the literature \cite{Pohl:2022nnd, McDermott:2022zmq, Song_2024} are not compatible with our current approach. The models have been created to describe gamma-ray emission in the $40^{\circ} \times 40^{\circ}$ window centered at the GC, whereas we require an $80^{\circ} \times 80^{\circ}$ window in order to break degeneracies between the templates and the GP.
Due to this limitation, in this work we do not perform fits with the set of ring-based models discussed recently in Refs.~\cite{Pohl:2022nnd, McDermott:2022zmq, Song_2024}. Since the templates of the ring-based models are fit energy bin-by-energy bin, it is unclear if spectral information would break the degeneracy between the GP and the ring-based templates.

We do use one example of a ring-based model to understand whether we should expect qualitative differences with respect to \textsc{Galprop}-based models, although we work with a model that has already been fitted to data and does not retain the full freedom associated with the rings. This is the ``Model O'' introduced in Ref.~\cite{Buschmann_2020}. We refer the reader to that reference for a full description of the model, but in brief, the authors employ separate gas maps for ${\rm H}_{\rm I}$ and ${\rm H}_{\rm II}$  obtained from a suite of hydrodynamical simulations \cite{Pohl:2007dz}, two residual ``dark gas'' templates (which are not subdivided into rings), and a \textsc{Galprop}-based model for ICS. The gas and ICS templates are subdivided into four Galactocentric rings and fitted to the data. Combining the gas and residual templates, with normalizations determined by the fit, then yields a single gas-correlated template (physically corresponding to $\pi^0$ + bremsstrahlung emission) and similarly combining the ICS rings yields a single ICS template. It is these combined templates that we employ in our analysis, as analogues to the \textsc{Galprop}-based \texttt{pib} and \texttt{ics} templates.

We will use different sets of templates throughout this work to determine how our inferences for the GCE vary with respect to our choices for diffuse modeling. When fitting our models to synthetic data in Section~\ref{sec:syndata} and the Appendices we use Model O as discussed above, in addition to two \textsc{Galprop}-based models which were identified in Ref.~\cite{Calore:2014xka} as providing particularly good fits to the data. These models are denoted models A and F following the convention of Ref.~\cite{Calore:2014xka}.\footnote{The exact templates for Models O, A, and F were taken from Ref.~\cite{List_2021}.}
When fitting to the real data, in addition to Model O we use the \textsc{Galprop}-based templates from Ref.~\cite{Cholis_2022} to study the systematic uncertainties associated with a more updated set of models.

In addition to these background sources of gamma-rays, previous analyses have found the GCE to be made up of one (or both) of the following distinct components: a spherically symmetric signal that decreases in intensity consistent with a generalized NFW profile squared (the expectation of a dark matter annihilation signal~\cite{Murgia_2020}), and a non-spherical component that is identified with the stellar bulge. Most of these previous studies attribute the bulk of the GCE to one of the two components\footnote{Though these results may differ based on their background model and stellar bulge model, a majority of these studies do not fit the two components alongside one another. As shown by Ref.~\cite{McDermott:2022zmq}, fitting both components together can flip the preference.} (\textit{e.g.,} Refs.~\cite{Bartels_2018, Macias_2018, Macias:2019omb, Pohl:2022nnd} support a bulge, Refs.~\cite{Di_Mauro_2021, Cholis_2022} support the NFW, and Ref.~\cite{McDermott:2022zmq} found a preference for an NFW only when it was fit alongside a bulge). Though we model the GCE with a GP, we will find it useful to compare our results with these existing signal templates. 

\begin{figure}[t!]  
    \centering
    \includegraphics[width = 1\columnwidth]{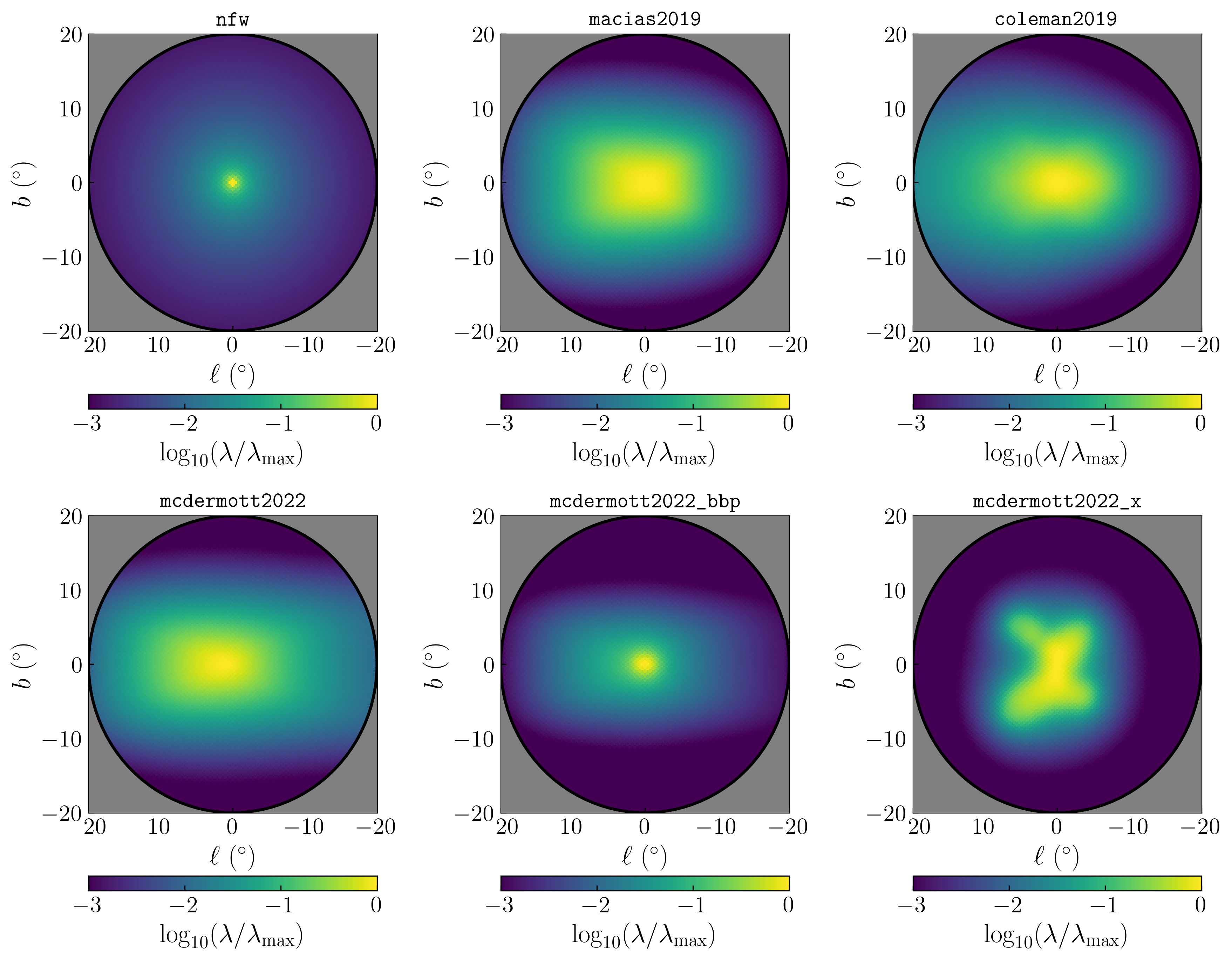}   
    \caption{Templates corresponding to GCE models. \textit{(Upper Left)} The template with radial emissions going as a squared NFW profile with $\gamma = 1.2$ and $R_{S} = 20$~kpc (\texttt{nfw}). 
    The stellar bulge templates are \textit{(Upper Center)} \texttt{macias2019} \cite{Macias_2019}, \textit{(Upper Right)} \texttt{coleman2019} \cite{Coleman_2020}, \textit{(Lower Left)} \texttt{mcdermott2022}, \textit{(Lower Center)} \texttt{mcdermott2022\_bbp}, and \textit{(Lower Right)} \texttt{mcdermott2022\_x} \cite{Cholis_2022, McDermott:2022zmq}.} \label{fig:gce_temps}
\end{figure}

The generalized NFW profile~\cite{navarro1997, ishiyama_2020_2} is given by the following smoothly broken power law:
\begin{equation} \label{eq:gNFW}
    \rho_{\rm gNFW}(r) = \frac{\rho_{0}}{(r/R_{S})^{\gamma}(1 + r/R_{S})^{3 - \gamma}},
\end{equation}
where the density parameter $\rho_{0}$ determines the profile's enclosed mass, the scale radius $R_{S}$ determines the characteristic radial distance at which the power law behavior changes, and $\gamma$ determines the power law behavior for $r \ll R_{S}$ and $r \gg R_{S}$. The gamma-ray map corresponding to this profile is the square of this density centered at the GC, integrated along the line of sight of \Fermi{}-LAT. 

In this work, we take the commonly assumed value for the scale radius, $R_{S} = 20$~kpc~\cite{Murgia_2020}. The density parameter $\rho_{0}$ floats and is set by the value of the normalization of the gamma-ray signal corresponding to the squared NFW profile. Hence, the NFW signal morphology is controlled by the slope of the power law $\gamma$. Multiple bulge component morphologies have been used in fitting to the GCE, and we consider several options in this work.

We show the GCE templates in Fig.~\ref{fig:gce_temps}. In the upper left, we show the template associated with the squared generalized NFW profile (\texttt{nfw}), with $\gamma=1.2$ and $R_{S}=20$~kpc. 
We then show the bulge templates we consider in this work (from left to right, starting on the upper row): \texttt{macias2019}~\cite{Macias_2019}, \texttt{coleman2019}~\cite{Coleman_2020}\footnote{\url{https://github.com/chrisgordon1/galactic_bulge_templates}}, \texttt{mcdermott2022}\footnote{We note this boxy bulge is a non-standard Galactic bulge template~\cite{Song_2024} and the process for creating it was not included in Ref.~\cite{McDermott:2022zmq}.}, \texttt{mcdermott2022\_bbp}, and \texttt{mcdermott2022\_x}~\cite{Cholis_2022, McDermott:2022zmq}.\footnote{\url{https://github.com/samueldmcdermott/gcepy/tree/main/gcepy/inputs/excesses}} 
Previous studies have shown that a model describing the GCE with a \texttt{coleman2019} bulge and/or an \texttt{nfw} component gives the best fit to the GCE of the options tested. Ref.~\cite{Song_2024} and Ref.~\cite{Zhong_2024} show  a strong preference for the \texttt{coleman2019} bulge over \texttt{nfw} when only one of the templates is allowed to fit the GCE. However, Ref.~\cite{Zhong_2024} also finds better fits are realized with a linear combination of these two templates. Though the preference of one over the other when fit together is dependent on diffuse modeling and masking, the \texttt{nfw} remains a significant, if not the dominant component of the GCE, for energies $E > 0.7 \ {\rm GeV}$.

\section{Gaussian Processes} \label{sec:gp}

Our goal is to investigate the GCE morphology by describing it using a flexible model -- the Gaussian process. In this section, we first introduce the Gaussian process and its basic properties. Then, we incorporate it into a Poissonian model describing the gamma-ray data using the background templates constructed in the previous section. We will describe how the model is fit to the data using stochastic variational inference (SVI) in the next section. 

\subsection{A Review of Gaussian Processes} \label{sec:gpreview}

A Gaussian process (GP) is a distribution of functions $f$ over inputs $x$, which generalizes the multivariate normal distribution (MVN) of finite-dimensional vectors~\cite{rasmussen}. Like the MVN, it is specified by a mean function $\mu$ and a covariance function (kernel) $k$:
\begin{align}
f &\sim \mathrm{GP}(\mu, k) \nonumber \\
\mu(x) &= \mathbb{E}_{f}[f(x)]  \\
k(x, x') &= \mathrm{Cov}_{f}[f(x),f(x')]. \nonumber 
\end{align}
Here, the expectation value $\mathbb E_{f}[f(x)]$ and covariance $\mathrm{Cov}_{f}[f(x),f(x')]$ at input $x$ (and $x'$) are taken over the statistical ensemble of functions $f$. The kernel $k$ encodes all the intrinsic characteristics about the functions we are modeling ({\it e.g.}, smoothness and correlation structure), whereas the mean function only affects the distribution of the family of functions determined by the kernel.\footnote{This follows from the fact that $f \sim \mathrm{GP}(\mu, k)$ is equivalent to $f = \mu + \varepsilon$, where $\varepsilon \sim \mathrm{GP}(0, k)$, so the mean function acts only as an additive offset to functions generated by the zero-mean GP, $\mathrm{GP}(0, k)$.}

If we restrict ourselves to any finite subset of inputs ($\{x_i\}_{i=1}^{N}$, written as $\mathbf{x}$), then the individual function-values evaluated at those points, $\mathbf{f} = f(\mathbf{x})$, are distributed according to the corresponding MVN:
\begin{align}
\mathbf{f} &\sim \mathcal{N}(\mathbf{\mu}(\mathbf{x}), k(\mathbf{x}, \mathbf{x}^{T}))  \nonumber \\
\mu(\mathbf{x}) &= \mathbb{E_{\mathbf{f}}[\mathbf{f}]} \\
k(\mathbf{x}, \mathbf{x}^{T}) &= \mathrm{Cov}_{\mathbf{f}, \mathbf{f'}}[\mathbf{f},\mathbf{f'}], \nonumber
\end{align}
where $\mu(\textbf{x})$ is the result of applying the mean function $\mu$ element-wise to $\mathbf{x}$ and $k(\mathbf{x}, \mathbf{x}^{T})$ is the covariance matrix calculated by applying the kernel element-wise on the matrix of pairs $(x_i, x_j)$. This \emph{marginalization} to a subset of finite-dimensional normally-distributed variables is a defining property of GPs and serves as the infinite-dimensional analogue of integrating out a subset of normally-distributed variables. 

A specific setting in which GPs are useful is in defining a prior $p(f)$ over the space of functions, $f \sim \mathrm{GP}(\mu, k)$. In this class of problems, there is a set of data $\textbf{y} = y(\textbf{x})$, which are interpreted as the output of a function $f$ at a set of inputs $\textbf{x}$ plus some additional noise. The distribution governing the relationship between $f$ and $\mathbf{y}$ is the likelihood $p(\mathbf{y} | f)$. In the problem at hand, the $\textbf{x}$ are two-dimensional spatial coordinates in our ROI, the $\textbf{y}$ are the observed counts of gamma-rays in the corresponding pixels, and so we assume the likelihood is Poisson.

In this setting, a central problem is to infer the posterior distribution $p(f|\textbf{y})$, related to the likelihood $p(\mathbf{y} | f)$ and the prior $p(f)$ via Bayes' Theorem:
\begin{align} \label{eq:bayes}
    p(f|\textbf{y}) = \frac{p(\mathbf{y} | f)\,p(f)}{p(\mathbf{y})}.
\end{align}
Here $p(\mathbf{y}) = \mathbb{E}_{p(f)}[p(\textbf{y} | f)]$ is the evidence (also known as the marginal likelihood). 
With the posterior distribution, it is then possible to sample functions $f$ whose action on the inputs $\textbf{x}$ produces outputs that are statistically compatible with the data $\textbf{y}$. 

Robust inference requires that a model accurately describe the data-generating process. For GPs, this amounts to choosing a mean function and kernel appropriate for the process being modeled, as these two functions determine all characteristics of a GP. The mean function and kernel can be chosen based on \textit{a priori} knowledge of the process being modeled. However, these functions can also be learned by posing the selection of mean and kernel as an optimization problem. 

In this work, we parameterize the mean functions and kernel with hyperparameters $\theta$ and optimize them using an objective function. Each particular set of hyperparameters $\theta$ corresponds to a particular model. To put the models on equal footing, we should compare their performances \textit{in the absence of information beyond what has been specified by the models themselves}. In our case, this means the models share the same likelihood and are distinguished only by the priors $p_{\theta}(f)$. 

We seek a model that accurately describes the data-generating process. As such, the model would generate the observed data $\textbf{y}$ with high probability compared to other models, in the absence of any additional information. The probability of generating the data $\textbf{y}$ is a distribution depending on the functions $f$, via the likelihood $p(\textbf{y} | f)$. The only information that we have on $f$ is the prior $p_{\theta}(f)$, so the probability of the data $\textbf{y}$ being generated by this model is the expectation value
\begin{equation} \label{eq:evidence}
    p_{\theta}(\textbf{y}) 
    = \mathbb{E}_{p_{\theta}(f)}[p(\textbf{y} | f)].
\end{equation}
This is none other than the evidence introduced above in Eq.~\eqref{eq:bayes}. Thus, the model with maximum evidence is the model that best describes the data-generating process and so Eq.~\eqref{eq:evidence} is our objective function for optimizing GP hyperparameters. Using the evidence for hyperparameter selection also implicitly guards against overfitting, \textit{e.g.} via a small correlation length. This is because GPs with a smaller correlation length kernel can describe many possible data well, so their density on our particular dataset will be smaller than GPs with an intermediate correlation length. This is sometimes referred to as the Bayesian Occam's Razor~\cite{WilsonA.G.2014Ckff}.

The evidence is the canonical objective function for optimizing GP hyperparameters
\cite{rasmussen}. Thus, our objectives are to
\begin{enumerate}
    \item choose a set of GP hyperparameters $\theta$ that maximize the evidence $p_{\theta}(\textbf{y})$, and 
    \item infer the posterior $p_{\theta}(f | \textbf{y})$ using the optimized GP prior $p_{\theta}(f)$.
\end{enumerate} 
When we say we \textit{train} the model on data, we mean we work towards accomplishing the above two objectives. Though the evidence and the posterior of a model are generally analytically intractable quantities, most GP applications render both quantities analytically tractable by assuming a Gaussian likelihood: 
\begin{align}
    \mathbf{y} \sim  \mathcal{N}(\mathbf{f},\sigma_{n}^{2} \,\mathbf{I}), 
\end{align}
where $\sigma_{n}^{2}\, \mathbf{I}$ denotes independent and identically distributed Gaussian noise added to each $\textbf{f}$ when generating the targets $\textbf{y}$. In this case, it is simple to train the GP model using standard gradient descent techniques. However, as mentioned before, we will be adopting a Poisson likelihood when incorporating the GP to a model of gamma-ray counts. This calls for a different approach to training the model, which we discuss in Section~\ref{sec:methodology}.

\subsection{Gaussian Processes Applied to the GCE}

We turn now to the GP application of interest: modeling the morphology of the GCE. 
Our data consists of gamma-ray counts $\mathbf{y}$ in spatial bins centered at positions $\mathbf{x}$. The count in each bin is assumed to have been drawn independently from a Poisson distribution, whose underlying rate $\lambda(\mathbf{x})$ is sourced by the various mechanisms of gamma-ray emission: the GCE and the backgrounds discussed in Section~\ref{sec:templates}. That is, 
\begin{align}
    \mathbf{y} & \sim \mathrm{Pois}[\lambda(\mathbf{x})].
\end{align}

The GP is incorporated into the model as the prior probability of the GCE contributing a log-rate $f(\mathbf{x})$ of gamma-rays. That is, if we denote the GCE rate by $\exp (f(\mathbf{x}))$, then the prior probability of $f(\mathbf{x})$ is given as
\begin{equation}
    f(\mathbf{x}) \sim \mathcal{N}(\mu(\mathbf{x}) , k(\mathbf{x}, \mathbf{x^{T}})),
\end{equation}
for a GP with mean function $\mu$ and kernel $k$. Describing the log-rate as a GP ensures that the rate is positive. A function $f$ drawn from the GP then contributes a rate $\lambda_{f}(\textbf{x})$ to the total rate $\lambda(\textbf{x})$ given by 
\begin{align}
    \lambda_{f}(\textbf{x}) = \exp (f(\mathbf{x})). 
\end{align}
 
All the known (background) gamma-ray sources are modeled by templates $T_j$ ($j = \texttt{bub}, \texttt{pib},$ {\it etc.}) with fixed morphologies and floating normalizations $S_j$. The template functions $T_j$ are each normalized to have a mean pixel count of 1 over the pixels in the inner 20$^{\circ}$ disk centered at the GC with a $|b| \leq 2^{\circ}$ band mask applied.\footnote{The pixels over which the templates are normalized are those of the Inner ROI, but without the point source masks.} Each source of gamma-rays then contributes $S_j T_j(\mathbf{x})$ to the total rate of emission $\lambda(\mathbf{x})$, and the total contribution of these background templates to the rate is 
\begin{align} \label{eq:first-rate-eqn}
    \lambda_{T}(\mathbf{x}) & = \sum_{j} S_j T_j(\mathbf{x}).
\end{align}

In the Inner ROI, the total rate is the sum of the background templates and the rate drawn from the GP:
\begin{equation}
    \lambda(\mathbf{x}) = \lambda_{f}(\textbf{x}) + \lambda_{T}(\mathbf{x}).
\end{equation}
By assumption the GP does not contribute to the rate in the Outer ROI, and so for those pixels, $\lambda(\mathbf{x}) = \lambda_{T}(\mathbf{x})$.

We now set our priors for the model and list them in Table~\ref{tab:priors}. As discussed in the previous section, we specify the GP prior on $f$ by defining a mean function and kernel. We take the GP to have zero mean. As the GP is modeling the log-rate, this corresponds to a constant rate of one count per pixel.\footnote{We tested different values for the mean and found no significant difference in the results of the fit.} We use the Matern32 kernel~\cite{rasmussen} defined with respect to the Euclidean norm in $(\ell, b)$ coordinates:\footnote{We expect negligible differences between kernels using Euclidean distances in $(\ell,b)$ versus ,\textit{e.g.}, great circle distances, as our $20^{\circ}$ ROI is approximately flat.
}
\begin{equation} \label{eq:matern}
    k^{\rm Matern32}_{\theta}(x_1, x_2) = \sigma^{2} (1 + \sqrt{3} r) \exp(-\sqrt{3} r),
\end{equation}
where 
\begin{equation} \label{eq:dist}
    r = | (x_1 - x_2) / \rho |
\end{equation}
is a rescaled version of the Euclidean distance between $x_1$ and $x_2$. Here, $\sigma$ sets the overall variance of the GP and $\rho$ sets the characteristic length scale over which the function values sampled from the GP are correlated. Their numeric values are chosen so that they maximize the evidence $p_{\theta}(\textbf{y})$ of this model. Hence, for this GP, $\theta = \{\sigma, \rho\}$ and $\rho > 0$. This choice of kernel gives a highly flexible distribution of functions that are first-order differentiable. We compare the performance of different kernel choices in Appendix~\ref{app:kernels}. 

We choose uniform priors for the template normalizations $S_j$. The boundaries for each template are listed in Table~\ref{tab:priors}. The limits on each prior are chosen to be uninformative (that is, the boundary values have minimal impact on the results of inference).

\begin{table}[]
\begin{tabular}{@{}c@{}}
        \begin{tabular}{cc}
            \toprule
            \textbf{Quantity} & \textbf{Prior} \\
            \midrule
            $S_\texttt{iso}$ & \text{Uniform}([0.001, 50]) \\
            $S_\texttt{bub}$ & \text{Uniform}([0.001, 50]) \\
            $S_\texttt{psc}$ & \text{Uniform}([0.001, 50]) \\
            $S_\texttt{ics}$ & \text{Uniform}([0.001, 100]) \\
            $S_\texttt{pib}$ & \text{Uniform}([0.001, 100]) \\
            $f(x)$ & $\mathrm{GP}\big(0,k^{\rm Matern32}_{\theta}(x,x')\big)$ \\
            \bottomrule
        \end{tabular}
    \end{tabular}

\caption{Priors for the template normalizations $S_j$ and the function $f$ determining the GCE log-rate $\textbf{f} = f(\textbf{x})$. The GP prior on $f$ is specified by defining a mean function and a kernel. All $S$-priors are uniform over the specified ranges.} \label{tab:priors}
\end{table}

\section{Fitting Methodology}\label{sec:methodology}

We must now fit our model to the data. There are two major differences in how we perform this fit, as compared to the typical approach used in most GP applications. 

First, our GP is part of a larger probabilistic model. Within our model, the GP provides only part of the rate of the gamma-ray data. As a result, the observations $\mathbf{y}$ are determined by a linear combination of the GP and the templates of the other known gamma-ray sources (and in the Outer ROI, the rate is fit only by the templates). We refer to the combination of GP and templates as the ``GP model."

Second, we adopted a Poisson likelihood for our count data. Unlike training a model consisting of a GP prior and a Gaussian likelihood, performing inference on the GP model and calculating its evidence are both analytically intractable. Instead, we must rely on approximate inference to train the GP. Variational techniques lend themselves to this problem, as sampling approaches like Markov Chain Monte Carlo (MCMC) scale poorly with the number of latent variables in a model and are challenging to perform correctly for models whose posteriors exhibit strong and non-trivial correlations (as it is then difficult for the Markov chains to trace the shape of the posterior distribution efficiently~\cite{gelmanbda04, kennedy14}). Both of these conditions hold for the GCE analysis. The GP model has many dimensions, one for each pixel in $\mathbf{x}$ ($\textbf{f} = f(\mathbf{x})$), and one for each of the template normalizations $S_j$. Further, we expect the posterior to have non-trivial correlations because we expect the true GCE to be smooth, which requires the log-rate $f(\mathbf{x})$ to be highly correlated at neighboring points. 

In addition to having better performance than sampling, variational inference (VI) allows us to bypass both the intractability of performing inference on the GP model and calculating the evidence simultaneously, as we discuss in Section~\ref{sec:svi}. As the memory requirements for our initial VI pipeline scale with the number of data points cubed, we use inducing points and subsampling to reduce the training complexity, as described in Section~\ref{sec:scaling}. We validate our model on simulated data in Section~\ref{sec:syndata}.

\subsection{Stochastic Variational Inference} \label{sec:svi}

In variational inference, the problem of inference is cast as an optimization problem, where the posterior distribution is approximated using a parametric class of functions for which accuracy of the approximation is improved through optimization of an appropriate objective. We set up this problem by approximating the posterior distribution $p(\mathbf{z} | \mathbf{y})$ of our model's latent variables, $\mathbf{z}=(\mathbf{f}, S_j)$, 
using a variational distribution (the approximate posterior) $q_{\phi}(\mathbf{z})$, where $\phi$ parameterizes a class of functions of our choice. 

The posterior function $p(\mathbf{z} | \mathbf{y})$ is not known, but (by construction) the joint distribution $p(\mathbf{y},\mathbf{z})$ is calculable, as is the variational distribution $q_{\phi}(\mathbf{z})$. We can therefore learn the best approximation of the true posterior by maximizing the \textit{Evidence Lower BOund} (ELBO), a theoretical lower-bound of the log-evidence, $\log p_{\theta}(\mathbf{y})$:
\begin{align} \label{eq:elbo_def}
    {\rm ELBO}(\theta, \phi) & = \mathbb{E}_{q_{\phi}(\mathbf{z})}[\log p_{\theta}(\textbf{y}, \mathbf{z}) - \log q_{\phi}(\mathbf{z})] .
\end{align}
Here, $p_{\theta}(\textbf{y}, \mathbf{z})$ is the joint model built from the likelihood and the prior: $p_{\theta}(\textbf{y},\textbf{z}) = p(\textbf{y}|\textbf{z})p_{\theta}(\textbf{z})$. 
To reduce memory requirements for training the GP, Eq.~\eqref{eq:elbo_def} will eventually be modified to incorporate inducing points. We describe this process in Section~\ref{sec:scaling}.

By approximating the posterior with a variational distribution obtained by maximizing the ELBO, we have solved the problem of performing inference on the GP model. In addition, this also bypasses our inability to directly maximize the evidence to optimize the GP hyperparameters. Since the ELBO is a theoretical lower-bound to the log-evidence, 
increasing the ELBO during optimization forces the evidence of the model up as well. Thus, maximizing the ELBO indirectly maximizes both the accuracy of the approximate posterior $q_{\phi}(\textbf{z})$ and the evidence $p_{\theta}(\textbf{y})$. See Appendix \ref{app:svi} for more details on variational inference.

The first term of the ELBO is calculated using the likelihood and the prior probabilities, $p_{\theta}(\textbf{y},\textbf{z}) = p(\textbf{y}|\textbf{z})p_{\theta}(\textbf{z})$, and the variational distribution $q_\phi(\mathbf{z})$ is chosen to be calculable. Since our model is differentiable, the ELBO is itself differentiable, allowing us to optimize it using gradient descent. However, the ELBO is a difficult quantity to compute exactly. In practice, the ELBO is instead estimated with a sample average: 
\begin{equation}
    \mathbb{E}_{q_{\phi}(\mathbf{z})}[\log p_{\theta}(\textbf{y}, \mathbf{z}) - \log q_{\phi}(\mathbf{z})] \rightarrow \frac{1}{N_{p}} \sum_{n}[\log p_{\theta}(\textbf{y}, \mathbf{z}_{n}) - \log q_{\phi}(\mathbf{z}_{n})].
\end{equation} 
Here, $N_p$ is the number of samples drawn to calculate the average.
This approximation is a stochastic estimate of the ELBO for optimization, and so we must rely on stochastic gradient descent (SGD) for optimization. This version of variational inference is referred to as stochastic variational inference (SVI)~\cite{hoffman2013stochastic}. 

To perform SVI as efficiently as possible, we built our model using the GCE modeling probabilistic-programming library presented in Ref.~\cite{Sun2024}. It leverages the high-level probabilistic programming capabilities of \texttt{NumPyro} to allow for easy creation of models and a streamlined approach to performing inference \cite{phan2019composable, bingham2019pyro}. Once a model is created, various types of inference can be performed with a few lines of code, including SVI. SVI runs seamlessly, since \texttt{NumPyro} runs on a \textsc{JAX} \cite{jax2018github} backend, allowing for computations to be sped up by JIT-compilation (Just-In-Time-compilation or dynamic compilation), GPU acceleration, and automatic differentiation.

In addition to specifying the GP model itself, we must select an appropriate class of variational distributions $q_\phi(\mathbf{z})$ to train the model with SVI. We require a flexible variational distribution that handles correlations between the latent variables $\textbf{z}$ since -- as mentioned previously -- we expect the posterior samples of the GP log-rate $\textbf{f}$ to be highly correlated. A variational distribution consisting of MVN distributions parameterized by their means $\mathbf{m}$ and covariance matrices $\boldsymbol{\Sigma}$: $q_{\phi}(\textbf{z}) = \mathcal{N}(\textbf{m}, \boldsymbol{\Sigma})$, would satisfy our desired properties. However, as we are using uniform priors for the template normalizations $S_j$, we are required to adopt a variational distribution whose support is consistent with that of the priors. Otherwise, the variational distribution will draw samples where the prior is zero and the ELBO will be undefined. 

To make the MVN (with unbounded support) compatible with the model priors, we define the MVN in an~unconstrained space of auxiliary variables $\tilde{\textbf{z}}$ and relate it to the constrained space of actual latent variables $\textbf{z} = (\textbf{f}, S_j)$ with the appropriate bijective map. For our specific case, where some of the $z_i$-priors have support in a finite interval $[z_{i, {\rm min}}, z_{i,{\rm max}}]$ (\textit{i.e.,} the priors for the template normalizations $S_j$), we use the bijective map $t_{i}:\tilde{z}_i  \rightarrow z_i$, where
\begin{equation}
    t_{i}( \tilde{z}_i ) = z_{i, {\rm min}} + (z_{i, {\rm max}} - z_{i, {\rm min}}) \ \mathrm{expit}(\tilde{z}_i ),
\end{equation}
where $\mathrm{expit}(x) = 1/(1+e^{-x})$ denotes the logistic (sigmoid) function.
If a latent variable $z_i$ is defined everywhere for all real numbers, 
$t_{i}(\tilde{z}_i)$ is simply the identity map. 

Thus, we take the following as our variational distribution:
\begin{align} \label{eq:q-mvn}
    q_{\phi}(\tilde{\textbf{z}}) & = \mathcal{N}(\mathbf{m}, \mathbf{\Sigma}) \\
    \textbf{z} & = \textbf{t}(\tilde{\textbf{z}})
\end{align}
where the trainable hyperparameters are $\phi = (\mathbf{m}, \mathbf{\Sigma})$ and $\textbf{t}(\tilde{\mathbf{z}})$ is the (vectorized) image of the bijective maps $t_i$ applied to each $\tilde{z}_i$. This choice of $q_\phi(\mathbf{z})$ is available in \texttt{NumPyro} through its \texttt{autoguide} module.\footnote{The \texttt{autoguide} module contains a set of variational distributions that are automatically generated once a model is created and automatically determines the bijective transformation required to make the variational distribution consistent with the model's prior.}

\subsection{Scalable Inference} \label{sec:scaling}

Exact GP inference does not scale well with large datasets, as its computational complexity goes as $\mathcal{O}(N^{3})$ for a dataset of size $N$ \cite{rasmussen}. Our model suffers from a similar degree of complexity: the gradient of the estimated ELBO across the entire space $\mathbf{x}$ must be computed at each training step, requiring $\mathcal{O}(N^{2})$ parameters (mainly from the variational covariance matrix $\mathbf{\Sigma}$) to be updated $N$ times (for each of the inputs $\mathbf{x}$). For our GP model, we have an overall complexity of $\mathcal{O}(N^{3})$, where $N = 4,525$ is the number of pixels $\textbf{x}$.

To reduce the complexity associated with training, we leverage the fact that the Gaussian process can interpolate between a fixed set of input points by conditioning on the values at those input points. In our case, we first extend the space of discrete inputs $\mathbf{x}$ of bin centers within the $20^{\circ}$ radius disk of the Inner ROI to the entire continuous disk.\footnote{As the GP is not defined on the Outer ROI, we do not have to repeat this procedure across the bins in that region.} Then, we define a set of inducing points $\mathbf{u} = f(\mathbf{x_u})$, that is, the function values of the GP evaluated at a set of pseudo-inputs $\mathbf{x_u}$. In this way, we define the variational distribution so that it describes the posterior of the inducing points $\mathbf{u}$ evaluated at $\mathbf{x_u}$ instead of the function values of the GP $\textbf{f}$ at the pixel positions $\mathbf{x}$. The variational distribution of the GP log-rate $\mathbf{f}$ at the actual bin centers $\mathbf{x}$ is defined such that it is the result of conditioning the GP on the inducing points $\mathbf{u}$:
\begin{equation} \label{eq:qf}
    q_{\theta, \phi}(\mathbf{f}) = \int p_{\theta}(\mathbf{f} | \mathbf{u}) \ q_{\phi}(\mathbf{u}) \ d\mathbf{u}.
\end{equation}
A GP whose corresponding variational distribution is defined in this way is called a sparse and variational Gaussian process (SVGP) \cite{hensman2014scalable, leibfried2022}.\footnote{A GP defined with inducing points is referred to as a sparse GP. A GP whose posterior is approximated with variational inference is referred to as a variational GP.} 

The ELBO for the SVGP model takes on the same form as Eq.~\eqref{eq:elbo_def}, with the latent variables using the inducing points $\textbf{u}$ rather than $\textbf{f}$. That is, the latent variables are now  $\mathbf{z} = (\mathbf{u}, S_j)$ (see Appendix \ref{app:svgp} for derivation). We rewrite Eq.~\eqref{eq:elbo_def} to separate the physical log-rate $\mathbf{f}$ from the pseudo-log-rate $\mathbf{u}$: 
\begin{align}
    {\rm ELBO}(\theta, \phi) & = \mathbb{E}_{q_{\phi}(\mathbf{z})}[\mathbb{E}_{p_{\theta}(\mathbf{f}|\mathbf{u})}[\log p(\textbf{y}|\mathbf{f}, S_j)]] + \mathbb{E}_{q_{\phi}(\mathbf{z})}[\log p_{\theta}(\textbf{z}) - \log q_{\phi}(\mathbf{z})]. \label{eq:svgp_elbo}
\end{align}

In this sparse implementation, the variational distribution $q_{\phi}(\textbf{z})$ has been made less complex, as it now tracks the inducing points $\textbf{u}$ which are chosen to consist of fewer points than $\mathbf{f}$. For $M$ inducing points, this reduces the number of free parameters in our model to $\mathcal{O}(M^{2})$ and hence reduces training complexity to $\mathcal{O}(NM^{2})$. However, to estimate the ELBO, we need to draw $\mathbf{f}$ from the joint distribution $q_{\theta,\phi}(\mathbf{f},\textbf{u})$ or by sampling $\mathbf{u}$ first and then $\mathbf{f}$ from the conditional GP with PDF $p(\mathbf{f} | \textbf{u})$. The \texttt{NumPyro} implementation of SVGP forces us to take the latter approach. In doing so, we incur the cost of conditioning the GP every time we sample $\mathbf{f}$ given $\mathbf{u}$. This increases the duration of training, but not its complexity. 

We further reduce training complexity by leveraging the fact that $\textbf{f}$ is only required to compute the log-likelihood term of the ELBO (in the term $p_{\theta}(\textbf{y}, \textbf{z}) = p(\textbf{y} | \textbf{z}) p_{\theta}(\textbf{z})$ of Eq.~\eqref{eq:svgp_elbo}). Rather than using all $N = 4,525$ of pixels $\mathbf{x}$, we estimate the log-likelihood term of the ELBO on a subsample of length $K$, randomly selected at each step of the SGD optimization: 
\begin{equation}
    \sum_{k=1}^{N} \log p(y_k | \mathbf{z}) \approx \frac{N}{K} \sum_{k \in \mathcal{I}_{K}} \log p(y_k | \mathbf{z}),
\end{equation}
where $\mathcal{I}_{K}$ is a randomly generated set of $K$ indices of the data. This procedure is valid, provided that our observations (the binned counts) are statistically independent. This is assumed in our joint Poisson likelihood model. Subsampling reduces training complexity to $\mathcal{O}(KM^{2})$ and boosts the speed of the fit by reducing the size of the conditional GP $p_{\theta}(\textbf{f} | \textbf{u})$ required for drawing samples from $q_{\theta, \phi}(\textbf{f})$.

The inducing point locations $\mathbf{x_u}$ are parameters built into our model. They are distributed evenly throughout the $20^{\circ}$ disk centered at the GC so that our GP interpolates -- as opposed to extrapolates -- the values $\textbf{f}$ from $\textbf{u}$.\footnote{Extrapolated values will not be reliable because the effect inducing points have on neighboring function values diminishes with distance.} In this way, the variational distribution for distant input points will be distributed according to the GP prior. Placing the inducing points roughly equidistant from one another also helps ensure that every $\mathbf{x_u}$ has a comparable effect on inference. In the absence of prior knowledge of the GCE morphology, there is no reason to allow a specific region of the inner disk to have a stronger effect on inference than other regions. To satisfy these two requirements, we distribute the inducing points in a sunflower configuration (see Appendix~\ref{app:inducing} for more details). In this work, we take 300 inducing points, corresponding to an average spacing of roughly $1.15^{\circ}$.\footnote{This sets a minimum length scale for the kernel length scale $\rho$ since the inducing points become roughly uncorrelated for kernel length scales smaller than this value and no meaningful functional behavior can be learned by the SVGP.} 

To summarize, our SVGP model is:
\begin{align}
    {\rm Counts:}~ & \quad \textbf{y} \sim {\mathrm {Pois}}[ \lambda(\mathbf{x}) ] \nonumber \\
    {\rm Rate:}~ & \quad \lambda(\mathbf{x}) = \sum_{j} S_jT_j(\mathbf{x}) + \exp(f(\mathbf{x})) \nonumber \\
    \nonumber \\
    {\rm Objective:}~ & \quad {\rm ELBO}(\theta, \phi) = \mathbb{E}_{q_{\phi}(\mathbf{z})}[\mathbb{E}_{p_{\theta}(\mathbf{f}|\mathbf{u})}[\log p(\textbf{y}|\mathbf{f}, S_j)]] + \mathbb{E}_{q_{\phi}(\mathbf{z})}[\log p_{\theta}(\textbf{z}) - \log q_{\phi}(\mathbf{z})] \nonumber \\
    {\rm Variational \ Distribution:}~ & \quad q_{\phi}(\tilde{\mathbf{z}}) = \mathcal{N}(\mathbf{m}, \mathbf{\Sigma}) \nonumber \\
    {\rm Latent \ Variables:}~ & \quad \textbf{z} = \mathbf{t}(\tilde{\mathbf{z}}) = (\mathbf{u}, S_j) \nonumber \\
    {\rm Sparse \ \& \ Variational \ GP:}~ & \quad q_{\theta, \phi}(\mathbf{f}, \mathbf{u}) = p_{\theta}(\mathbf{f} | \mathbf{u}) \ q_{\phi}(\mathbf{u}). \nonumber
\end{align}
The GP prior on the function $f$ and the prior on the template normalizations $S_j$ are listed in Tab.~\ref{tab:priors}.

\subsection{Model Validation with Simulated Data}  \label{sec:syndata}
Before training the GP model on real data, we assess model performance using a single synthetic dataset created with template models for all the gamma-ray sources (including a simulated GCE). We generate our synthetic data using all the templates we introduced in Section~\ref{sec:templates}: isotropic emission, point sources, the \Fermi{} bubbles, $\pi^{0}$+bremsstrahlung, ICS, and the GCE. 
The diffuse templates are taken from Model O. We simulate the GCE using the \texttt{coleman2019} template and an NFW template with $\gamma = 1.2$ and $R_{S} = 20 \ $kpc. To obtain realistic normalizations for these templates, we fit the observed \Fermi{} data $\mathbf{y}$ with the fixed templates and generate the synthetic data using the mean posterior sample of the template model. The fit is performed using SVI as presented above using only templates for each component, including for the GCE (that is, no GP is involved). The results of this fit are shown in Appendix~\ref{app:performance} (in particular see Fig.~\ref{fig:syndata}). 

From these templates (with normalizations inferred from the gamma-ray data), we generate a synthetic set of gamma-ray observations by Poisson-sampling the combined template map. We then fit the GP model to the Poisson pseudodata. For this GP fit, we use the constrained MVN as our approximate posterior, 15,000 optimization steps, $N_p =8$ for the ELBO sample averaging, a subsample size $K=500$, and an Adam optimizer~\cite{kingma2017, dozat2016} with a scheduled learning rate. The learning rate starts with an initial value of 0.005, warms up by increasing linearly to a peak value of 0.05 for 1,000 steps, remains at the peak value until step 5,000, and then exhibits exponential decay for the rest of the fit with a decay ``time constant" of 1,000 steps. The warm-up phase ensures that the optimizer does not enter a pathological region of the parameter space, the constant phase ensures the fit converges in a reasonable amount of time, and the decay phase ensures that the fitter gets progressively closer to the optimal state. Given these settings, each fit takes $\sim$4 minutes and drawing $\mathcal{O}(10^{4})$ samples takes ${\cal O}($seconds) when running on a NVIDIA GeForce GTX 1080 Ti GPU. 

In this example, our model fits the simulated data with the same background templates used to generate the data (see Appendix~\ref{app:mismodel} for fits where we fit the background components with different templates than those used to generate the data). Therefore, the only gamma-ray source whose morphology can differ from the simulated truth is the GCE, which is fit by the flexible GP rather than a template. 

\begin{figure}[t!] 
    \centering
    \includegraphics[width=0.8\columnwidth]{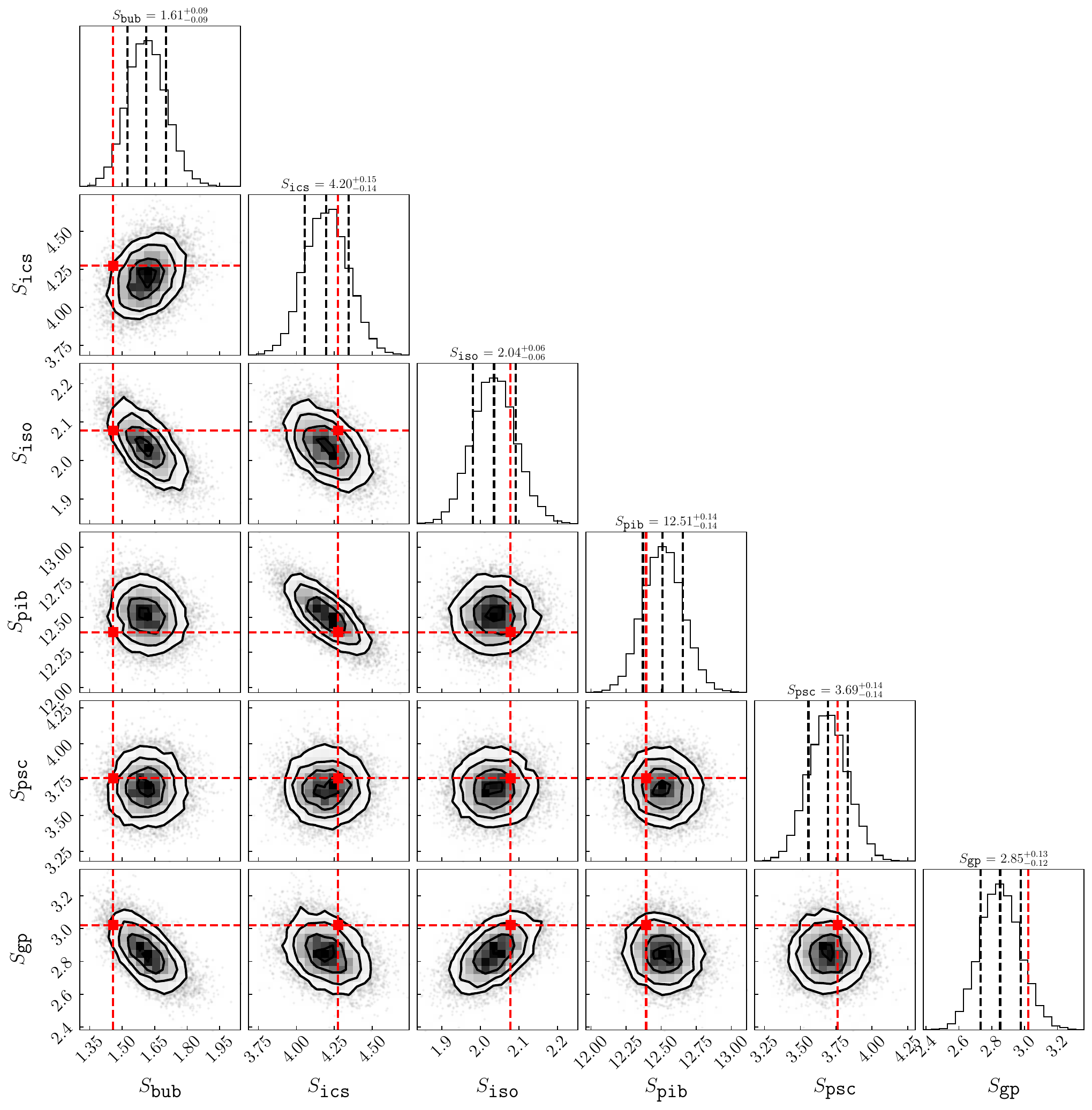}
    \includegraphics[width=0.8\columnwidth]{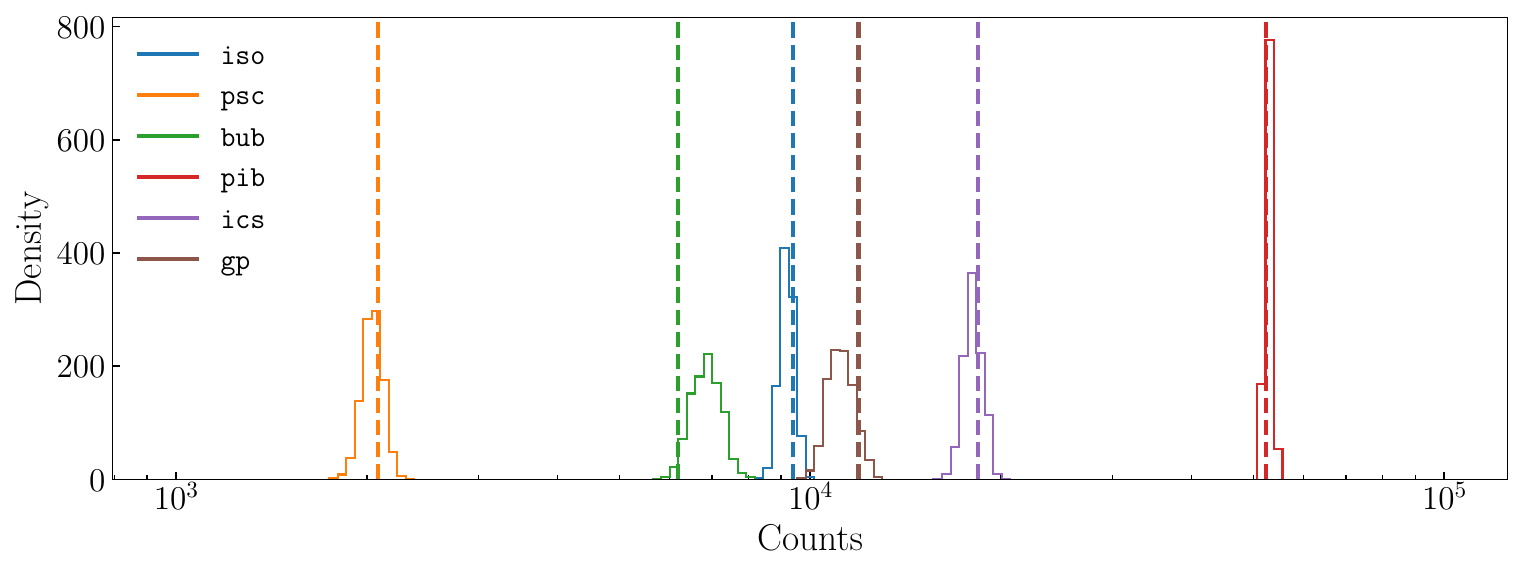}
    \caption{Posterior distribution after fitting model to synthetic data. \textit{(Upper Panel)} The posterior distribution of $S$-normalizations. True values are denoted as red dashed lines. \textit{(Lower Panel)} The expected total number of counts of each gamma-ray source in the Inner ROI and are binned logarithmically. True normalizations are denoted by dashed lines.} \label{fig:post_norms}
\end{figure}

The posterior distribution for each of the templates is summarized in Fig.~\ref{fig:post_norms}. The corner plot shows the approximate posterior distribution of the template normalizations and the posterior distribution of the expected total number of counts over the Inner ROI corresponding to each template. The template normalization posteriors (lower panel) are in agreement with their true values, as are the GP posterior and the true GCE. The biases in the fit are partly due to Poisson noise inherent in the pseudodata. We show that the bias associated with the Poisson noise averages out over repeated sampling of the synthetic data in Appendix~\ref{app:accuracy}.

\begin{figure}[t!] 
    \centering
    \includegraphics[width = 0.8\columnwidth]{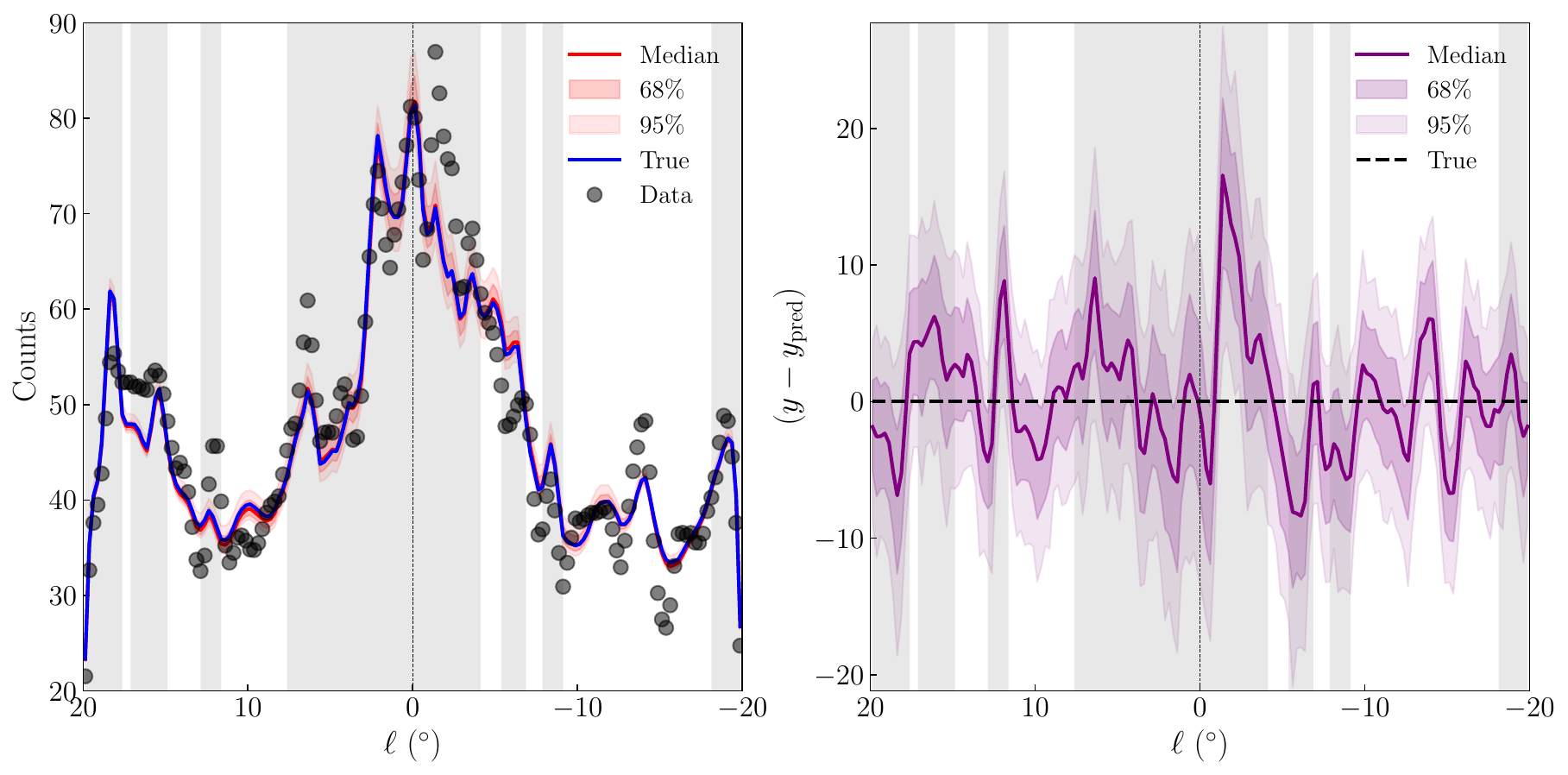}   
    \caption{Horizontal 1D slice of posterior samples of counts on Cartesian bins with lengths of $0.25^{\circ}$ for our fit to synthetic data. This slice is situated at $b = 3.375^{\circ}$. Gray bands in the background indicate where masking is present. \textit{(Left)} Posterior predictive samples of the total rate of emission compared to the true total rate of emission and the measured counts per bin. \textit{(Right)} Posterior predictive samples of the count residuals per bin $(y - y_\mathrm{pred})$. As can be seen, the observed data is a likely outcome of the model.} \label{fig:sim_1d_slices}
\end{figure}

To demonstrate that our GP model captures the morphological features of the data, we plot in Fig.~\ref{fig:sim_1d_slices} the posteriors in 1D slices through the $20^{\circ}$ disk in $(\ell,b)$ space. In this and similar plots in this work, we use a bilinear interpolation to project the posterior predictive samples from HEALPix bins to Cartesian (square) bins with side lengths of $0.25^{\circ}$. 
For comparison purposes, throughout this work we will show slices of the posteriors through either the $b = 3.375^{\circ}$ horizontal line or $\ell = -3.375^{\circ}$ vertical line. In the left panel of Fig.~\ref{fig:sim_1d_slices}, we show a horizontal slice of the posterior distribution of the total rate $\lambda(\textbf{x})$ along with the synthetic data and the underlying true rate. The right panel of Fig.~\ref{fig:sim_1d_slices} shows the residual between Poisson samples drawn from our posterior predictive distribution of the total rate $\lambda(\textbf{x})$ and the raw simulated counts $\mathbf{y}$. As can be seen, the simulated data is a likely outcome of the model, as expected given that our fitting procedure maximizes the likelihood of the data given the model.

As our GP prior is defined over the entire continuous $20^{\circ}$ disk containing the Inner ROI, we can use Eq.~\eqref{eq:qf} to infer a distribution of the GCE log-rate across the entire disk. In particular, we can interpolate the GCE emission in the pixels within $20^{\circ}$ from the GC that were masked in the Inner ROI, allowing us to make some inference of the morphology of the GCE in this region. 

\begin{figure}[t!] 
    \centering
    \includegraphics[width = 0.67\columnwidth]{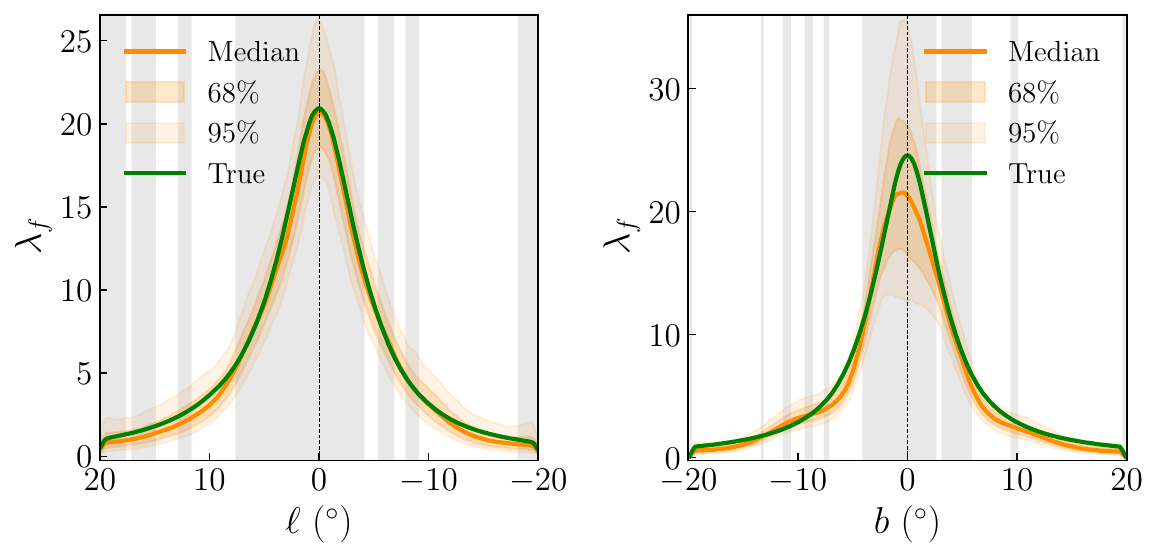}
    \includegraphics[width = \columnwidth]{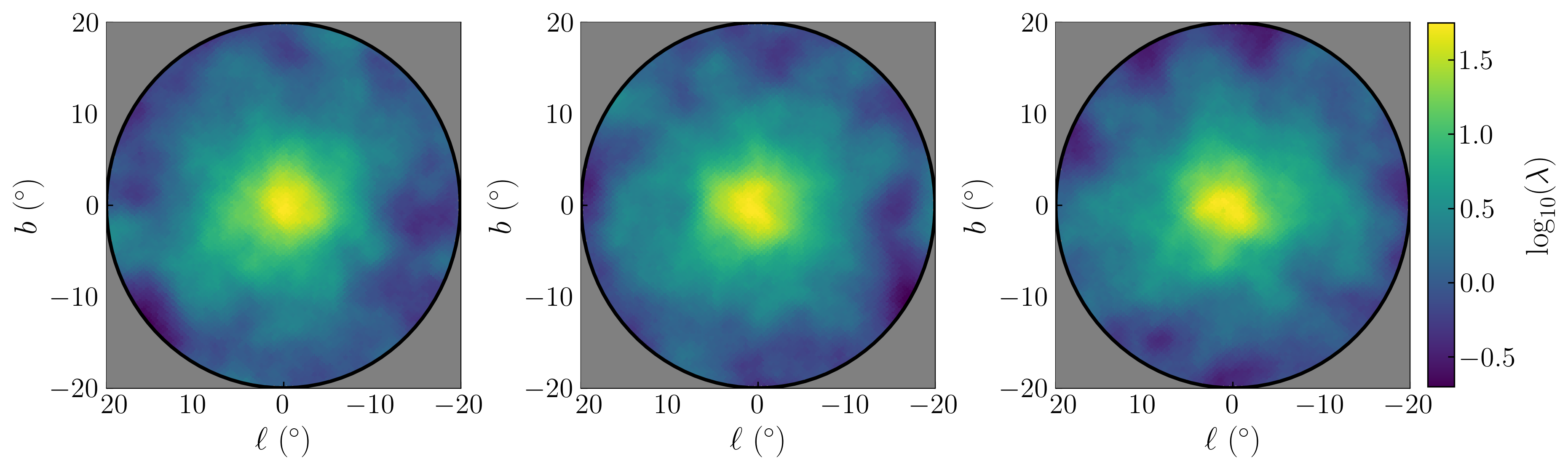}
    \caption{GP morphology for model fit to synthetic data composed of an \texttt{nfw} template ($\gamma = 1.2$) and a \texttt{coleman2019} bulge. \textit{(Top Left)} Horizontal 1D slice of the GP posterior distribution of counts on Cartesian bins with lengths of $0.25^{\circ}$ for our fit to synthetic data. This slice is situated at $b = 3.375^{\circ}$. \textit{(Top Right)} Vertical 1D slice of the GP posterior distribution of counts on Cartesian bins with lengths of $0.25^{\circ}$ for our fit to synthetic data. This slice is situated at $\ell = -3.375^{\circ}$. Gray bands in the background indicate where masking is present. \textit{(Bottom Row)} Three posterior samples of the GCE rate.} \label{fig:gce_sim_samples}
\end{figure}

In Fig.~\ref{fig:gce_sim_samples}, we visualize the alignment of the GP posterior with the true morphology of the templates used to generate the synthetic data, using the 1D slices of the GCE posterior rate $\lambda_{f}(\textbf{x})$. In the leftmost panel of Fig.~\ref{fig:gce_sim_samples} we plot the horizontal slice ($b = 3.375^{\circ}$) through the Inner ROI, and to the right of that panel we show the vertical slice ($\ell = -3.375^{\circ}$). 

In addition to comparing slices of the GP posterior to the simulated data, we study the GP's morphology across the entire $20^{\circ}$ disk centered at the GC. To better emphasize the variation in the best-fit GP's morphology, we show three samples randomly drawn from the GP posterior in the remaining three panels of Fig.~\ref{fig:gce_sim_samples}. The GP posterior is highly correlated at scales smaller than the optimal length scale $\rho \sim 10^{\circ}$, as previously discussed.

Our results show that the GP accurately interpolates from the unmasked pixels (Inner ROI) to the masked pixels (the entire $20^{\circ}$ disk). This is a consequence of the GP's kernel length scale being fit to $\rho \sim 10^{\circ}$, implying that training the GP on the unmasked pixels was enough for the GP to learn that the GCE rate is highly correlated at scales smaller than $\sim 10^{\circ}$. These correlations suppress random variations of the GP's samples over scales smaller than $\sim 10^{\circ}$, and so the GP can accurately interpolate the values in the masked regions. For example, the vertical posterior slice in Fig.~\ref{fig:gce_sim_samples} shows that the GP accurately interpolates the rate inside the band mask at $|b| \leq 2^{\circ}$ -- the largest mask on the disk. Though accurate, the lack of data in these regions results in high predictive variance in this region. 

Though slices of the posterior crossing the center are not shown, the model underestimates the signal at the center due to the model's inability to learn the sharp central cusp of the \texttt{nfw} component of the GCE, a feature that is hidden by the mask and not present anywhere else. In general, the features that our model can learn are limited to features present in the data.

Lastly, we note that the underlying GCE template components used to generate the synthetic GCE can be recovered from the GP. We demonstrate this in Appendix~\ref{app:kl-poisson}, where we fit an accurate GCE template model (\textit{i.e.,} describing the GCE with the \texttt{coleman2019} bulge template and the \texttt{nfw} template) to the best-fit GP posterior of counts (see left panel of Fig.~\ref{fig:poiss_summary_gp}). 

A more detailed assessment of the model's performance is provided in Appendix~\ref{app:performance}. In Appendix~\ref{app:accuracy}, we continue the GP model validation by comparing the best-fit GP to the GCE template used to generate the synthetic data~(Appendix~\ref{app:extract}) and fitting to many Poisson draws of the synthetic emission rate to remove the effect of Poisson noise and expose our model's systematics (Appendix~\ref{sec:poiss}). In Appendix~\ref{app:stability}, we assess how our results change when we either change the GP or fit to different synthetically-generated datasets. Appendix~\ref{app:kernels} considers different GP kernels, Appendix~\ref{app:Nu} different values for the number inducing points,  Appendix~\ref{app:blg_scan} synthetic datasets made using different GCE models, Appendix~\ref{app:blg_scan} fits to synthetic datasets with no GCE components, and Appendix~\ref{app:ps} fits using different point source masks. In Appendix~\ref{app:mismodel}, we study how mismodeling the diffuse emission affects our interpretation of the GCE. Appendix~\ref{sssec:mismodel-single} compares fits employing a single diffuse emission model, Appendix~\ref{sssec:mismodel-multi} compares fits using two erroneous models to study how additional flexibility in the diffuse modeling may improve our fits, and Appendix~\ref{sssec:mismodel-all} compares fits using multiple diffuse models, including the model correctly describing the diffuse emission in the synthetic data.

\section{Results on \Fermi Data} \label{sec:dataresults}

\subsection{Baseline Model}

\begin{figure}[t!] 
    \centering
    \includegraphics[width=\columnwidth]{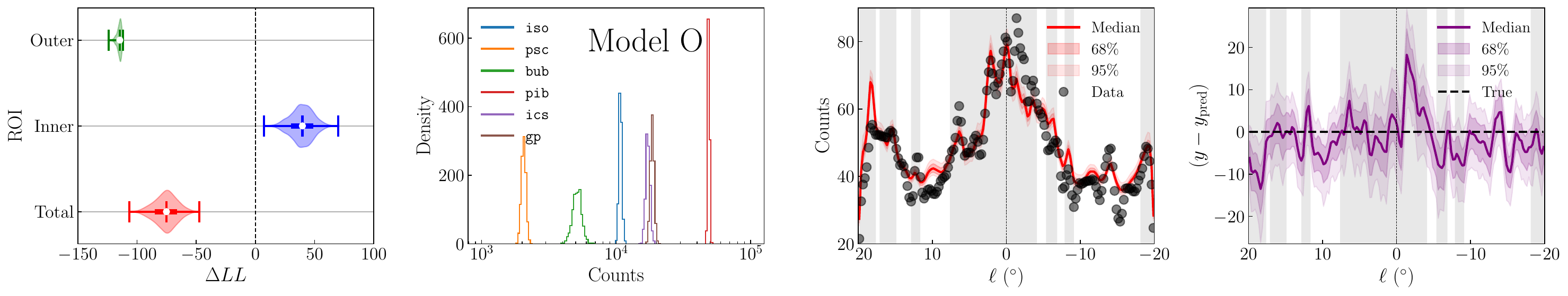}
    \includegraphics[width=\columnwidth]{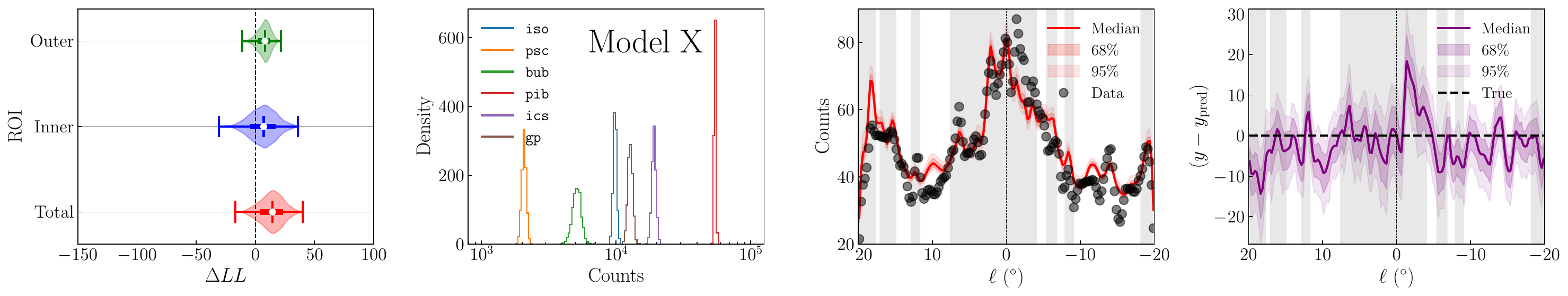}
    \includegraphics[width=\columnwidth]{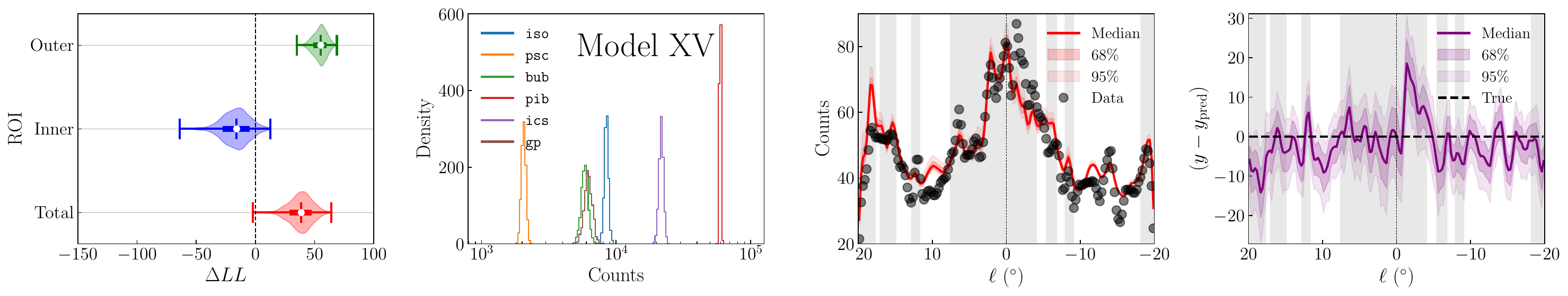}
    \includegraphics[width=\columnwidth]{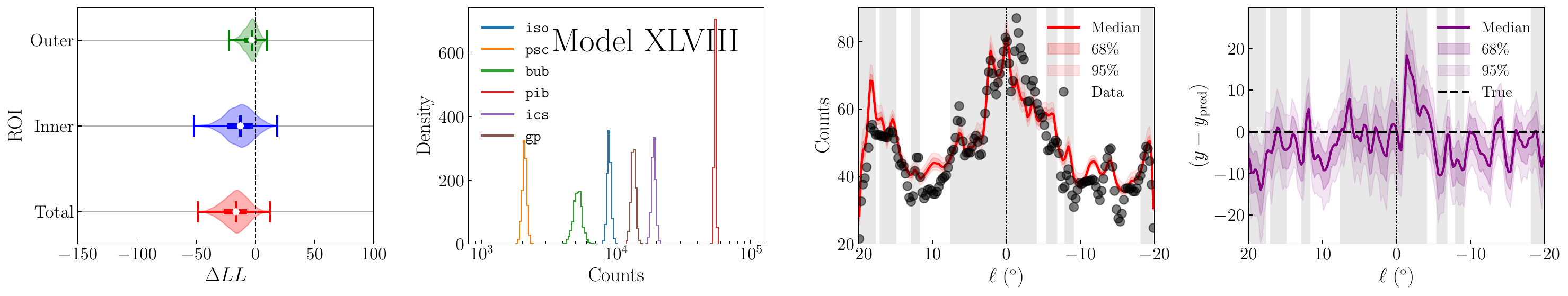}
    \includegraphics[width=\columnwidth]{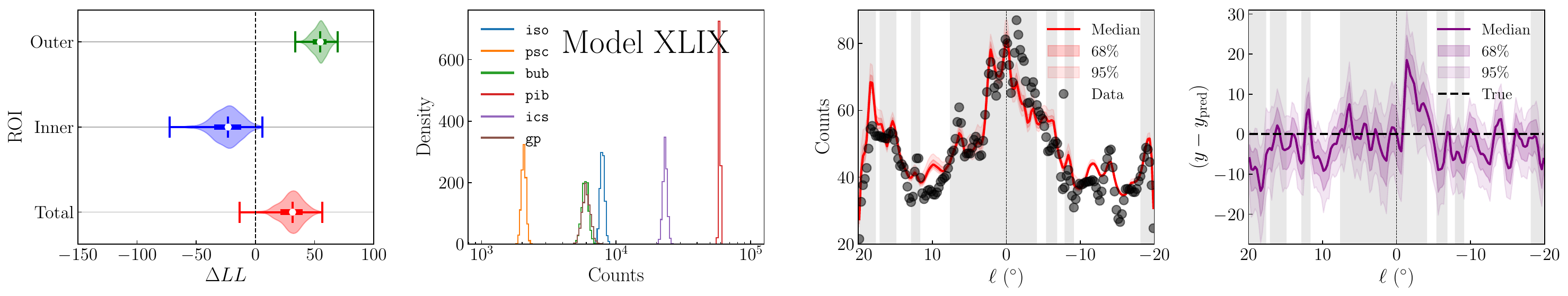}
    \includegraphics[width=\columnwidth]{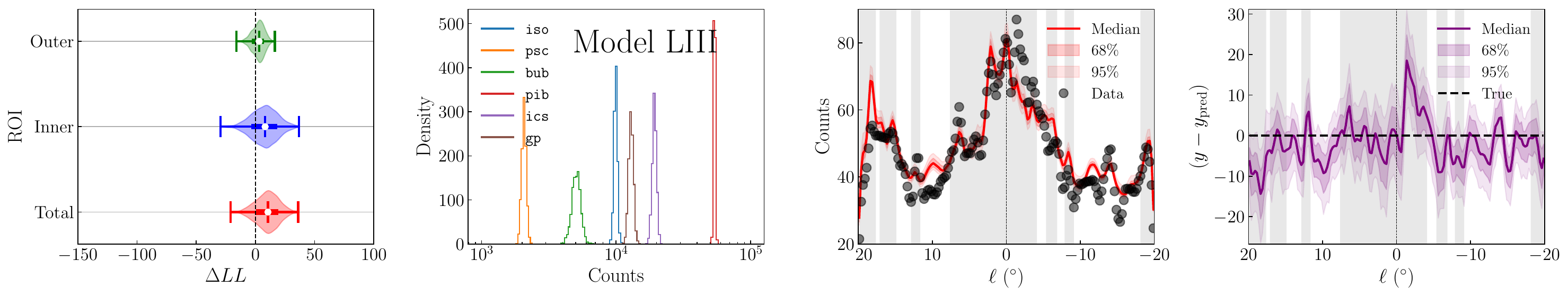}
    \caption{Summary of fits corresponding to different diffuse modeling (\texttt{pib} and \texttt{ics}). Each row corresponds to a fit using either Model O \textit{(Top Row)} or one of the top five \textsc{Galprop}-based models from Fig.~12 in Ref.~\cite{Cholis_2022}. Model names are located at the top of the plots in the second column. \textit{(Column 1)} Posterior predictive distribution of the log-likelihoods (relative to the average log-likelihood corresponding to all posterior samples from all the models $\langle {\cal L} \rangle$) of the observed data in different regions of the sky: the Complete ROI, the Inner ROI, and the Outer ROI. 
    \textit{(Column 2)} Posterior distributions of the total expected number of counts of the various gamma-ray sources over the Inner ROI. \textit{(Column 3)} Horizontal 1D slice at $b = 3.375^{\circ}$ of posterior samples of Cartesian-binned counts 
    for our fit to synthetic data. Gray bands in the background indicate where masking is present. \textit{(Column 4)} Posterior predictive samples of the total rate of emission compared to the true total rate of emission and the measured counts per bin. 
    } \label{fig:galprop_spread}
\end{figure}

We now use our model -- validated in the previous section -- to fit the real {\it Fermi} data. The fit settings are identical to those used in the fit to synthetic data, described in Section~\ref{sec:syndata}. We provide a summary of the model performance in the top row of Fig.~\ref{fig:galprop_spread} (labeled ``Model O''). The first column provides a comparison of the likelihood of the model fit in comparison with other models, which will be discussed (along with other template models, which comprise the rest of the rows in Fig.~\ref{fig:galprop_spread}) in Section~\ref{sec:template_comparison}. The second column of Fig.~\ref{fig:galprop_spread} shows the posterior normalizations of the templates and GP over the Inner ROI (2nd column). We also show the horizontal slice of the posterior and data through $b = 3.375^{\circ}$ (3rd column), and the posterior predictive distribution of residuals (4th column).

\begin{figure}[t!]  
    \centering
     \includegraphics[width=\columnwidth]{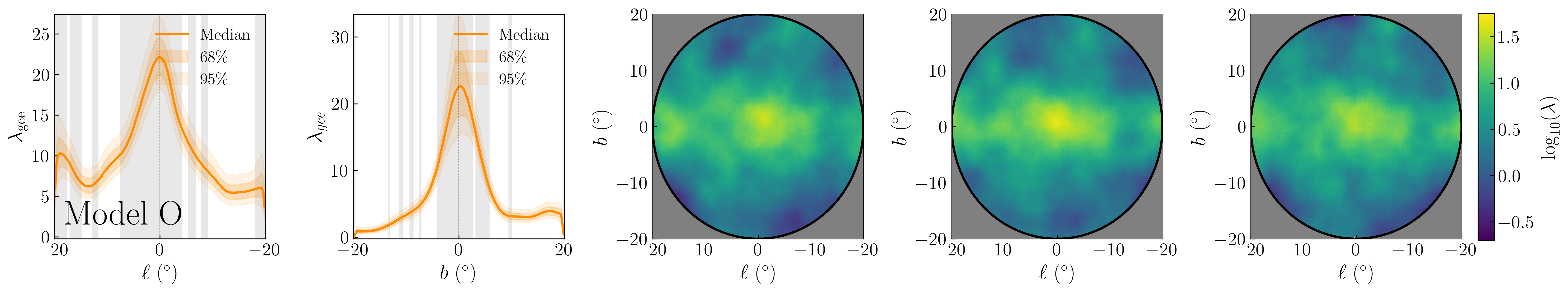}
    \includegraphics[width=\columnwidth]{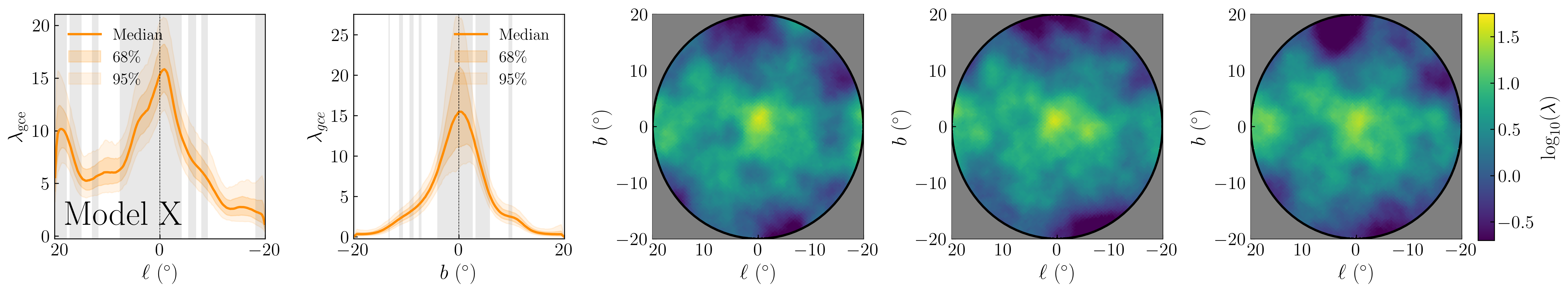}
    \includegraphics[width=\columnwidth]{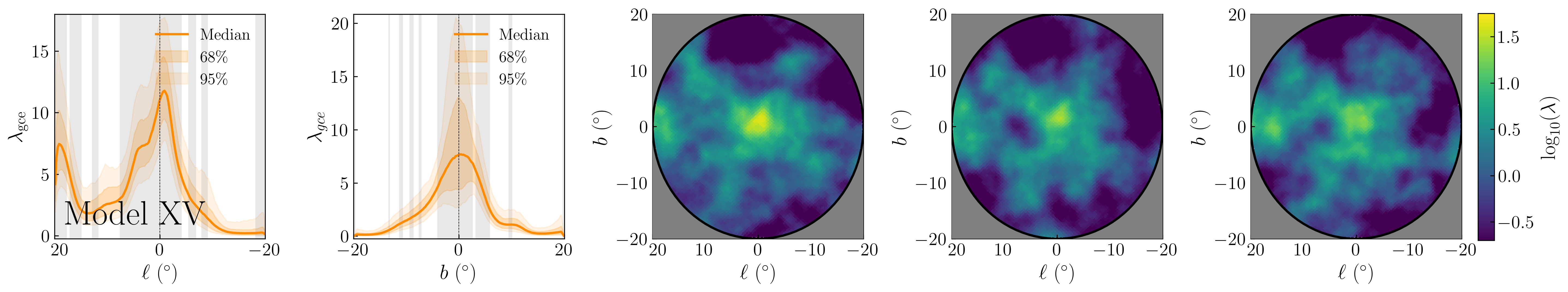}
    \includegraphics[width=\columnwidth]{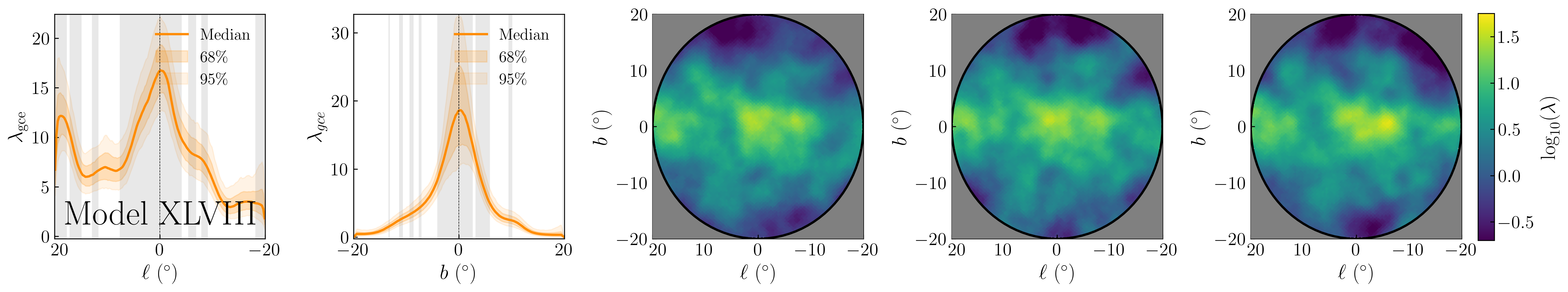}
    \includegraphics[width=\columnwidth]{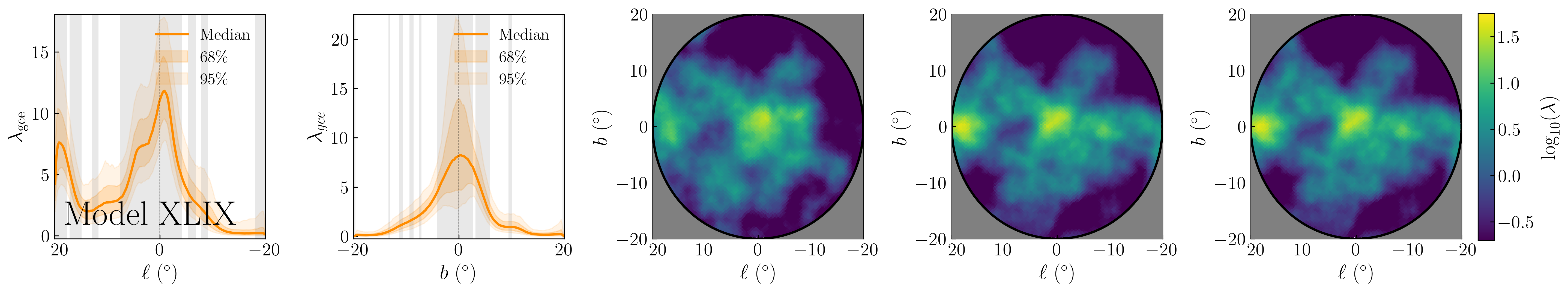}
    \includegraphics[width=\columnwidth]{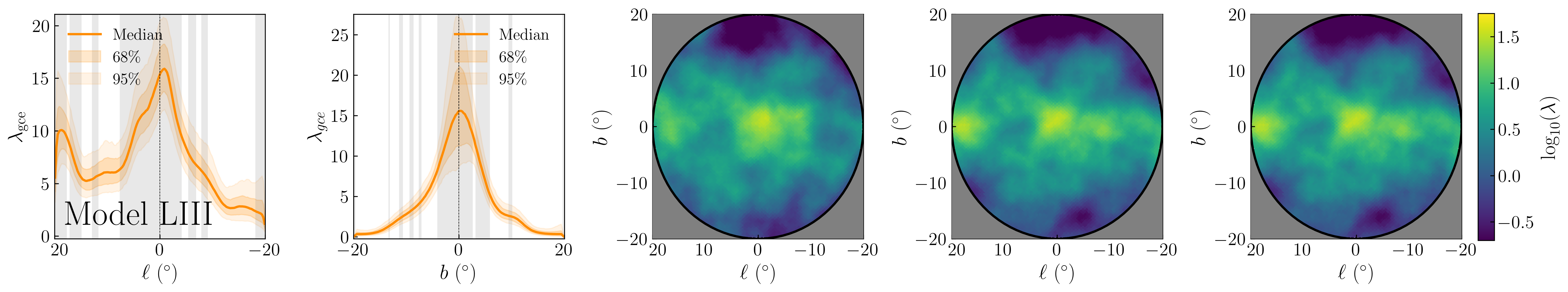}
    \caption{Fits to the \textit{Fermi} data using Model O \textit{(Top Row)} or one of the top five best-performing \textsc{Galprop}-based models from Fig.~12 in \cite{Cholis_2022} \textit{(Rows 2-6)}. Ordering of models is as in Fig.~\ref{fig:gce_sim_samples}, and corresponds to the same fits. \textit{(Column 1)} Horizontal 1D slice at $b = 3.375^{\circ}$of the GP posterior distribution of counts on Cartesian bins.  \textit{(Column 2)} Vertical 1D slice at $\ell = -3.375^{\circ}$ of the GP posterior distribution of counts on Cartesian bins.  Gray bands in the background indicate where masking is present. \textit{(Columns 3-5)} Three posterior samples of the GCE rate. \label{fig:data-gp-samples}}
\end{figure}

We show the inferred GCE morphology of our fit to Model O in the top row of Fig.~\ref{fig:data-gp-samples} (the later rows show the morphology for other template models, and are discussed in Section~\ref{sec:template_comparison}). As we did for the fit to synthetic data, we show the GP posterior at both a horizontal slice ($b = 3.375^{\circ}$) and a vertical slice ($\ell = -3.375^{\circ}$). Additionally, we show three samples from the GP posterior to demonstrate the variation in the GP morphology. As expected for a GCE signal, our GP-derived excess peaks towards the Galactic Center. However, the GP has features that are not usually associated with the GCE: in particular the bright localized source to the far left of the Inner ROI ($\ell \sim 20^{\circ}$ and $b \sim 0^\circ$), and the diagonal ``arm'' extending far out towards the northwest from the GC. Note that the bright source near $b \sim 0^\circ$ would largely be covered by the Galactic disk mask, and so our inferred morphology is an interpolation from the gamma-ray data from outside the disk.

These additional features are present because our GP is a flexible model capable of capturing the mismodeling that occurs when fitting the background with fixed spatial templates. 
The presence of an Outer ROI constrains the templates in our fit, allowing the GP to fill in the signal in the Inner ROI that cannot be accommodated by the templates. As our fit takes both the Inner and Outer ROIs into account simultaneously, so the template fits are not completely determined by their ability to fit the Outer ROI and the GP is not completely determined by its ability to correct for template mismodeling. In this way, the GP corrects background mismodeling in addition to describing unknown physical sources contributing to the measured gamma-ray signal. The effects of mismodeling, as demonstrated on synthetic data, are more carefully considered in Appendix~\ref{app:mismodel}. 

\subsection{Systematic Uncertainties Associated with Diffuse Emission Modeling} \label{sec:template_comparison}

In addition to Model O -- a ring-based model of the diffuse emission by $\pi$+bremsstrahlung and ICS -- to capture the systematic uncertainties associated with our particular choice of diffuse model, we fit our GP model to the data using the five best-performing \textsc{Galprop}-based models from Ref.~\cite{Cholis_2022} (Models X, XV, XLVIII, XLIX, and LIII from Fig.~12 of that reference). The fits to these five additional models make up the rest of the rows in Fig.~\ref{fig:galprop_spread} and Fig.~\ref{fig:data-gp-samples}.

The first column in Fig.~\ref{fig:galprop_spread} compares the goodness-of-fit of each GP model by making use of the posterior predictive distribution of the log-likelihood $q_{\cal L}({\cal L})$. The log-likelihood is defined as 
\begin{equation} \label{eq:ll_def}
    {\cal L}(\textbf{z}) = \log p(\textbf{y} | \textbf{z}),
\end{equation}
and from this, the distribution of ${\cal L}$ is calculable in terms of the best-fit approximate posterior $q_{\theta,\phi}(\textbf{z})$:
\begin{equation} \label{eq:ll}
    q_{\cal L}({\cal L}) = \int q_{\theta, \phi}(\textbf{z}) \ \delta({\cal L}-{\cal L}(\textbf{z})) \ d\textbf{z}.
\end{equation}

In practice, the distribution is estimated by drawing samples of $\textbf{z}$ from the approximate posterior $q_{\theta,\phi}(\textbf{z})$, calculating the log-likelihoods corresponding to each sample ${\cal L}(\textbf{z})$, and generating an empirical distribution from the samples. 

Rather than comparing the log-likelihood of each model directly, in Fig.~\ref{fig:galprop_spread} we compare the difference between the log-likelihood of a particular model to the log-likelihood averaged over both $\textbf{z}$ and model choice: $\Delta {\cal L} = {\cal L} - \langle {\cal L} \rangle$. Under this definition, models with a distribution of log-likelihoods greater than zero fit the data better than the mean expectation across all models and models with $\Delta {\cal L} <0$ perform worse than the average.

The first column of Fig.~\ref{fig:galprop_spread} shows $\Delta {\cal L}$ for each model, calculated using three different sets of data: the Outer ROI, the Inner ROI, and both ROIs (note that these choices correspond to different $\langle {\cal L} \rangle$ as they are computed using different sets of data). Model O is particularly interesting because it gives the best fit to the Inner ROI, but it fits very poorly to the Outer ROI. This is expected because Model O was constructed to fit the data in a ROI with radius $25^\circ$; its lack of calibration to the Outer ROI explains its poor performance in this region. Models XV and XLIX are similar in that they give the best fit to the Outer ROI and the total signal, but perform worse than the mean expectation in the Inner ROI. In these cases, the alignment between the Outer ROI and the Complete ROI is expected because the Outer ROI has roughly $1.75\times$ the amount of pixels that the Inner ROI has, and so it contributes more than the Inner ROI to the log-likelihood measured in the Complete ROI.

The second column of Fig.~\ref{fig:galprop_spread} shows the distribution of the total expected number of counts across the Inner ROI for all models. The GP posterior varies more with respect to model choice than the other templates, with the ICS posterior varying the second-most. This suggests that the GP and the ICS models share the highest degree of degeneracy compared to other possible pairings of gamma-ray sources. 

The third and fourth columns of Fig.~\ref{fig:galprop_spread} illustrate how each model's total rate compares to the measured data in a horizontal 1D slice. As we see, though the gamma-ray sources vary in relative normalization, the total rates predicted by all models in the slice are consistent. There are subtle deviations in the morphology of the total rate, but these deviations are mainly a result of the assumed \texttt{pib} template, as it accounts for most of the predicted signal across all the fits.

The morphology of the GCE inferred by the GP for each diffuse model is summarized in Fig.~\ref{fig:data-gp-samples}. Across all the fits, the GP posteriors share morphological features. Among the most prominent of these are the roughly central peak near the GC, the bright localized source to the far left ($\ell \sim 20^{\circ}$ and $b\sim 0^\circ$), and the diagonal arm stretching far out towards the northwest from the GC. Note that some of these features (the shape of the central peak, some of the emission near $b\sim 0^\circ$, and the source at $\ell \sim 20^\circ$) appear at least partly in masked regions that require interpolation from the GP.

To cast these results in terms of bulge plus NFW template models, we fit a GCE template model made up of an \texttt{nfw} and a bulge component to each of the GP posterior distributions corresponding to different diffuse model templates over the unmasked pixels, using the method described in Appendix \ref{app:kl-poisson}.\footnote{Due to the lack of flexibility of the template model, we have found this approach to be consistently overconfident. However, it is theoretically motivated and can recover the true gamma-ray sources, provided we fit many times to different Poisson draws of the true sources and our template models are accurate.} 

In performing a template fit to the GP, the NFW $\gamma$ parameter is a latent variable with a uniform prior from 0.2 to 2. For our bulge template, we allow all five templates from Fig.~\ref{fig:gce_temps} to contribute to the GCE in a linear combination. The sum of the bulge templates has its own $S_{\rm blg}$ normalization parameter, while each individual bulge template has a relative normalization $\theta_{\rm blg}$, whose sum is normalized to 1. We choose a prior for these parameters that penalizes mixed template models, imposing the prior belief that only one bulge model can be realized. See Appendix~\ref{app:multi-temp} for more details.

Ideally, we would perform this fit across the entire Inner ROI to train the GCE template model. However, the bulge plus NFW model can only describe a gamma-ray source that is bright around the GC and decays smoothly. It cannot describe additional features such as the bright, disconnected source that all the GP fits find near $\ell \sim 20^{\circ}$, and so would fit the regions containing such features poorly. To mitigate this mismodeling, we fit the GCE template model to the GP using unmasked pixels only within $10^{\circ}$ from the GC. 

\begin{figure}[t!] 
    \centering
    \includegraphics[width = \columnwidth]{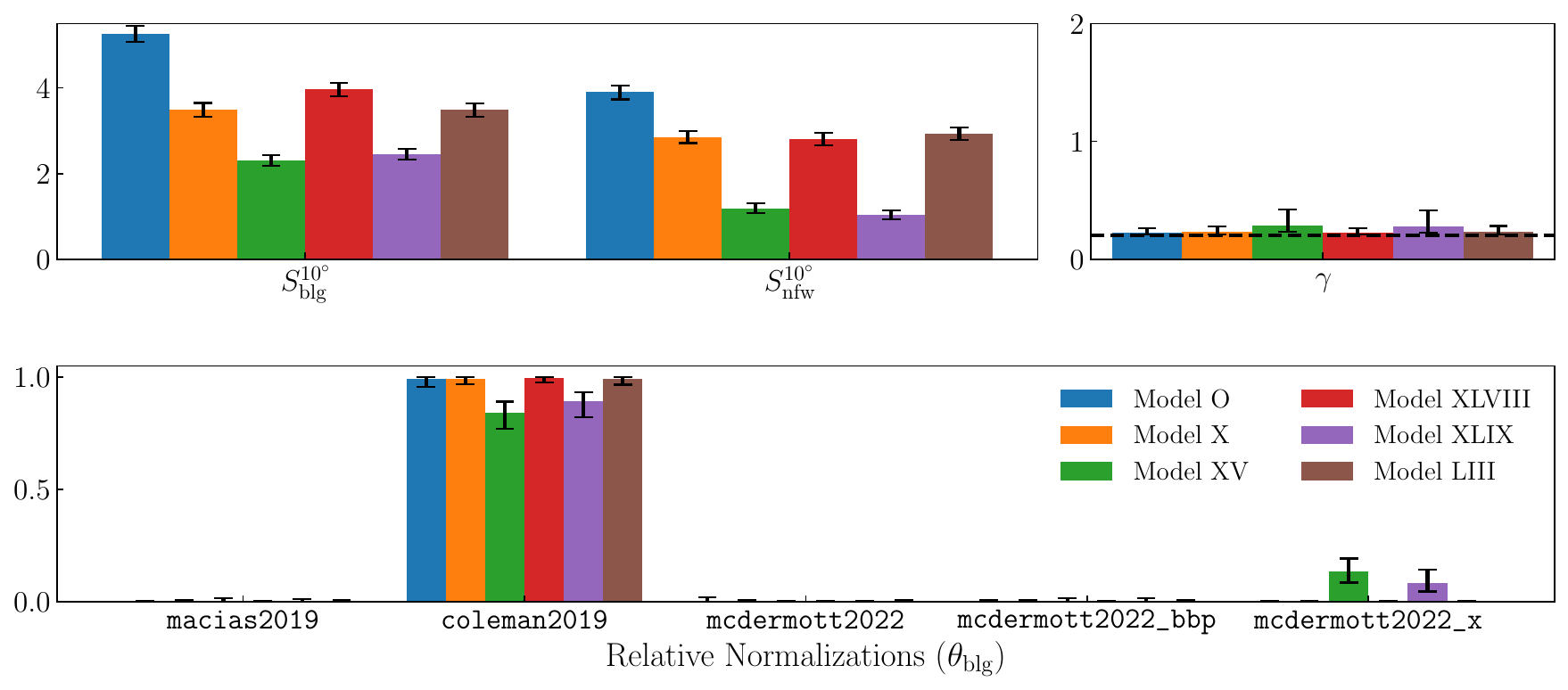}   
    \caption{Results of fitting an aggregate GCE template model to the GP posteriors obtained from fits to data (\textit{i.e.,} the GPs from Fig.~\ref{fig:galprop_spread}) within $10^\circ$ of the GC. The aggregate GCE template model describes the GCE as a linear combination of an \texttt{nfw} template (with $\gamma$-parameter which can be fit to data) and the five bulge templates introduced. Each color corresponds to a fit to a specific GP posterior, differing by the diffuse model (Model O or the top five models from Fig.~12 in Ref.~\cite{Cholis_2022}) used to fit the corresponding GP model to the data. \textit{(Top Left)} Median normalization posterior samples of the nfw and bulge templates with 68\% probability regions denoted by error bars. \textit{(Top Right)} Median normalization posterior sample of the NFW $\gamma$ parameter with 68\% probability regions denoted by error bars. \textit{(Bottom)} Median posterior samples of the relative normalizations of each candidate bulge model included in the fit, with 68\% probability regions denoted by error bars.} \label{fig:blg-fit}
\end{figure}

\begin{figure}[t!] 
    \centering
    \includegraphics[width = \columnwidth]{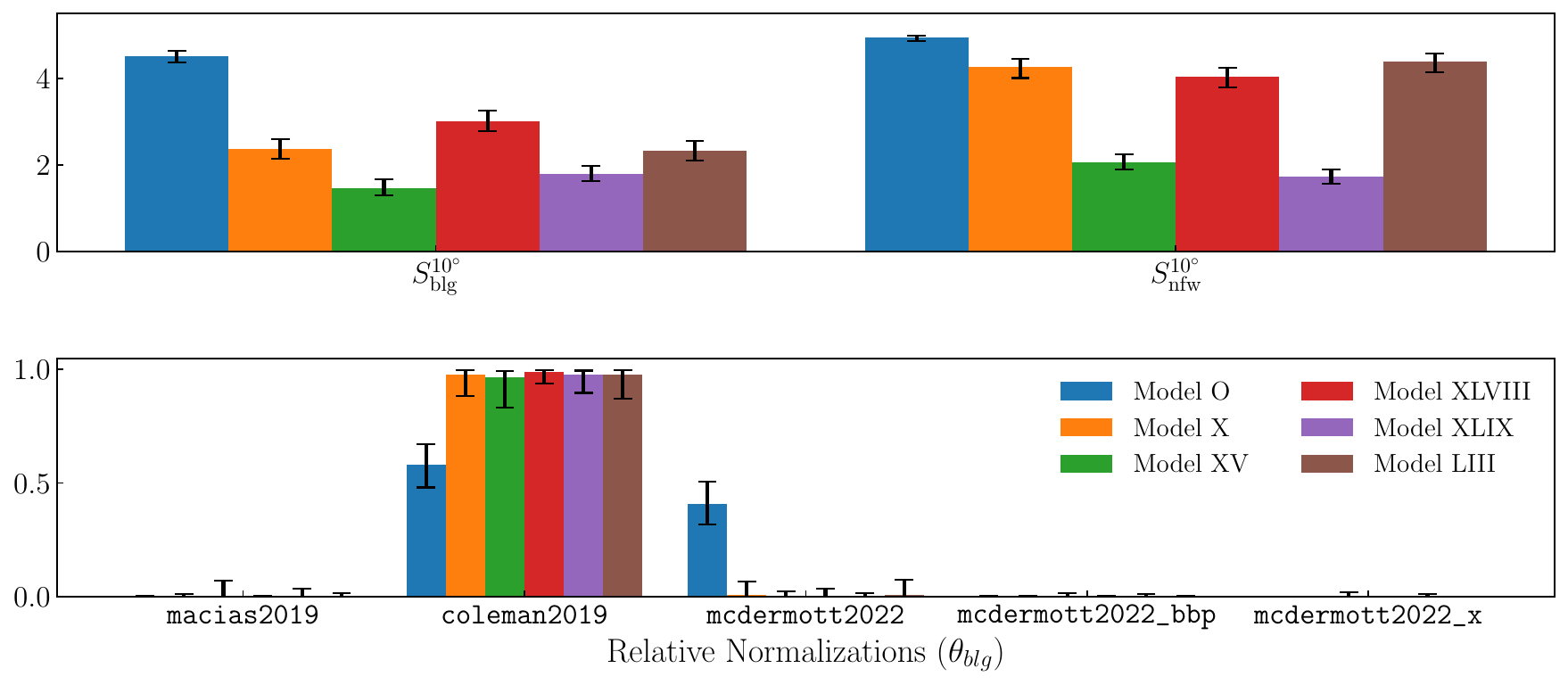}   
    \caption{As in Fig.~\ref{fig:blg-fit}, except we assume $\gamma = 1.2$ for the \texttt{nfw} template when we fit the GCE template model to the GP.} \label{fig:blg-fit-g1p2}
\end{figure}

The results of each fit are shown in Fig.~\ref{fig:blg-fit}. At the bottom, we compare the relative normalizations $\theta_{\rm blg}$ corresponding to each bulge template. These results show that, within $10^{\circ}$ and across all the diffuse models we have considered, the \texttt{coleman2019} bulge is the preferred bulge among the choices of bulge models that we fit to the GP. This is likely due to the \texttt{coleman2019} peaking at the center and exhibiting more signal at $\ell > 0^\circ$ than at $\ell < 0^\circ$, matching what the GP prefers in all fits. We show the variation in the normalizations of the \texttt{blg} and \texttt{nfw} templates and the NFW $\gamma$-parameter for comparison. Here, $S_{j}^{10^{\circ}}$ denotes the mean number of counts in the $10^{\circ}$ disk with the $|b| \leq 2^{\circ}$ band mask applied. As we fit only in the inner $10^\circ$, we normalize the templates using only this inner region (rather than normalizing to the counts within $20^\circ$ as was done previously).

We find that the $\gamma$-parameter is pushed to the prior's edge of 0.2 for all diffuse models. This is perhaps unexpected compared to previous studies of the GCE, which found more peaked profiles ($\gamma \sim 1-1.2$), at least when the GCE was fitted solely with a NFW profile. This appears to be a result of the GP being significantly more spread out across the inner $10^\circ$ than the GCE template models fitting to the GP; because the NFW template in this case has more flexibility than the bulge model, and in particular the bulge model has no capacity to absorb this more extended emission, the presence of this component drives $\gamma$ to small values. Recall also that we mask the region with $|b|<2^\circ$, so there is no impetus to closely match the distribution of emission at the Galactic Center itself. For these reasons, in these fits, the flux assigned to the NFW template is a better proxy for the amount of extended emission than for the amount of centrally-peaked emission traditionally associated with the GCE. 

There are some cases where the GP is relatively sparse, such as the one corresponding to Model XV and Model XLIX, as shown in Fig.~\ref{fig:data-gp-samples}. Moreover, the preference for the \texttt{mcdermott2022\_x} bulge in Fig.~\ref{fig:blg-fit} is a result of this sparsity, as it makes the GP appear more like a localized, X-shaped source of emission.

The fraction of the GP's signal that is fit by the \texttt{nfw} template versus the \texttt{blg} template differs across diffuse models. Due to the posterior peaking at small $\gamma$, \texttt{nfw} has a roughly flat morphology, so the \texttt{nfw} contributes more to the signal that the GP provides at large radii than near the center. In contrast, the \texttt{blg} template contributes most of the signal near the center. Thus, the ratio of $S_{\rm blg}^{10^{\circ}}$ to $S_{\rm nfw}^{10^{\circ}}$ is roughly a measure of the ratio of signal near the center to the signal far from the center. Using this metric, the fainter GPs coming from fits using Model XV or Model XLIX have almost twice the signal coming from the best-fit $\texttt{blg}$ component than the $\texttt{nfw}$ component. 

These results depend on the prior we choose for $\gamma$. In Fig.~\ref{fig:blg-fit-g1p2} we fit the GCE template model to the GP posterior fixing $\gamma = 1.2$, which forces the NFW component to be associated with centrally peaked rather than extended emission. In this case, the slight preference for the X-shaped bulge when using Model XV and Model XLIX and allowing $\gamma$ to vary disappears. However, fixing $\gamma$ results in a roughly equal preference for either a \texttt{coleman2019} bulge or \texttt{mcdermott2022} bulge when using Model O diffuse templates. Moreover, the ratio of $S_{\rm blg}^{10^{\circ}}$ to $S_{\rm nfw}^{10^{\circ}}$ now prefers more signal in the \texttt{nfw} component then when $\gamma$ is allowed to float.  This is a result of the $\texttt{nfw}$ template now having a peaked profile (comparable to the peak of the bulge models), as opposed to the flat profile when $\gamma \sim 0.2$.  

Since we allow the templates to be constrained by their values in the Outer ROI, the GP is effectively filling in the residuals of the templates in the Inner ROI, and hence depends on the ability for the diffuse templates to fit the data. From this behavior (demonstrated by the differences of the GPs in Fig.~\ref{fig:data-gp-samples} and studied in more detail in Appendix~\ref{app:mismodel}) the dependence of the best-fit GP morphology on the diffuse model adopted supports suggestions from earlier work~\cite{Bartels_2018, Caron_2023, Song_2024} that previous studies fitting the GCE as a template model may have arrived at conflicting results due to differences in background modeling.

We end this analysis by comparing the performance of Model O and all the 80 \textsc{Galprop}-based models from Ref.~\cite{Cholis_2022} when fitting our model to the data. We do this by performing a fit that uses a linear combination of all the tested diffuse template $\pi^{0}+$bremsstrahlung and ICS models. Similar to the GCE template model from the previous section, each model-specific \texttt{pib} template has relative normalizations $\theta_{\rm pib}$ that sum to one. A similar relative normalization parameter $\theta_{\rm ics}$ applies to each \texttt{ics} template. Though each model corresponds to a pair of \texttt{pib} and \texttt{ics} templates, the relative normalizations of each pair $\theta_{\rm pib}$ and $\theta_{\rm ics}$ are not required to be the same in order to maximize flexibility. Unlike the GCE template model, we adopt a prior for the relative normalizations that has no preference for either mixed or unmixed diffuse models (see Appendix~\ref{app:multi-temp} for more details). We reduced the peak learning rate to 0.025 in this fit, as our previous peak learning rate of 0.05 leads the model to pathological regions in the parameter space during training. 

\begin{figure}[t!]  
    \centering
    \includegraphics[width = \columnwidth]{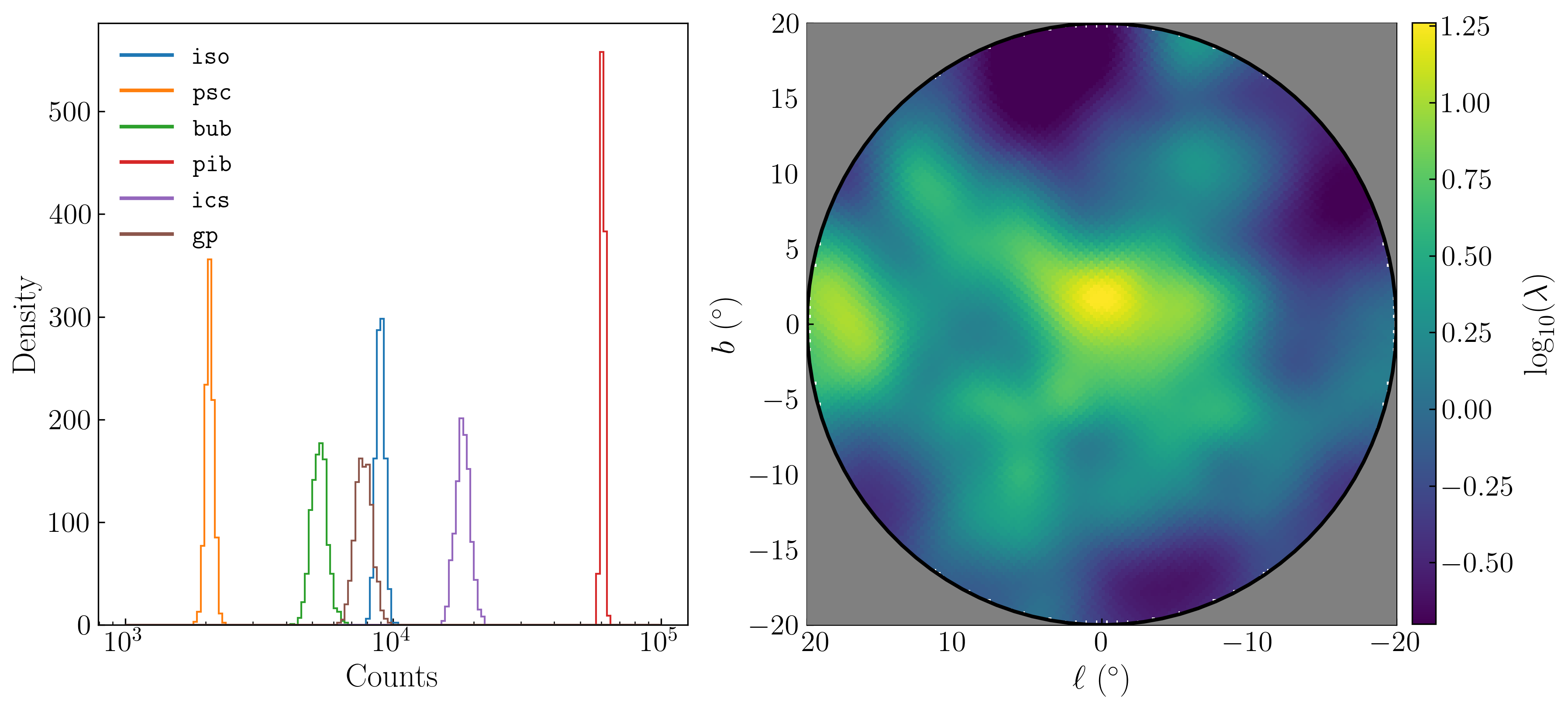}
    \includegraphics[width = \columnwidth]{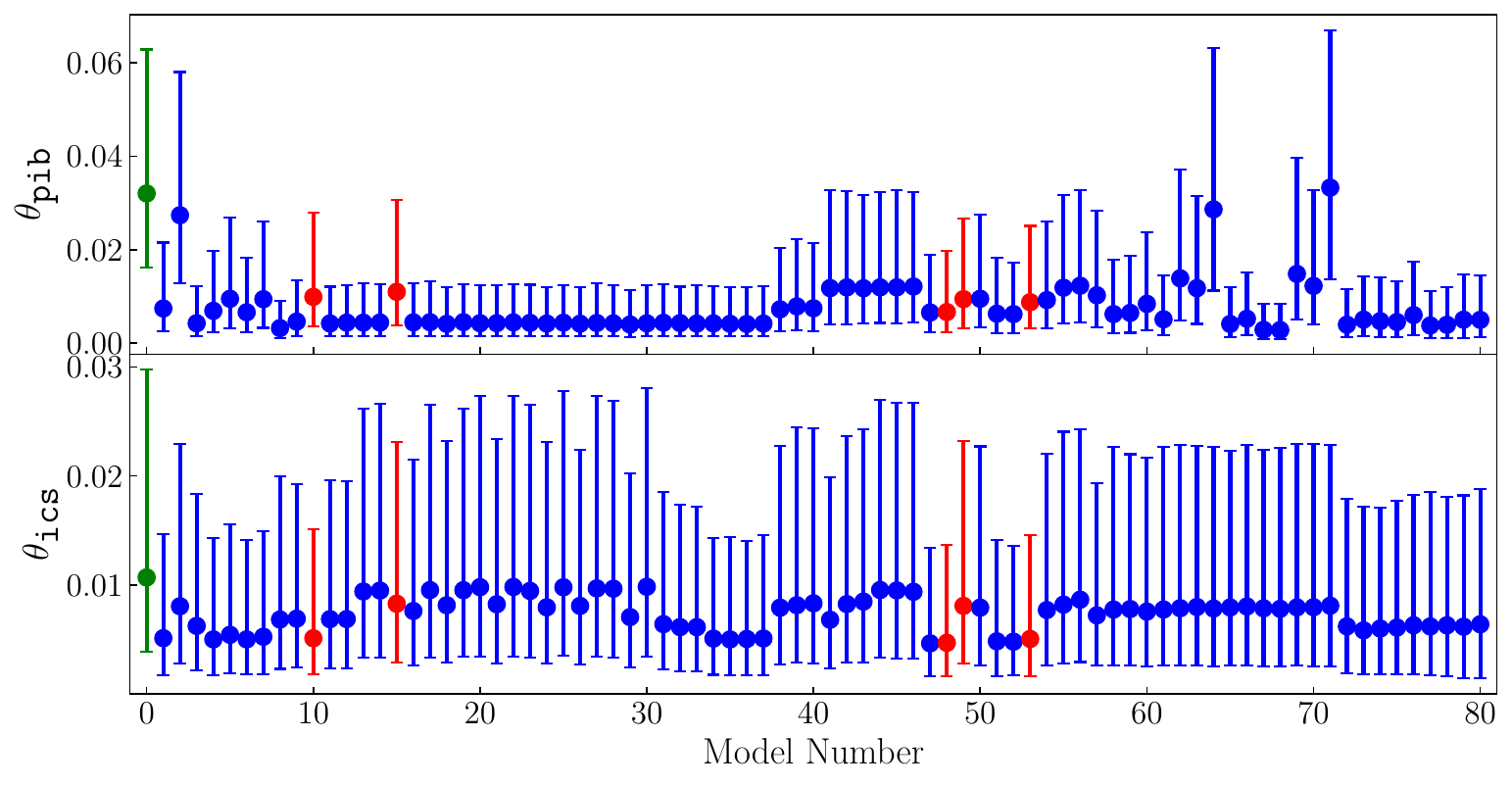}   
    \caption{Fit using a diffuse template taken as a linear combination of Model O and all the 80 galprop-based diffuse models from Ref.~\cite{Cholis_2022}. \textit{(Top Left)} Posterior of the total expected number of counts of the templates over the Inner ROI. \textit{(Top Right)} Median Gaussian process sample ($\log_{10}$-units). \textit{(Middle)} Median and 95\% probability interval of relative normalizations $\theta_{pib}^{m}$ of each pib model $m$. The top five best models from are red. \textit{(Bottom)} Median and 95\% probability interval of relative normalizations $\theta_{ics}^{m}$ of each ics model $m$. Model numbers from $1 - 80$ in the bottom two plots follow the same convention as in Ref.~\cite{Cholis_2022}, except we do not indicate them with Roman numerals. The top five best models from Fig.~12 in Ref.~\cite{Cholis_2022} are shown in red. Model number 0 corresponds to Model O, denoted in green.} \label{fig:galprop_all}
\end{figure}

The best-fit GP model resulting from this fit to the combined templates is summarized in Fig.~\ref{fig:galprop_all}. We find a preference for a few select models, albeit with a high degree of variance. Model O is somewhat preferred in both modeling $\pi^{0}+$bremsstrahlung and ICS emission. The best-five models identified by Ref.~\cite{Cholis_2022} are not highly preferred to fit either ICS or $\pi^{0}+$bremsstrahlung. Regardless of these results, the median GP in the top right exhibits similar characteristics to the GPs that we identified in the previous section. This indicates that the best-performing diffuse models have similar morphological assumptions, and so they are unable to fit the various previously-noted regions of the sky modeled by the GP.

We again fit the GP model here with the GCE template model, as was done to fit the GP posteriors corresponding to Model O and the five best models from Ref.~\cite{Cholis_2022}. In Fig.~\ref{fig:all_gp}, we show the result of the fit for a model that uses a uniform prior on $\gamma$ from 0.2 to 2. In the upper panel, we show a corner plot of the fits for the normalizations $S^{10^{\circ}}$ and the $\gamma$ parameter. 

As in the individual diffuse models, the $\gamma$ parameter peaks at the prior edge 0.2. As explained above, we believe this occurs because the GP finds spatially extended emission that cannot be absorbed by any of the bulge templates (\textit{e.g.,} unexplained ICS emission from a central source of electrons). This can be seen explicitly in the bottom panels of Fig.~\ref{fig:all_gp}, where we also note that the best-fitting models undershoot the GP at small Galactic longitude, in favor of a better fit at higher longitude and latitude (recall that we mask $|b|<2^\circ$ and so the low-Galactic-latitude results are only an extrapolation that does not contribute to the fit).

The preference for a small $\gamma$ in turn results in a preference for the \texttt{blg} template,  because the \texttt{nfw} template is roughly flat for this range of $\gamma$. In the bottom-left panel, we show the relative normalizations of the \texttt{blg} templates fit to the data. Here, the \texttt{coleman2019} bulge contributes $\sim 90\%$ of the total bulge signal and \texttt{mcdermott2022\_x} makes up most of the remaining $\sim 10\%$. In the remaining two panels, we show horizontal (bottom center) and vertical (bottom right) 1D slices of various posteriors. We show slices of the posterior rates from the GP component, the GCE template~(\texttt{nfw} + \texttt{blg}), the \texttt{nfw} template, the \texttt{blg} template, and the dominant \texttt{coleman2019} bulge template. 

As before, we consider how our results change when setting $\gamma = 1.2$. The results are shown in Fig.~\ref{fig:all_gp_g1p2}. In the upper-left panel, we show a corner plot of the normalizations $S^{10^{\circ}}$. As before, the \texttt{nfw} template provides more signal due to the presence of its central peak. In the bottom-left panel, we show the relative normalizations of the \texttt{blg} templates being fit to the data. Here, the \texttt{coleman2019} bulge makes up most of the bulge contribution. In the right panel, we show horizontal and vertical 1D slices of various posteriors. We show slices of the posterior rates from the GP component, the GCE template (\texttt{nfw} + \texttt{blg}), the \texttt{nfw} template, the \texttt{blg} template, and the dominant \texttt{coleman2019} bulge template. Note, since the GCE template posteriors in Figs.~\ref{fig:all_gp} and~\ref{fig:all_gp_g1p2} were trained using unmasked pixels only, the fits are significantly different in the Galactic Center, as the fits are not significantly penalized by the differences between the GCE template and the GP in this region, but rather by their differences in the unmasked pixels prevalent farther away from the center. However, it is interesting to note that while being a worse fit overall, fixing $\gamma=1.2$ does lead to a better description of the longitudinal profile of the GP, in particular at small Galactic longitudes.

\section{Conclusions}
\label{sec:conclusions}

In this work, we have inferred the GCE's signal and morphology from the \textit{Fermi} data by incorporating the GCE into our fit as a GP. In our aggregate Poisson model for the measured gamma-ray emission, the GP models the component of the data that cannot be explained by the background templates describing known sources. In this way, the GP describes both unknown physical sources of gamma-radiation in our ROI and corrects for mismodeling of the background templates. However, we cannot distinguish which aspects of the GP are due to unknown sources of gamma-rays and which are due to template mismodeling in the current version of our pipeline.

The range of morphologies of the GP that we find show additional features that are not described by GCE candidates studied previously, such as a localized bright source at around $(\ell,b) = (20^{\circ}, 0^{\circ})$ and a diagonal arm extending northwest from the GC. These features appear across fits involving a variety of morphologically distinct background templates. By fitting the GP with a GCE template model consisting of a generalized NFW squared component and several bulge components taken from the literature, we found that the preferred GCE template model was one composed primarily of the \texttt{coleman2019} bulge \cite{Coleman_2020} and a generalized NFW squared component, across all the diffuse models considered. This is consistent with recent template-based studies on inferring the GCE morphology \cite{Zhong_2024, Song_2024}.

\begin{figure}[t!]  
    \centering
    \includegraphics[width = 0.6\columnwidth]{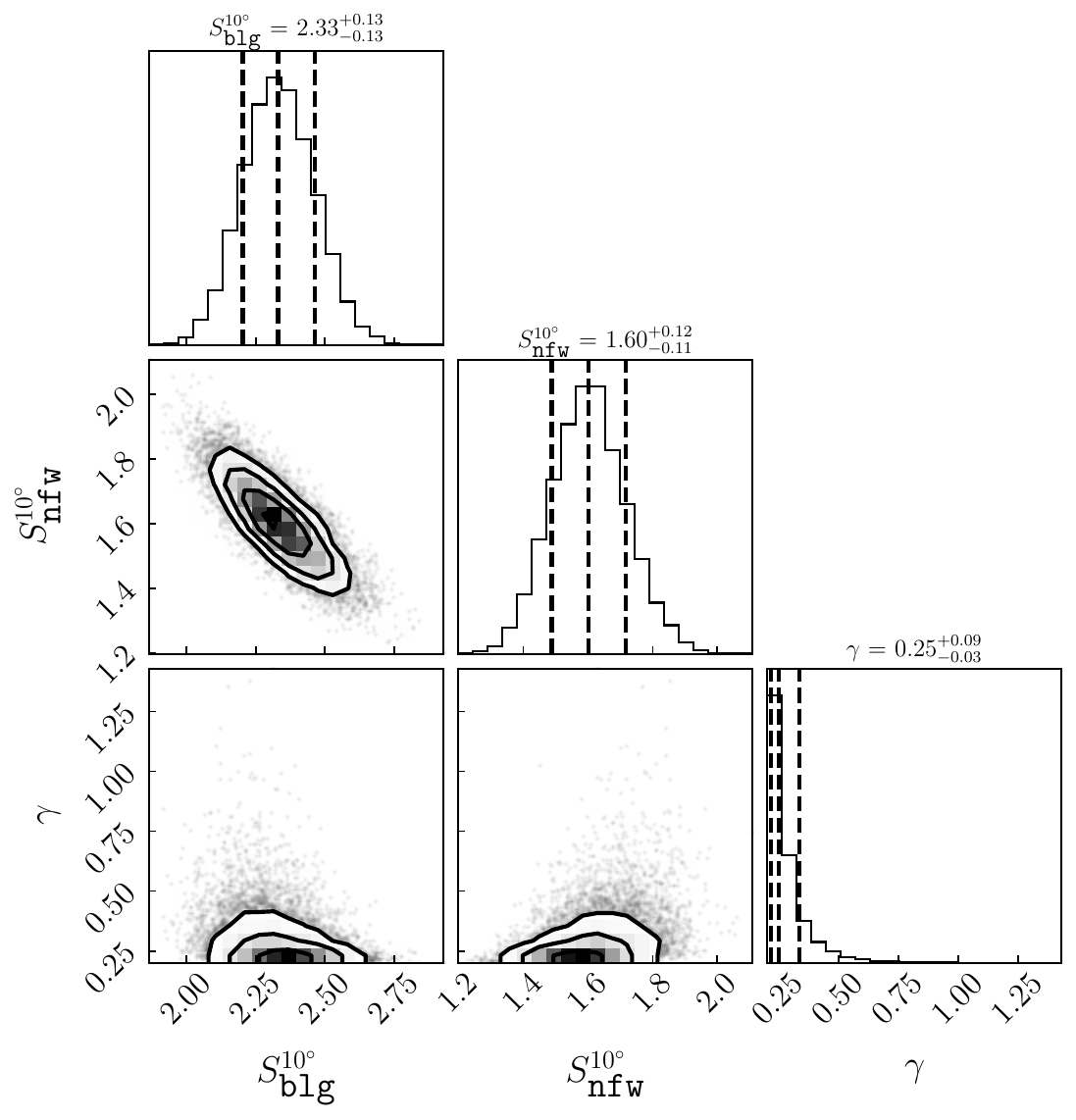} 
    \includegraphics[width = \columnwidth]{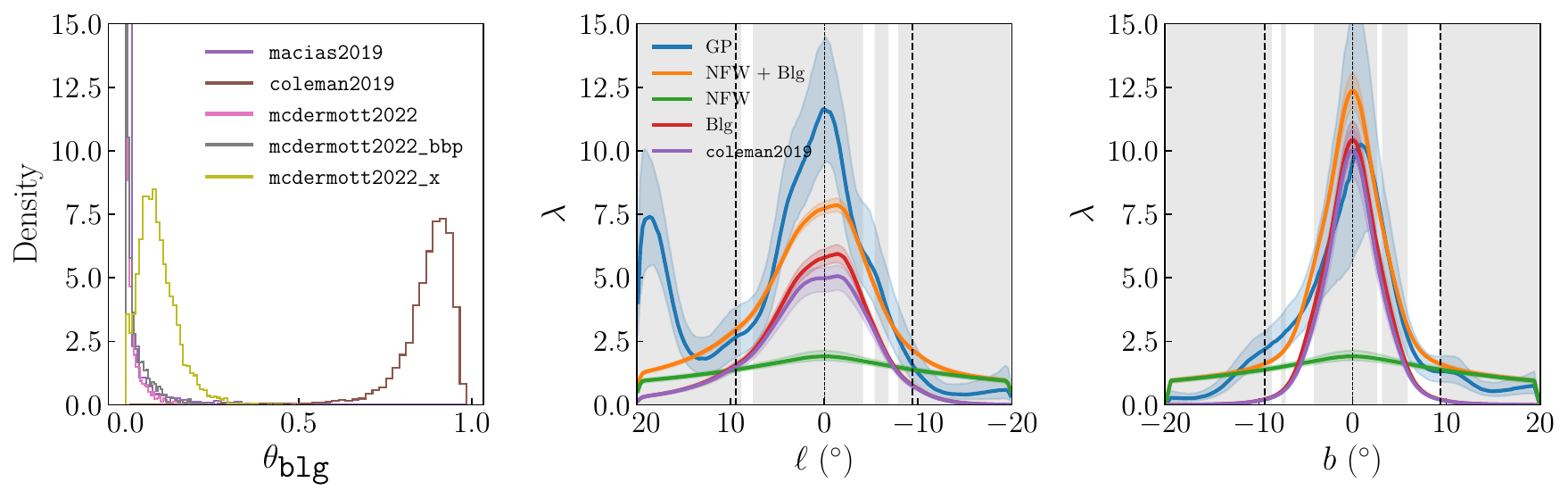} 
    \caption{Result of fitting a GCE template model to the GP posterior obtained by fitting the \Fermi{} data with a model that takes its diffuse template taken as a linear combination of Model O and all the 80 \textsc{Galprop}-based diffuse models from Ref.~\cite{Cholis_2022}. The aggregate GCE template model describes the GCE as a linear combination of an \texttt{nfw} template (with flexible $\gamma$-parameter) and the five bulge templates introduced. \textit{(Top)} Posterior of GCE template model parameters $S_{\rm nfw}^{10^\circ}$, $S_{\rm blg}^{10^\circ}$ and $\gamma$. \textit{(Bottom Left)} Posteriors of relative normalizations corresponding to each bulge template. \textit{(Bottom Center)} Horizontal 1D slice at $b = 3.375^{\circ}$ of posterior samples of counts drawn from the GP on Cartesian bins. 
    \textit{(Bottom Right)} Vertical 1D slice at $\ell = -3.375^{\circ}$ of posterior samples of counts drawn from the GP on Cartesian bins. Gray bands in the background indicate where masking is present. Dashed vertical lines correspond to $10^{\circ}$ circular boundary where the GCE template model was fit to the GP posterior.} \label{fig:all_gp}
\end{figure}

Our fits to the GP generally have the somewhat surprising feature of preferring a very small value for the NFW slope profile $\gamma$, implying a nearly isotropic contribution in the region of interest. We attribute this result to the presence of extended emission (away from the Galactic Center) in our inferred GP morphology, which cannot be absorbed into any of the bulge templates we consider, and so is instead assigned to the more flexible NFW template. However, this in turn means the NFW profile cannot rise steeply toward the Galactic Center to absorb emission there. Because this extended emission may well be due to pervasive diffuse mismodeling, we also considered the effects of fixing the NFW slope parameter to $\gamma=1.2$, consistent with previous GCE studies. When we employ a maximally flexible diffuse model consisting of a linear combination of 81 models from the literature, requiring $\gamma=1.2$ provides a better description of the low-longitude profile of the GP (as shown in Figs.~\ref{fig:all_gp}$-$\ref{fig:all_gp_g1p2}), although it remains an overall worse fit to the GP. For this diffuse model choice, in the case where $\gamma$ is floated (and becomes near-zero) the flux attributed to the bulge template with $10^\circ$ of the Galactic Center is roughly $60\%$ of the total (vs $40\%$ for the flat NFW template), with these ratios being reversed when $\gamma$ is fixed to 1.2. These results suggest that the bulge/NFW separation may be quite sensitive to how these components are modeled, and also to the choice of ROI (due to the presence of extended emission not well-described by the diffuse model), both in our analysis and in the broader literature.

The fits here do not take into account spectral information, as we fit the data to a single energy bin of $2 - 20$ GeV. In the future, we will generalize our pipeline to include energy as a feature. This will allow us to infer the spectrum of the GCE and the variation of the GCE's morphology with energy. 
In addition, adding energy information will provide another handle to break the degeneracies between the background templates and the GP in the Inner ROI. In the current analysis, these degeneracies are broken by fitting the templates in the Outer ROI, where the GCE is not expected to be present. Removing the Outer ROI from a future analysis (as may be possible through including spectral information) is desirable, as 
this would mitigate the impact of mismodeling on our GP fit.
In turn, this would allow us to better determine whether a signal from the GP arises from an unknown gamma-ray source or as a correction for the mismodeling of known gamma-ray sources.

\section{Acknowledgements}
We thank Ilias Cholis, Chris Gordon, Dan Hooper, Sung Hak Lim, Oscar Macias-Ramirez, Samuel D.~McDermott, Martin Pohl, Nicholas Rodd, and Yiming Zhong for useful discussions. We thank Ilias Cholis, Oscar Macias-Ramirez, Samuel D. McDermott, and Yiming Zhong for providing access to some of the templates used in this work. We thank the {\em Fermi} Collaboration for the use of {\em Fermi} public data.

YS was supported by a Trottier Space Institute Fellowship.
% SM acknowledgments
This work was partly supported by the U.S. Department of Energy, Office of Science, Office of High Energy Physics of U.S. Department of Energy under grant Contract Numbers DE-SC0012567 and DOE-SC0010008. 
This work is supported by the National Science Foundation under Cooperative Agreement PHY-2019786 (The NSF AI Institute for Artificial Intelligence and Fundamental Interactions, \href{http://iaifi.org/}{http://iaifi.org/}).
This work was performed in part at the Aspen Center for Physics, which is supported by NSF grants PHY-2210452.
T.R.S.~was supported in part by a Guggenheim Fellowship; the Edward, Frances, and Shirley B.~Daniels Fellowship of the Harvard Radcliffe Institute; the Bershadsky Distinguished Fellowship of the Harvard Physics Department; and the Simons Foundation (Grant Number 929255, T.R.S). T.R.S.~thanks the Kavli Institute for Theoretical Physics (KITP), the Aspen Center for Physics, and the Mainz Institute for Theoretical Physics for their hospitality during the completion of this work; this research was supported in part by grant no.~NSF PHY-2309135 to KITP.

\section{Data Availability}
The code we used to perform our fits is publicly available in the GitHub repository \url{https://github.com/edwarddramirez/gce-gp}.

\begin{figure}[t!]  
    \centering
    \includegraphics[width = 0.4\columnwidth]{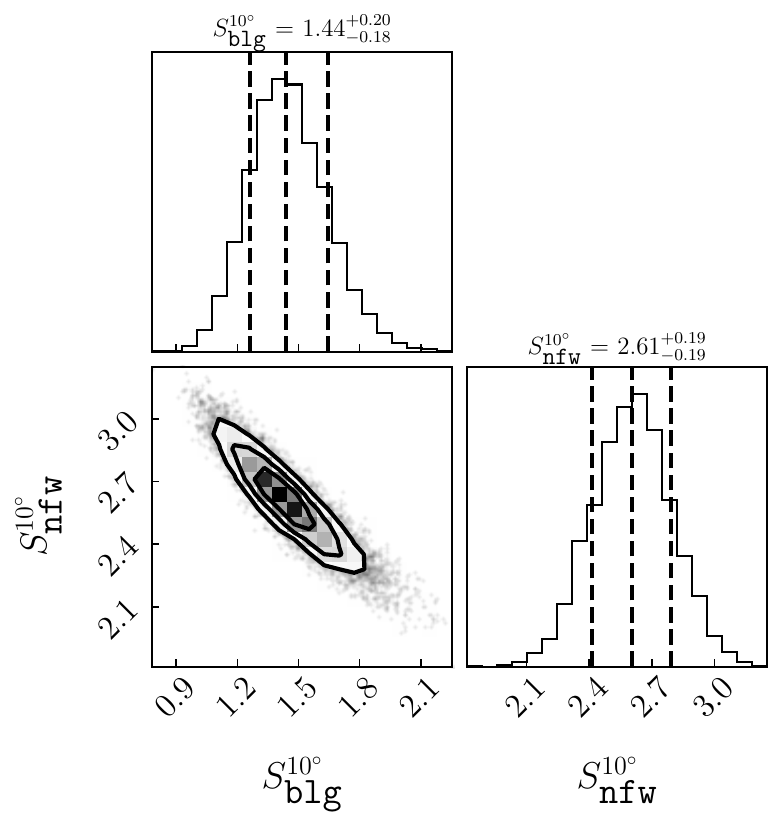} 
    \includegraphics[width = \columnwidth]{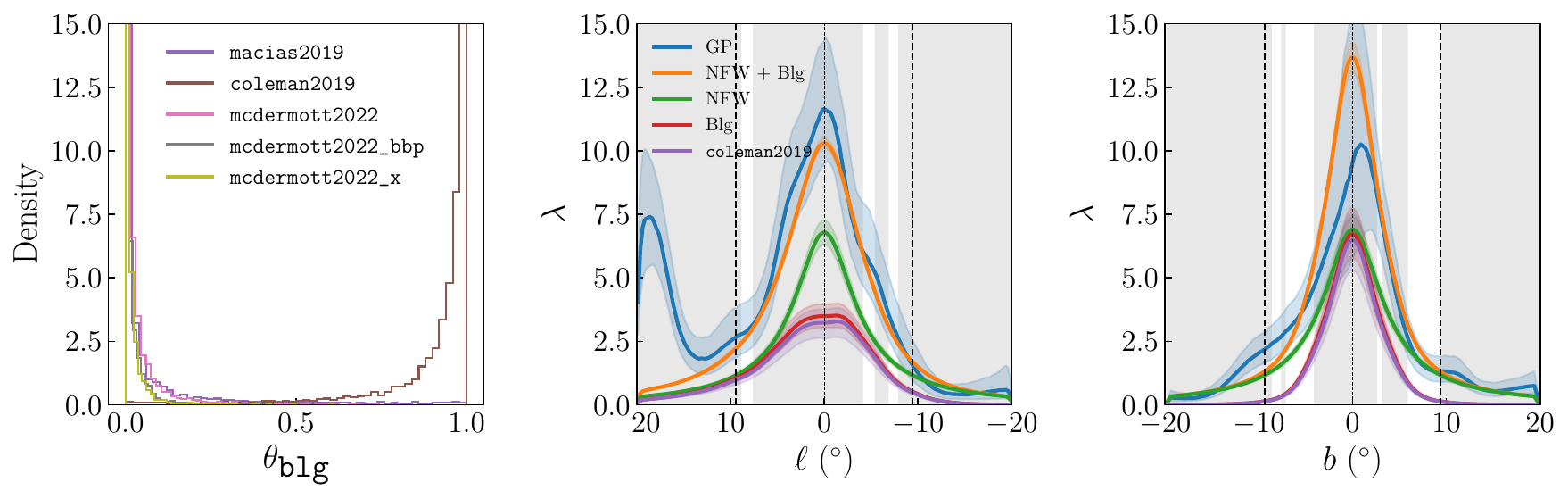} 
    \caption{Same as Fig.~\ref{fig:all_gp}, except we assume $\gamma = 1.2$ for the \texttt{nfw} template when we fit the GCE template model to the GP.} \label{fig:all_gp_g1p2}
\end{figure}

\bibliography{references}

\appendix

\section{Modeling Choices} \label{app:choices}

\subsection{Degeneracies in the Inner ROI} \label{app:degeneracy}

If the model is fit to data only in the Inner ROI, the GP and the other templates will be degenerate due to the flexibility of the GP. We show this using the synthetic data generated for Section~\ref{sec:syndata}, which uses the Model O diffuse templates and generates the GCE with a combination of the \texttt{coleman2019} bulge template and an \texttt{nfw} template with $\gamma = 1.2$ and $R_S = 20$~kpc. As in Section~\ref{sec:syndata}, we fit the synthetic data with the same background templates used to generate the data, except we now only use the Inner ROI during the fit. 

We find that the GP model can converge to two different optima during the fit, depending on the learning rate and the number of steps. The first fit result (Fit 1) is obtained with the scheduled learning rate with exponential decay described in Section~\ref{sec:syndata}, with 20,000 steps. The second fit (Fit 2), corresponds to a constant learning rate of $10^{-4}$ and 100,000 steps. 

\begin{figure}[t!] 
    \centering
    \includegraphics[width = 1\columnwidth]{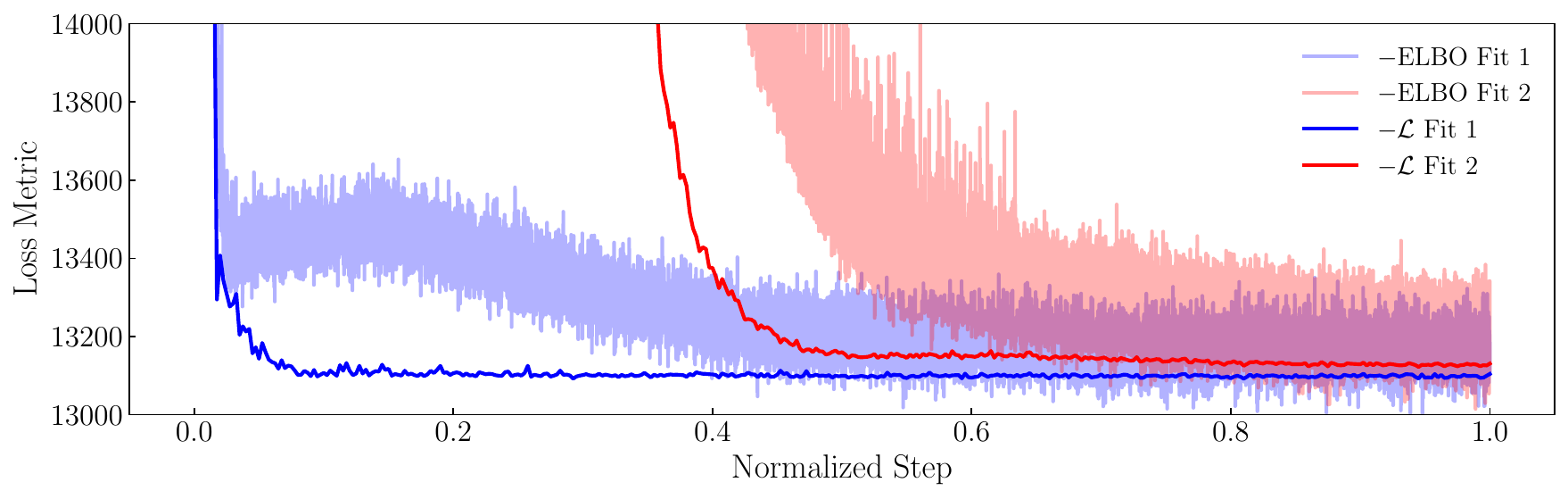}
    \includegraphics[width = 1\columnwidth]{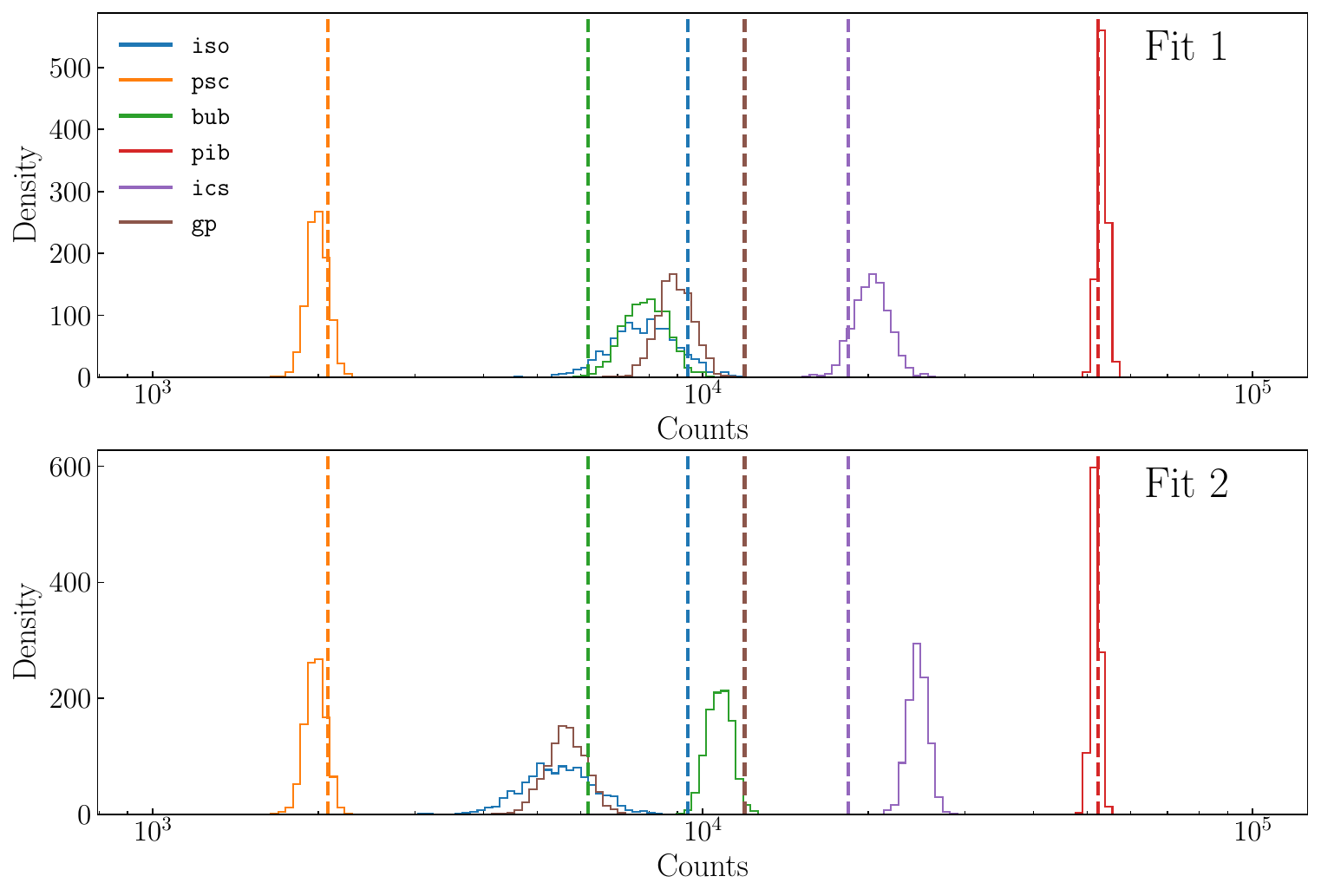}  
    \caption{Results of testing the degeneracy between the GP and templates by performing identical fits using different learning rates and steps for fitting. \textit{(Top)} Convergence of the loss function ($-$ELBO) and the posterior mean negative log-likelihood ($-\mathcal{L}$, see Eq.~\eqref{eq:ll}) to different optima, depending on fit settings. Since the fits use a different number of steps, we normalize the number of steps from 0 to 1. \textit{(Center and Bottom)} The expected total number of counts of each gamma-ray source in the Inner ROI for both fits, binned logarithmically. True normalizations are denoted by dashed lines.
} \label{fig:degeneracy}
\end{figure}

A comparison between the results from both fits is shown in Fig.~\ref{fig:degeneracy}. In the top panel, we show the evolution of both the loss function $-{\rm ELBO}$ (Eq.~\eqref{eq:elbo_def}) and the posterior predictive mean of the negative log-likelihood $-{\cal L}$ (Eq.~\eqref{eq:ll}) calculated using the pixels in the Inner ROI. The total number of steps are normalized to 1. The evolution of the $-{\rm ELBO}$ loss functions match closely for both fits, though Fit 2 exhibits a higher degree of variance. The $-{\cal L}$ of Fit 2 converges to a higher value than Fit 1, indicating a poorer fit at this local minimum. In the middle and bottom panels of Fig.~\ref{fig:degeneracy}, we plot the posteriors of the expected total number of counts sourced by the gamma-ray templates in the Inner ROI.

These two local minima exist due to the GP's flexibility. As a result, different sets of background templates can fit the data nearly equally well, as the GP corrects for any mismodeling from the background templates to fit the data well.\footnote{That is, the degeneracy is not an exclusive feature of the GP. Any flexible model for the GCE will be degenerate with the templates.} This leads to a loss function with many ``false" optima. 

In general, we have found that lower learning rates yield fainter GPs, which is corrected by modifying the normalizations of the other templates. This behavior suggests that the learning rate determines the ``speed" at which the templates fit the data, and thus which optimum the model converges to. For low learning rates, the GP is initially faint and the templates fit the data well before the GP can catch up, leading to the GP correcting for what the templates could not fit on their own. For a high learning rate, the GP is allowed to fit the data first and then the templates follow to fit the residuals. 

These results indicate that the GP's flexibility can be detrimental to our model if we have no additional information to constrain the model. In our work, we adopt the Outer ROI where the GP (GCE) is assumed to be negligible. This additional information constrains the templates, allowing the GP to infer the GCE in the Inner ROI, as opposed to just correcting for mismodeling of the templates. Spectral information may similarly constrain the templates without the need for the Outer ROI. We leave this analysis for future work.

\subsection{Choice of ROI} \label{app:roi}
As discussed in the previous section and in the main text, we adopt an Outer ROI to break the degeneracy between the flexible GP and the other gamma-ray source templates. In the main text, we take our Complete ROI to consist of two regions: the Inner ROI where both the GP and the templates contribute to the signal,  and the Outer ROI where the GP is assumed to be zero. The Inner and Outer ROIs are separated by a buffer, whose gamma-ray counts are not included in the fit. The Inner ROI we take to be a 20$^{\circ}$ disk (plus disk and point source masks), separated from the Outer ROI by a $20^{\circ}-30^{\circ}$ annulus acting as a buffer. The Outer ROI we take as a $30^{\circ}-40^{\circ}$ annulus (plus masks). We motivate these choices here.

\begin{figure}[t!] 
    \centering
    \includegraphics[width = \columnwidth]{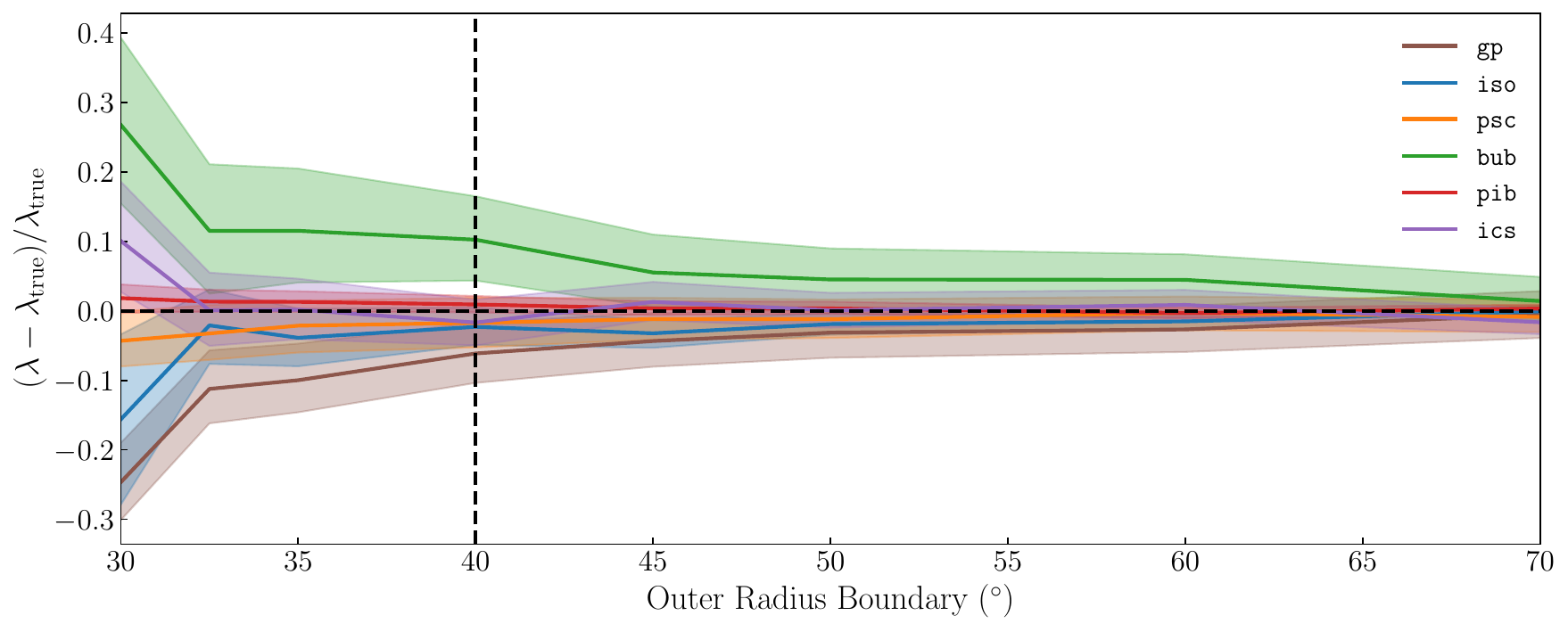}   
    \caption{Posteriors of the gamma-ray components' total rate summed over the Inner ROI versus the outer boundary of the Outer ROI, for an inner boundary of $30^{\circ}$ and a buffer spanning from $20^{\circ}-30^{\circ}$. The vertical axis shows the relative difference $(\lambda - \lambda_{\rm true}) / \lambda_{\rm true}$ corresponding to each gamma-ray component. Each posterior is the average of ten fits each using different Poisson draws of the synthetic rate as data. The choice that we make for our analysis ($40^{\circ}$) is denoted by the dashed vertical line.
} \label{fig:roi_scan_outer_roi}
\end{figure}

\begin{figure}[h!] 
    \centering
    \includegraphics[width = 1\columnwidth]{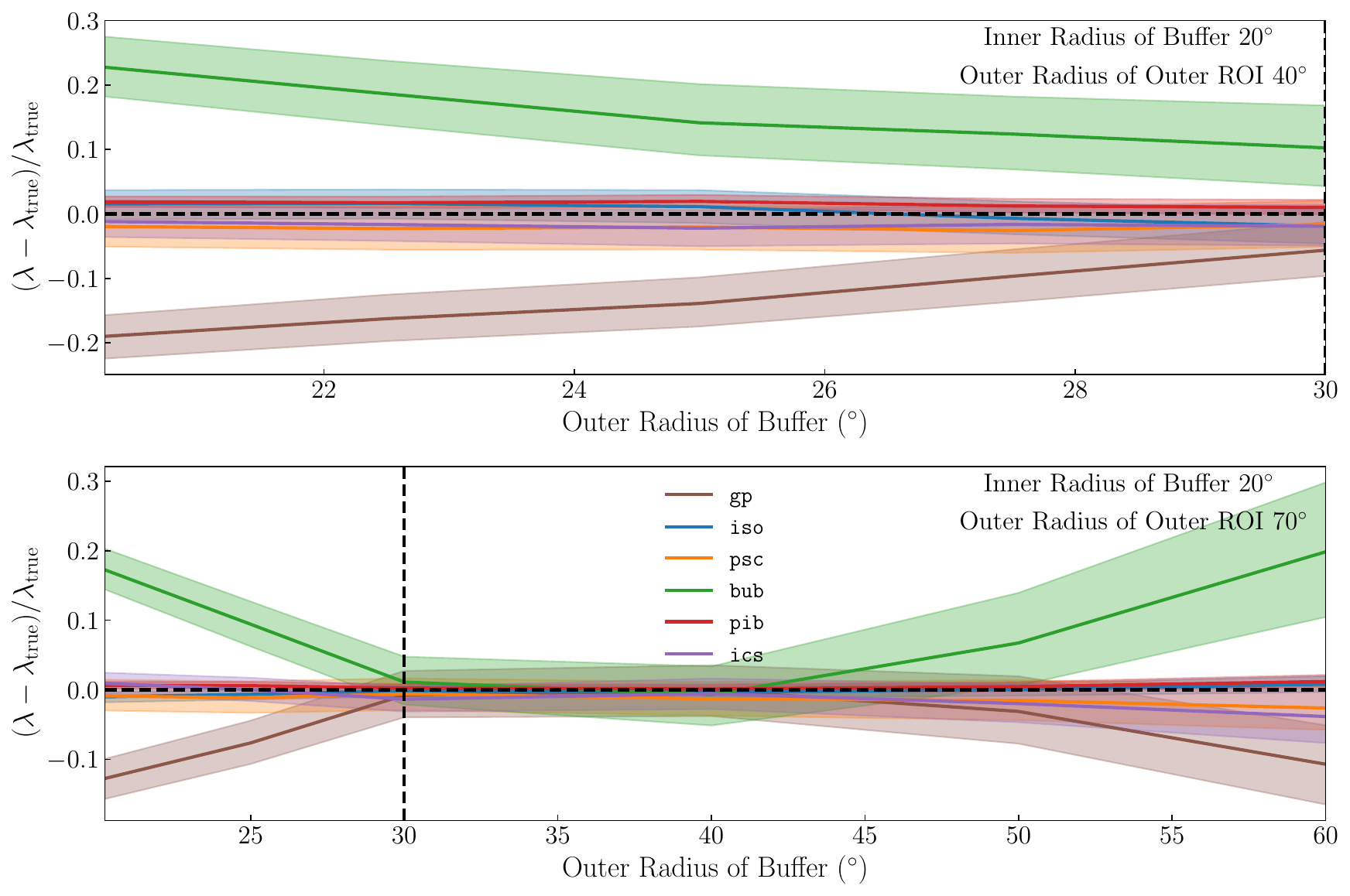}   
    \caption{Posteriors of gamma-ray components' total rate summed over the Inner ROI versus the location of the boundary separating the Buffer and the Outer ROI. The vertical axes show the relative difference $(\lambda - \lambda_{\rm true}) / \lambda_{\rm true}$ corresponding to each gamma-ray component. Each posterior is the average of ten fits each using different Poisson draws of the synthetic rate as data. \textit{(Top)} Scan over outer boundary of Buffer when Buffer's inner boundary is set to $20^{\circ}$ and the Outer ROI's outer boundary is located at $40^{\circ}$. \textit{(Bottom)} Scan over outer boundary of Buffer when Buffer's inner boundary is set to $20^{\circ}$ and the Outer ROI's outer boundary is located at $70^{\circ}$. 
} \label{fig:roi_scan_buffer}
\end{figure}

The buffer between the Inner and Outer ROI is the result of a compromise between two competing effects. First, we cannot extend the GP arbitrarily far from the Galactic Center, due to the limited memory of our GPUs. The number of parameters in the GP fit scales with the number of pixels in the region over which the GP is defined, which goes as the area of the Inner ROI. Defining the Inner ROI out to $20^\circ$ corresponds to the limit of our current available memory. Given this, we must deal with the second issue: an NFW$^{2}$ component to the GCE would extend somewhat further than $20^\circ$. Though the signal is weak at the pixel level, if the Outer ROI (without a GP model) began at the edge of the Inner ROI, the bleed-over of the GCE signal into this region would affect the inferred normalizations of the background templates.

In Fig.~\ref{fig:roi_scan_outer_roi} we show the difference between the expected total number of counts of the best-fit gamma-ray sources and the expected total number of counts of the true synthetic gamma-ray sources, across the Inner ROI, as a function of the outer radius of the Outer ROI. The simulation properties are otherwise the same as in the main text (see Section~\ref{sec:syndata}). In Fig.~\ref{fig:roi_scan_buffer}, we show the difference between the expected total number of counts of the best-fit gamma-ray sources and the expected total number of counts of the true synthetic gamma-ray sources, across the Inner ROI, as a function of the boundary separating the buffer and the Outer ROI, keeping the outer radii of the inner and outer ROIs fixed. 

\subsection{Extending Inpainted Bubbles Template to the Complete ROI} \label{app:bub}

\begin{figure}[t!] 
    \centering
    \includegraphics[width = \columnwidth]{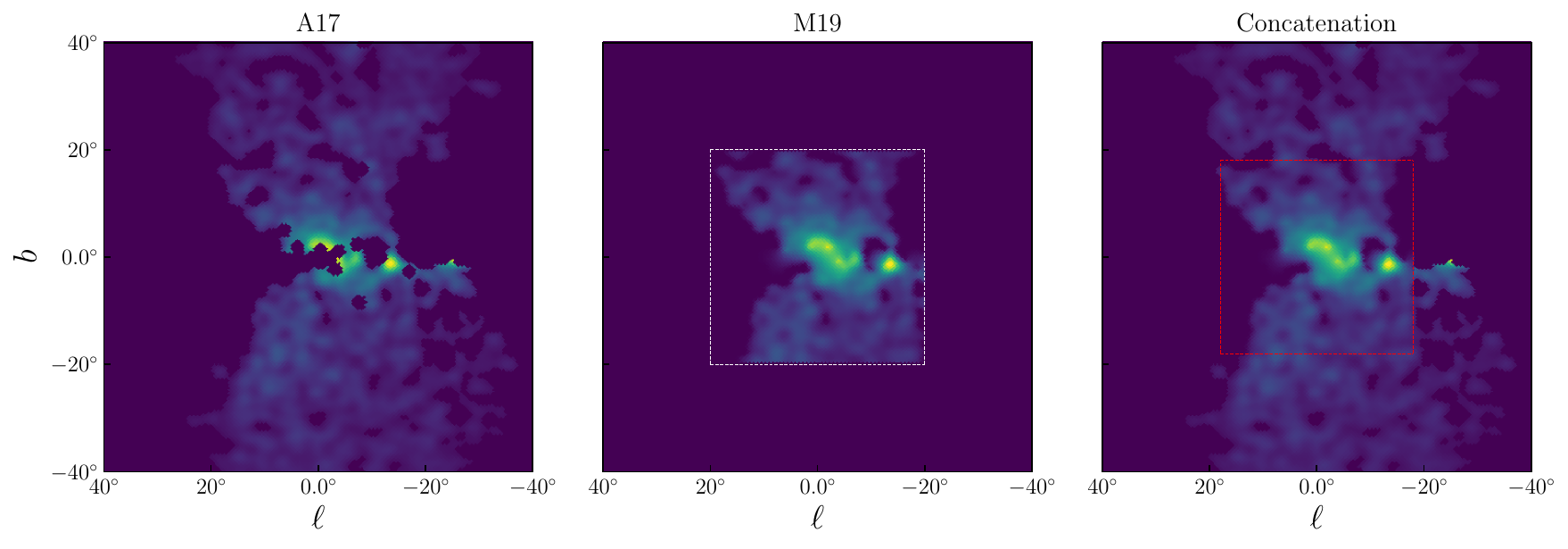}   
    \caption{Bubbles templates with consistent normalizations. \textit{(Left)} A17 Bubbles template. \textit{(Middle)} M19 Bubbles template with its boundary of definition denoted by the dashed white square ($40^{\circ} \times 40^{\circ}$). \textit{(Right)} Concatenated Bubbles template with the concatenation boundary denoted by the dashed red square ($36^{\circ} \times 36^{\circ}$). Inside this boundary the M19 template describes emission. Outside this boundary the A17 template describes emission.
    } \label{fig:app-bubs}
\end{figure}

We have two templates for the Fermi Bubbles from Ref.~\cite{Ackermann_2017} (A17) and Ref.~\cite{Macias_2019} (M19). The former is an older Bubbles template defined for $|b| < 60^{\circ}$ and $|\ell| < 45^{\circ}$, shown to the left of Fig.~\ref{fig:app-bubs}. The latter is an updated version of the A17 template, restricted to $-20^{\circ} \leq \ell, b \leq 20^{\circ}$, shown in the middle panel of Fig.~\ref{fig:app-bubs}. This template was processed with Laplacian inpainting by M19 to interpolate the rate of the Bubbles in regions that were covered by point source masking in the analysis of A17. To adopt the updated M19 template, we must extend it to our Outer ROI. 

To accomplish this, we combined both templates by ``concatenating" the inpainted M19 template with the A17 template where the former is not defined. This required the A17 and M19 bubbles templates to have a consistent normalization. Naively, we can normalize both templates so that they share the same total emission inside the inner $40^{\circ} \times 40^{\circ}$ square region. However, this is not the correct approach, as the M19 template necessarily has more signal than the A17 template due to the inpainting. Instead, we normalize the templates so that they have the same total emission over the unmasked region within the inner $40^{\circ} \times 40^{\circ}$ square, as this is the region where they are consistent.

As we did not have the point source masks from A17, we thresholded the A17 template to select the unmasked pixels. We also note that the A17 and M19 templates do not agree near the edges of the $40^{\circ} \times 40^{\circ}$ square region, so we take our concatenation region to be the slightly smaller $36^{\circ} \times 36^{\circ}$ square region.
Once we normalized the templates over the unmasked pixels inside the modified inner region, we combined the templates so that the M19 template describes the inner region and the A17 describes the outer region. The final result is shown in the right panel of Fig.~\ref{fig:app-bubs}.

\subsection{Inducing Point Configuration} \label{app:inducing}

\begin{figure}[t!] 
    \centering
    \includegraphics[width = 1\columnwidth]{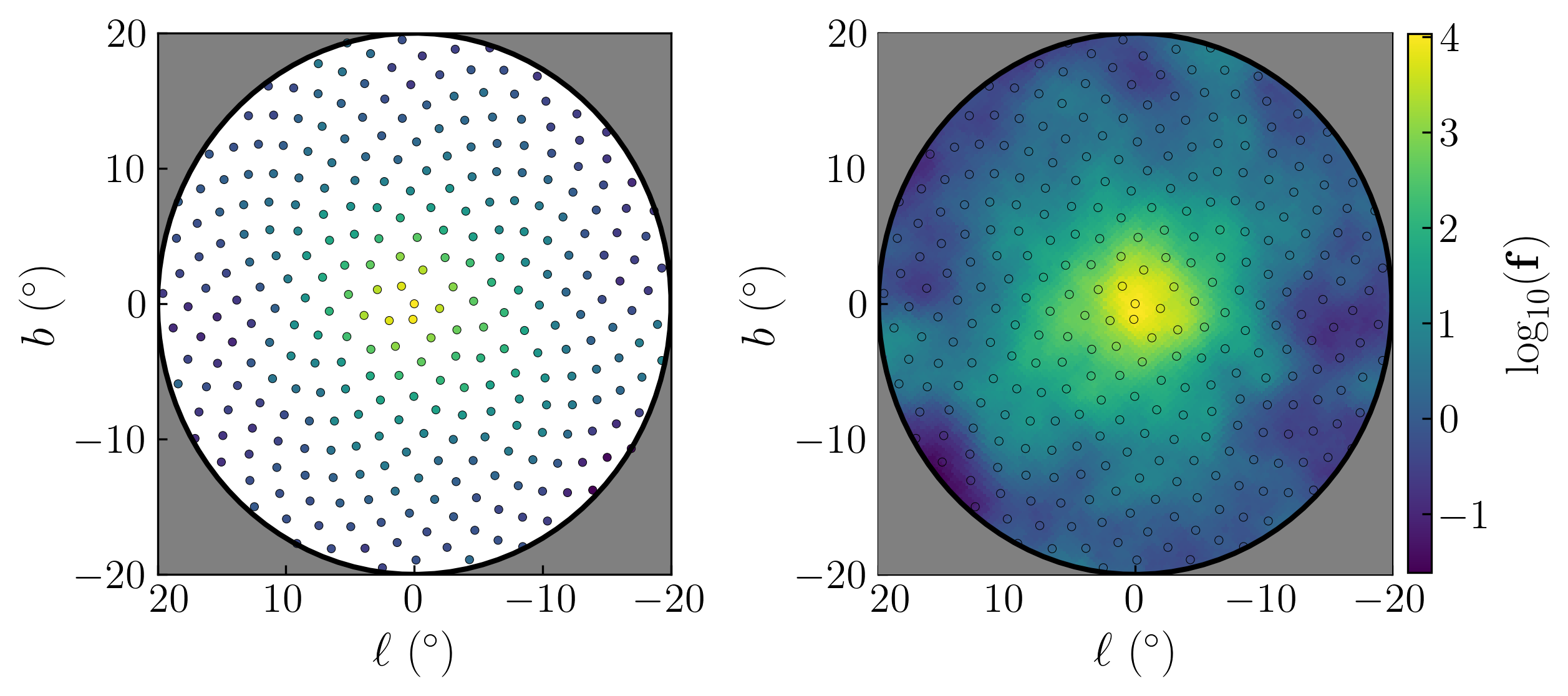}   
    \caption{\textit{(Left)} Inducing points $\mathbf{u} = f(\mathbf{x_{u}})$ sampled from the optimized approximate posterior $q_{\phi}(\textbf{u})$ of Section~\ref{sec:syndata}. \textit{(Right)} Function-values $\textbf{f} = f(\textbf{x})$ drawn from the optimized GP prior conditioned on $\textbf{u}$, $p_{\theta}(\textbf{f} | \textbf{u})$, as is done when sampling $\textbf{f}$ from the approximate posterior $q_{\theta,\phi}(\textbf{f})$.\label{fig:inducing_pts}}
\end{figure}

We train our sparse and variational GP model on a set of inducing points, $\mathbf{u} = f(\mathbf{x_{u}})$, where the pseudo-inputs $\mathbf{x_{u}}$ are distributed in a sunflower configuration. In the sunflower configuration, the $k^{\rm th}$ point (seed) is placed in the following point described in polar coordinates
\begin{align*}
    \theta & = \frac{2 \pi}{\phi^{2}} \ k \\
    r & = c \ \sqrt{k},
\end{align*}
where $\phi$ is the golden ratio $(1 + \sqrt{5})/2$ and $c$ is an arbitrary constant which controls the size of the outer radius into which the points are distributed \cite{VOGEL1979}. We set $c$ such that the last point is $20^{\circ}$ away from the center. For visualization, a sample of 300 inducing points $\textbf{u}$ from the optimized approximate posterior $q_{\phi}(\textbf{u})$ of Section~\ref{sec:syndata} is given in the left panel of Fig.~\ref{fig:inducing_pts}. To illustrate the relationship between $\textbf{u}$ and the physical log-rate $\textbf{f}$, we show in the right panel of Fig.~\ref{fig:inducing_pts} a sample $\textbf{f}$ drawn from the optimized GP prior conditioned on $\textbf{u}$, $p_{\theta}(\textbf{f} | \textbf{u})$ (following Eq.\eqref{eq:qf}).

As stated in the text, we chose this configuration to create pseudo-inputs which are roughly equidistant. Other configurations giving roughly equidistant configurations are placing the points in a mesh grid (square or hexagonal) around the $20^{\circ}$ disk. Given that our Inner ROI is a circular disk, we find the sunflower configuration to be a more appropriate distribution than these other options.
The sunflower configuration is also easily modified to accommodate any number of points $N_{u}$, whereas grid-based approaches are defined in the disk only for a specific set of integers set by the grid spacing. 

\subsection{Modeling a Single Gamma-Ray Source with Multiple Templates} \label{app:multi-temp}

For our fits, we need not restrict ourselves to modeling each background gamma-ray component with a single template. To implement multiple templates for the same gamma-ray source, we follow the same exact fit approach, except we now must include relative normalizations for each of the template models. For example, suppose we want to use $N_{j}$ templates for a component indexed by $j$ (see paragraph containing Eq.~\eqref{eq:first-rate-eqn} for definition). Then, the rate $\lambda_{j}(\textbf{x})$ sourced by the component is given by:
\begin{equation} \label{eq:multi-rate-eqn}
    \lambda_{j}(\mathbf{x}) = S_{j} \sum_{k=1}^{N_{j}}  \theta_{j}^{k} T_{j}^{k}(\mathbf{x}),
\end{equation}
where $\theta_{j}^{k}$ denotes the relative normalization of the $k^{th}$ template modeling gamma-ray component $j$. By being relative normalizations, $\sum_{k=1}^{N_{j}} \theta_{j}^{k} = 1$ and $S_{j}$ retains its original definition. To obey their unit total normalization, the priors on $\theta_{j}^{k}$ are assumed to follow a symmetric Dirichlet distribution with concentration $\alpha_{j}$. By taking a symmetric Dirichlet distribution, there is no preference for one template model over another. A concentration equal to 1 corresponds to a uniform prior on the relative normalizations. By taking a concentration less than 1, the prior is more sparse, so higher probability is assigned to states corresponding to a single dominant template model as opposed to states describing mixed template models. When we assume a single template model is dominant (\textit{e.g.,} Fig.~\ref{fig:blg-fit}), we take $\alpha_{j} = N_{j}^{-1}$.

\section{SVI Addendum} \label{app:addendum}

\subsection{Introduction to Variational Inference} \label{app:svi}

We discuss the foundations of variational inference, as discussed in Ref.~\cite{pyro_svi_i} and references therein. Our overarching goal is to develop the best model to explain the available data. The model is specified by observations $\textbf{y}$, latent variables $\textbf{z}$, and parameters $\theta$ with the probability of data being given by the evidence:
\begin{align}
    p_{\theta}(\textbf{y}) = \int p_{\theta}(\textbf{y},\textbf{z}) \ d\textbf{z}.
\end{align}
Therefore, the optimal model is achieved by finding the parameter that maximizes the log-evidence:
\begin{align} \label{eq:svi_thetastar}
    \theta^{*} = \argmax_{\theta} \log p_{\theta}(\textbf{y}).
\end{align}

We want to be able to make predictions and estimate uncertainties on the latent variables via inference, using the optimal model. That is, we want the posterior $p_{\theta}(\textbf{z}|\textbf{y})$. Calculating this is generally analytically intractable. Variational inference (VI) is an approximate inference technique that addresses this intractability. The objective in VI is to approximate the posterior $p_{\theta}(\textbf{z}|\textbf{y})$ with a variational distribution (approximate posterior) $q_{\phi}(\textbf{z}) \approx p_{\theta}(\textbf{z}|\textbf{y})$, parameterized by $\phi$. The condition corresponding to optimal approximation accuracy of the variational distribution is given by: 
\begin{equation} \label{eq:svi_phistar}
    \phi^{*} = \argmin_{\phi} {\rm KL}[ \ q_{\phi}(\textbf{z}) \ || \ p_{\theta}(\textbf{z}|\textbf{y}) \ ],
\end{equation}
where KL is the Kullback–Leibler (KL) divergence. 

By performing VI, we are attempting a double-optimization: simultaneously finding the model best supported by the data and the best approximation to the model's posterior. We begin by replacing the KL-term in Eq.~\eqref{eq:svi_phistar} with something more tractable. We note that
\begin{align} 
    {\rm KL}[  q_{\phi}(\textbf{z}) \ || \ p_{\theta}(\textbf{z}|\textbf{y}) ] & = \mathbb{E}_{q_{\phi}(\textbf{z})} \left [ \log \frac{q_{\phi}(\textbf{z})}{p_{\theta}(\textbf{z}|\textbf{y})} \right ] \nonumber \\
   & = \mathbb{E}_{q_{\phi}(\textbf{z})} \left [ \log q_{\phi}(\textbf{z}) \right ] - \mathbb{E}_{q_{\phi}(\textbf{z})} \left [ \log p_{\theta}(\textbf{z}|\textbf{y}) \right ] \nonumber \\
   & = \mathbb{E}_{q_{\phi}(\textbf{z})} \left [ \log q_{\phi}(\textbf{z}) \right ] - \mathbb{E}_{q_{\phi}(\textbf{z})} \left [ \log p_{\theta}(\textbf{y},\textbf{z}) \right ] + \mathbb{E}_{q_{\phi}(\textbf{z})} \left [ \log p_{\theta}(\textbf{y}) \right ] \nonumber \\
    & = \mathbb{E}_{q_{\phi}(\textbf{z})} \left [ \log q_{\phi}(\textbf{z}) - \log p_{\theta}(\textbf{y},\textbf{z}) \right ] +  \log p_{\theta}(\textbf{y}),
\end{align}
To get from the second to third line, we used $p_{\theta}(\textbf{y},\textbf{z}) = p_{\theta}(\textbf{z}|\textbf{y}) p_{\theta}(\textbf{y})$. The first term in the final line is the negative of the ${\rm ELBO}$ defined in Eq.~\eqref{eq:elbo_def}. Thus, the final expression shows that our two objective functions (Eq.~\eqref{eq:svi_thetastar}$-$\eqref{eq:svi_phistar}) are related through the ${\rm ELBO}$:
\begin{equation}
    \log p_{\theta}(\textbf{y}) = {\rm ELBO} + {\rm KL}[  q_{\phi}(\textbf{z}) \ || \  p_{\theta}(\textbf{z}|\textbf{y})  ].
\end{equation} 

The ELBO has the following two properties, which are useful for our double-optimization problem:
\begin{align}
    \log p_{\theta}(\textbf{y}) & \geq {\rm ELBO} \label{eq:pELBO} \\
    \log p_{\theta}(\textbf{y}) - {\rm ELBO} & = {\rm KL}[  q_{\phi}(\textbf{z}) ||  p_{\theta}(\textbf{z}|\textbf{y}) ]. \label{eq:ELBO_KL}
\end{align}
The first property follows from the non-negativity of the KL-divergence.\footnote{This property was first derived using Jensen's Inequality~\cite{jordan1999introduction}: let $g(X)$ be a probability density function (PDF), then
\begin{equation} \label{eq:jensen}
    \log \mathbb{E}[g(X)] \geq \mathbb{E}[\log g(X)].
\end{equation}
} It implies that maximizing the ELBO will either move the ELBO closer to the log-evidence or push the log-evidence up with it. If the former occurs, then the variational distribution becomes a better approximation of the posterior, by virtue of Eq.~\eqref{eq:ELBO_KL}. If the latter occurs, then the model becomes better at explaining the data.

VI can be thought of as a type of maximum likelihood estimation. We can see this by rewriting the ELBO as
\begin{align} \label{eq:elbo_mle}
    {\rm ELBO} = \mathbb{E}_{q_{\phi}(\textbf{z})}[\log p(\textbf{y}|\textbf{z})] - {\rm KL}[ q_{\phi}(\textbf{z}) || p_{\theta}(\textbf{z}) ]. 
\end{align}
Therefore, maximizing the ELBO is the same as maximizing the expected log-likelihood, $\mathbb{E}_{q_{\phi}(\textbf{z})}[\log p(\textbf{y}|\textbf{z})]$, with a regularization term, $ {\rm KL}[  q_{\phi}(\textbf{z}) \ || \ p_{\theta}(\textbf{z}) ]$, that penalizes deviations of the variational distribution from the prior.

\subsection{Sparse and Variational GP ELBO} \label{app:svgp}

In our sparse and variational GP model (SVGP), we have the physical log-rates $\mathbf{f} = f(\mathbf{x})$ defined at the pixel centers $\mathbf{x}$ and the inducing points $\mathbf{u} = f(\mathbf{x_{u}})$ defined at the pseudo-inputs $\mathbf{x_{u}}$. Here, we derive the ELBO for an SVGP model consisting of a GP (with no additional sources) and an arbitrary likelihood term $p(\mathbf{y}|\mathbf{f})$. 

Without inducing points, the model is described by the joint PDF $p(\mathbf{y}, \mathbf{f}) = p(\mathbf{y} | \mathbf{f})p(\mathbf{f)}$. With inducing points, the model is augmented with a larger latent space containing the inducing points $\mathbf{u}$, and so it has a PDF $p(\mathbf{y}, \mathbf{u}, \mathbf{f}) = p(\mathbf{y} | \mathbf{u}, \mathbf{f})p(\mathbf{u}, \mathbf{f)}$. 

To approximate the posterior of this augmented model using SVI, we require a variational distribution. The standard SVGP formulation uses the following variational distribution: $q(\mathbf{u},\mathbf{f}) = p(\mathbf{f}|\mathbf{u})q(\mathbf{u})$, where $q(\mathbf{u})$ is typically a MVN distribution to give the marginal distribution $q(\mathbf{f})$ a closed-form expression (since $p(\mathbf{f}|\mathbf{u})$ is the PDF of a conditioned GP). Therefore, the variational distribution is not entirely flexible with respect to the latent variables $(\mathbf{u},\mathbf{f})$, unlike in traditional SVI. In this way, the fundamental problem of SVI is now a constrained optimization problem in the SVGP framework:
\begin{align} \label{eq:svgp-constrained}
    q^{*} & = \argmin_{q} KL[ q(\mathbf{u},\mathbf{f}) \ || \ p(\mathbf{u},\mathbf{f}|\mathbf{y}) ] \\
    q(\mathbf{u},\mathbf{f}) & = p(\mathbf{f}|\mathbf{u}) q(\mathbf{u}) \nonumber
\end{align}
Here, it is implicit that the optimal variational distribution $q^{*}$ is selected from a class of functions parameterized by $\phi$. The standard ELBO takes the same form as in traditional SVI, except with an augmented latent space:
\begin{align}
    {\rm ELBO} & = \mathbb{E}_{q(\mathbf{u},\mathbf{f})}\left [ \log p(\mathbf{y},\mathbf{u},\mathbf{f}) - \log q(\mathbf{u},\mathbf{f}) \right ]
\end{align}
We arrive at the expression of the ELBO in Eq.~\eqref{eq:svgp_elbo} by making the following manipulations:
\begin{align}
    {\rm ELBO} & = \mathbb{E}_{q(\mathbf{u},\mathbf{f})}\left [ \log p(\mathbf{y},\mathbf{u},\mathbf{f}) - \log q(\mathbf{u},\mathbf{f}) \right ] \nonumber \\
    & = \mathbb{E}_{q(\mathbf{u},\mathbf{f})}\left [ \log p(\mathbf{y}|\mathbf{u},\mathbf{f}) \right] - \mathbb{E}_{q(\mathbf{u},\mathbf{f})}\left [ \log q(\mathbf{u},\mathbf{f}) - \log p(\mathbf{u},\mathbf{f}) \right ] \nonumber \\
    & = \mathbb{E}_{q(\mathbf{u})}\left [ \mathbb{E}_{p(\mathbf{f}|\mathbf{u})} \left [ \log p(\mathbf{y}|\mathbf{f}) \right ] \right ] - \mathbb{E}_{q(\mathbf{u},\mathbf{f})}\left [ \log p(\mathbf{f}|\mathbf{u}) q(\mathbf{u})  - \log p(\mathbf{f}|\mathbf{u}) p(\mathbf{u}) \right ] \nonumber \\
    & = \mathbb{E}_{q(\mathbf{u})}\left [ \mathbb{E}_{p(\mathbf{f}|\mathbf{u})} \left [ \log p(\mathbf{y}|\mathbf{f}) \right ] \right ] - \mathbb{E}_{q(\mathbf{u})}\left [ \log q(\mathbf{u}) - \log p(\mathbf{u}) \right ]. \label{eq:aug_elbo_deriv}
\end{align}

\subsection{Fitting a GCE Template Model to the GP} \label{app:kl-poisson}
We present a technique for casting our best-fit GPs in terms of GCE template models. This allows us to interpret our best-fit GP posteriors in terms of physically-motivated GCE models (as we show in Figs.~\ref{fig:blg-fit},~\ref{fig:blg-fit-g1p2},~\ref{fig:all_gp}, and~\ref{fig:all_gp_g1p2}). On the other hand, it allows us to validate our fits on synthetic data by checking that we can recover the true \texttt{blg} and \texttt{nfw} parameters from the fit to the GP (as we show in Figs.~\ref{fig:temp_extract} and~\ref{fig:poiss_summary_gp}).  

Suppose we have a best-fit GP posterior $P(\textbf{f}) = q_{\theta,\phi}(\textbf{f})$, where $\textbf{f} = f(\textbf{x})$ denotes the log-rate of the GP at each unmasked pixel $\textbf{x}$. To interpret the GP posterior in terms of a set of GCE templates, we first set up a GCE template model. 

Let $\textbf{y}$ denote a Poisson draw from the GP posterior of Poisson samples. \textit{I.e.,} 
\begin{align}
    \textbf{y} & \sim {\rm Pois}[\lambda_{f}(\textbf{x})] \nonumber \\
    \lambda_{f}(\textbf{x}) & = \exp (\textbf{f}) \nonumber \\
    \mathbf{f} & \sim P(\mathbf{f}).
\end{align}
This defines the posterior predictive distribution $P(\textbf{y})$. A template-based model to the GCE would describe the GCE rate $\lambda_{f}(\textbf{x})$ using fixed spatial templates, which would be expressed using an equation similar to Eq.~\eqref{eq:first-rate-eqn} (or Eq.~\eqref{eq:multi-rate-eqn} if we are using a linear combination of templates). That is, its latent variables $\textbf{z}$ consist of normalizations, relative normalizations, and other parameters affecting the shape of the template (\textit{e.g.,} the NFW $\gamma$ parameter). Let $\varphi$ denote parameters affecting the prior distribution of the latent variables $q_{\varphi}(\textbf{z})$. Such a model predicts a distribution of Poisson samples from a distribution of latent variables $\textbf{z}$ determined by $\varphi$: 
\begin{align} \label{eq:temp-model}
    q_{\varphi}(\textbf{y}) = \mathbb{E}_{q_{\varphi}(\textbf{z})}[p(\textbf{y} | \textbf{z})],
\end{align}
where $p(\textbf{y} | \textbf{z})$ is the (Poisson) likelihood of the model (since each $\textbf{z}$ corresponds to a rate $\lambda(\textbf{x})$).

To interpret the best-fit GP posterior in terms of the template-based model, we require a version of the template-based models whose predictions match that of the GP posterior. \textit{I.e.,} our objective is to maximize the similarity between $P(\textbf{y})$ and $q_{\varphi}(\textbf{y})$, which can be achieved with the KL-divergence:
\begin{equation} 
    \varphi^{*} = \argmin_{\varphi} {\rm KL}[P(\textbf{y}) \ || \  q_{\varphi}(\textbf{y})]. \label{eq:KLpoisson_phi}
\end{equation}

Note the similarity between this objective and the SVI objective Eq.~\eqref{eq:svi_phistar}. We will pose our current optimization problem in a way that can be solved with SVI. We do this by performing the following manipulations:
\begin{align} \label{eq:klpoisson-obj}
    {\rm KL}[P(\textbf{y}) \ || \ q_{\varphi}(\textbf{y})] & =  \mathbb{E}_{P(\textbf{y})}[\log P(\textbf{y}) - \log p_\varphi (\textbf{y})] \nonumber \\
    & \sim -\mathbb{E}_{P(\textbf{y})}[\log q_{\varphi}(\textbf{y})], \nonumber \\
    & = -\mathbb{E}_{P(\textbf{y})}[\log \mathbb{E}_{q_{\varphi}(\textbf{z})}[p(\textbf{y} | \textbf{z})]] \nonumber \\
    & \lesssim  
    -\mathbb{E}_{P(\textbf{y})}[\mathbb{E}_{q_{\varphi}(\textbf{z})}[\log p(\textbf{y} | \textbf{z})]].
\end{align}
In the second line, the $\sim$ means that we ignore the $\mathbb{E}_{P(\textbf{y})}[P(\textbf{y})]$ as it has no effect on the optimization of $\varphi$. To get to the third line, we used Eq.~\eqref{eq:temp-model}. To get to the fourth line, we used Jensen's Inequality (Eq.~\eqref{eq:jensen}). As we encountered in Eq.~\eqref{eq:elbo_mle}, our current optimization problem can be treated as a variant of maximum likelihood estimation: maximizing the expected log-likelihood, $\mathbb{E}_{P(\textbf{y})}[\mathbb{E}_{q_{\varphi}(\textbf{z})}[\log p(\textbf{y} | \textbf{z})]]$, minimizes the original KL-divergence, ${\rm KL}[P(\textbf{y}) \ || \ q_{\varphi}(\textbf{y})]$. 

Let's consider how our expression is related to SVI. In a traditional SVI problem, we have a set of data $\textbf{y}$ and our objective is to maximize the ELBO (see Eq.~\eqref{eq:elbo_mle}):
\begin{align}
    {\rm ELBO} = \mathbb{E}_{q_{\varphi}(\textbf{z})}[\log p(\textbf{y} | \textbf{z})]] - {\rm KL}[q_{\varphi}(\textbf{z}) \ || \ p(\textbf{z})],
\end{align}
where $p(\textbf{z})$ denotes the model's prior distribution. By maximizing the ELBO, we maximize the accuracy of $q_{\varphi}(\textbf{z})$ as an approximation to the posterior $p(\textbf{z}|\textbf{y})$. Suppose we know a distribution $P(\mathbf{y})$ instead of a particular outcome $\textbf{y}$, as in our current case. Consider the following expression:
\begin{align} \label{eq:klpoisson-elbo}
    \mathbb{E}_{P(\textbf{y})}[{\rm ELBO}] = \mathbb{E}_{P(\textbf{y})}[\mathbb{E}_{q_{\varphi}(\textbf{z})}[\log p(\textbf{y} | \textbf{z})]] - {\rm KL}[q_{\varphi}(\textbf{z}) \ || \ p(\textbf{z})].
\end{align}
By the non-negativity of the KL-divergence, this expected ELBO,  $\mathbb{E}_{P(\textbf{y})}[{\rm ELBO}]$, is a lower bound to the expected log-likelihood, $\mathbb{E}_{P(\textbf{y})}[\mathbb{E}_{q_{\varphi}(\textbf{z})}[\log p(\textbf{y} | \textbf{z})]]$, and hence can be used to minimize Eq.~\eqref{eq:klpoisson-obj}. Therefore, by choosing this expected ELBO as our objective to maximize, we have changed objectives: instead of maximizing the accuracy of $q_{\varphi}(\textbf{z})$ as an approximation to the posterior corresponding to a single dataset $p(\textbf{z} | \textbf{y})$, we maximize the accuracy of $q_{\varphi}(\textbf{y})$ as an approximation to the distribution $P(\textbf{y})$. Thus, Eq.~\eqref{eq:klpoisson-elbo} is the objective we use to fit GCE template models to GP posteriors. Note, this objective has the additional KL-divergence term ${\rm KL}[q_{\varphi}(\textbf{z}) \ || \ p(\textbf{z})]$ acting as a regularization, penalizing deviations of $q_{\varphi}(\textbf{z})$ from a chosen prior $p(\textbf{z})$. When we specify a template model prior in the main text (\textit{e.g.,} prior on the NFW $\gamma$ parameter), we are specifying $p(\textbf{z})$. We have verified that the fit is stable with respect to variations in the assumed prior.

The only difference between the new objective function Eq.~\eqref{eq:klpoisson-elbo} and the standard ELBO (Eq.~\eqref{eq:elbo_mle}) is the extra expectation on the likelihood term. Therefore, fitting the template model to the GP posterior only requires an additional step to traditional SVI: we draw samples of $\textbf{y}$ from the GP posterior $P(\textbf{y})$, in addition to samples from the template model variational distribution $q_{\varphi}(\textbf{z})$, to estimate the expected ELBO in Eq.~\eqref{eq:klpoisson-elbo} at each SVI step. Training (backpropagation) is unchanged. For all of our fits to GP posteriors, we take $q_{\varphi}(\textbf{z})$ to be a multivariate normal distribution as we defined in Eq.~\eqref{eq:q-mvn} to model the correlations between the various latent space variables in our models.

\begin{figure}[t!] 
    \centering
    \includegraphics[width = \columnwidth]{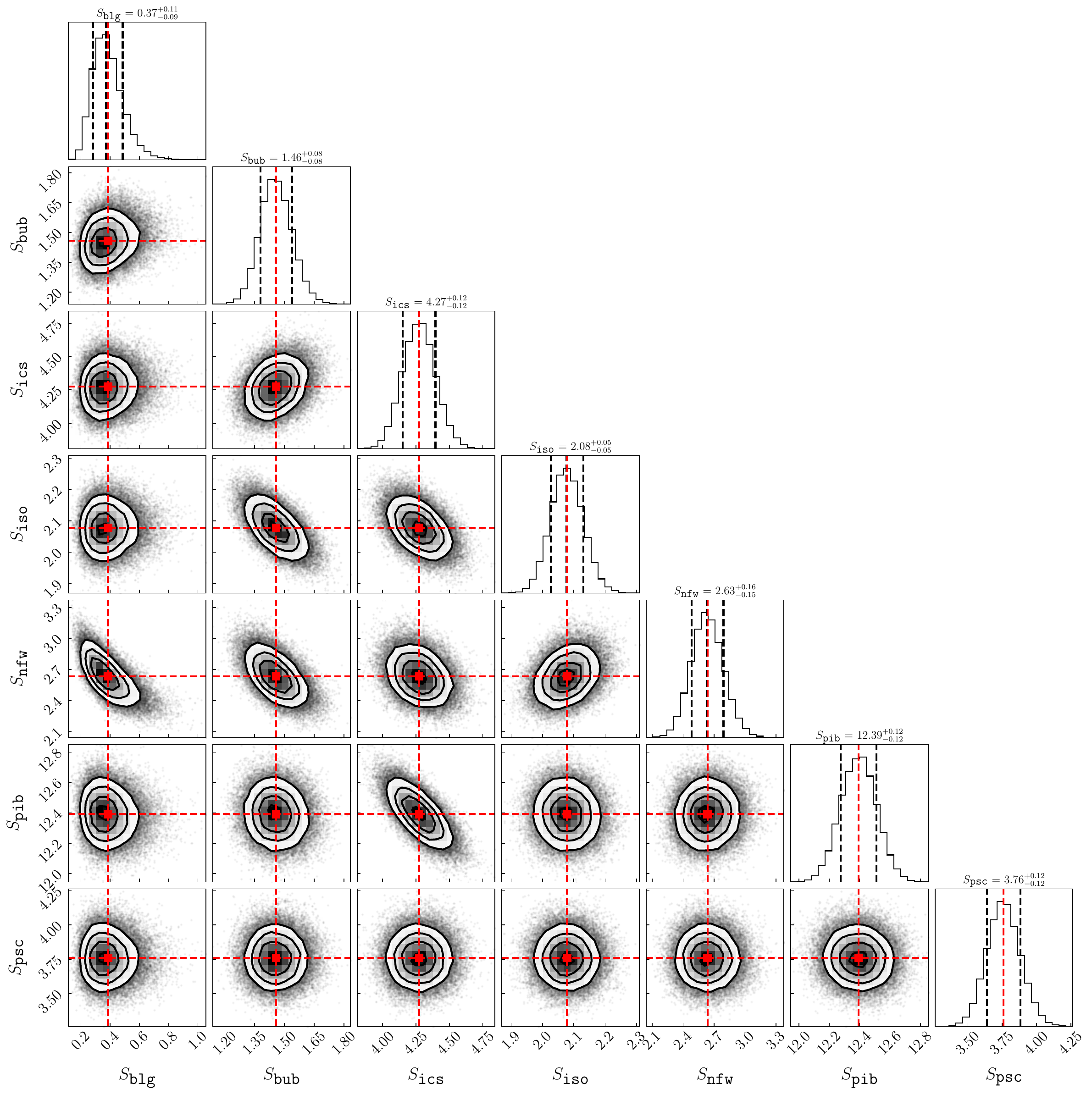}   
    \caption{Corner plot of samples from the approximate posterior distribution $q_{\phi}(\textbf{z})$ obtained by fitting the template-based model to the data. The black dashed lines denote 16\%, 50\%, and 84\% percentiles. The red dashed lines denote the means of the normalizations $S_{j}$ used to generate synthetic data.} \label{fig:syndata}
\end{figure}

\section{Model Performance on Synthetic Data} \label{app:performance}

We continue from the end of Section~\ref{sec:syndata}. This section serves to validate the model for the settings given in that section. In this appendix, we restrict ourselves to the specific ROI settings from that section (variations of the ROI are discussed in Appendix~\ref{app:roi}). In particular, we consider the case where all of our template models accurately model the corresponding source of gamma-ray emission. We study the variations of our fits with respect mismodeling in Appendix~\ref{app:mismodel}. 

Our synthetic data is generated from a purely template-based model of the gamma-ray emission in our Complete ROI, as described in the beginning of Section~\ref{sec:syndata}. We show the normalization posteriors obtained by fitting the template-based model of Galactic gamma-ray emission to the \textit{Fermi} data in Fig.~\ref{fig:syndata}. This is the result of an SVI fit corresponding to $10^5$ steps, an Adam optimizer with a constant learning rate of $2\times 10^{-4}$, and 8 samples used to estimate the ELBO. The normalizations used to generate our synthetic gamma-ray sources are chosen to be the means of these normalization posteriors.

\begin{figure}[t!] 
    \centering
    \includegraphics[width = 0.6\columnwidth]{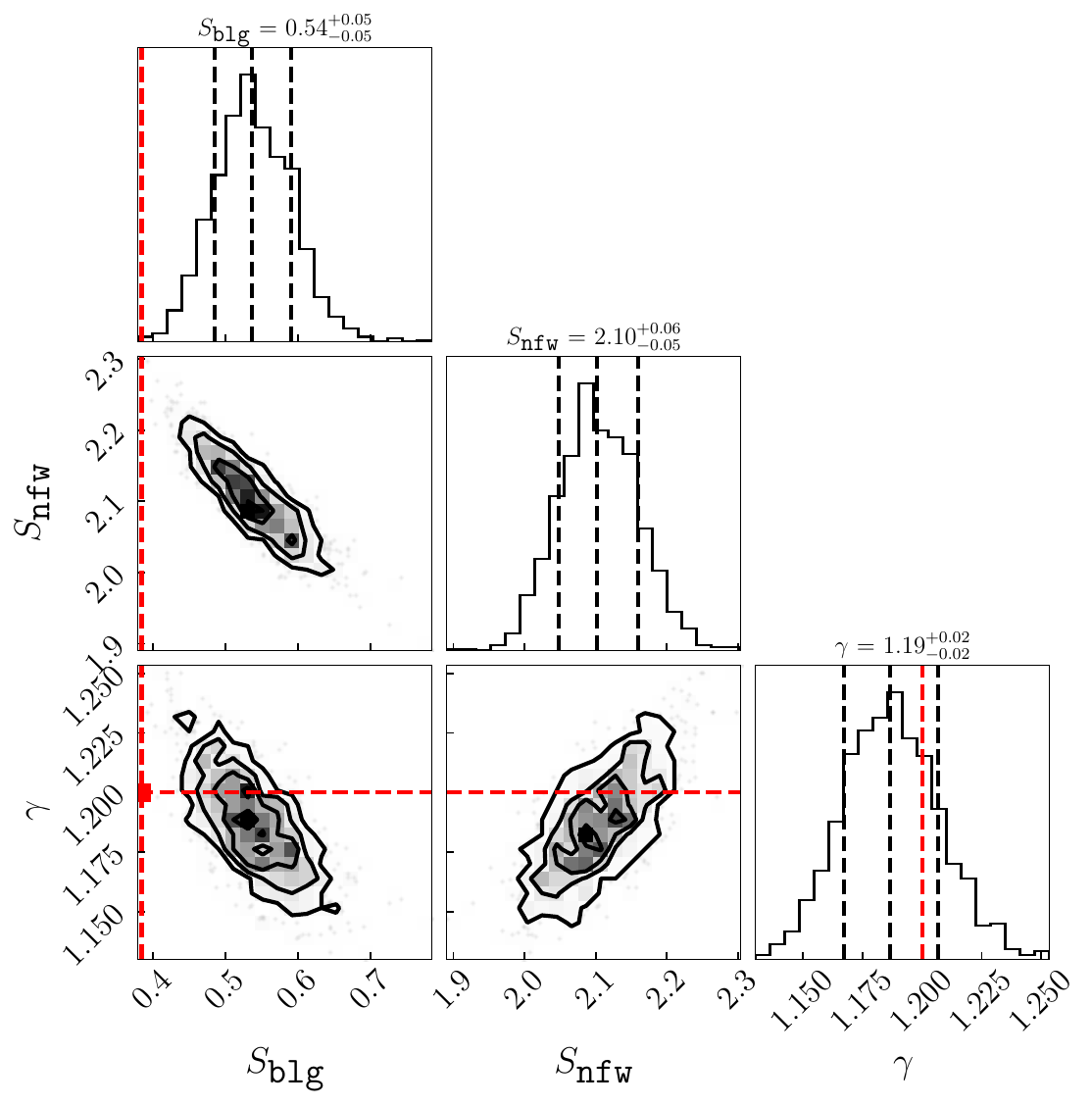}   
    \caption{Posterior of GCE template model fit to the GP posterior that was fit to synthetic data. True parameters are denoted by the red dashed lines.} \label{fig:temp_extract}
\end{figure}

\subsection{Accuracy Tests} \label{app:accuracy}

\subsubsection{Extracting Templates from GP Posterior} \label{app:extract}

Our results from Section~\ref{sec:syndata} demonstrate that the model contains the true rate $\lambda(\textbf{x})$ with high probability. 
In addition, we show that we can recover the template parameters from the GP after fitting the equivalent GCE template model to the best-fit GP posterior, verifying that the GP contains the rates sourced by the templates of the synthetic data. 
As per the discussion in Appendix \ref{app:kl-poisson}, we perform the fit by maximizing Eq.~\eqref{eq:klpoisson-elbo}. The results are shown in Fig.~\ref{fig:temp_extract}. The disagreement between the template model's posterior and the true GCE template parameters indicates that the template model's predictions are overconfident. The overconfidence is the result of the template model's lack of flexibility, limiting its ability to fit the GP posterior.

\begin{figure}[t!] 
    \centering
    \includegraphics[width = 0.8\columnwidth]{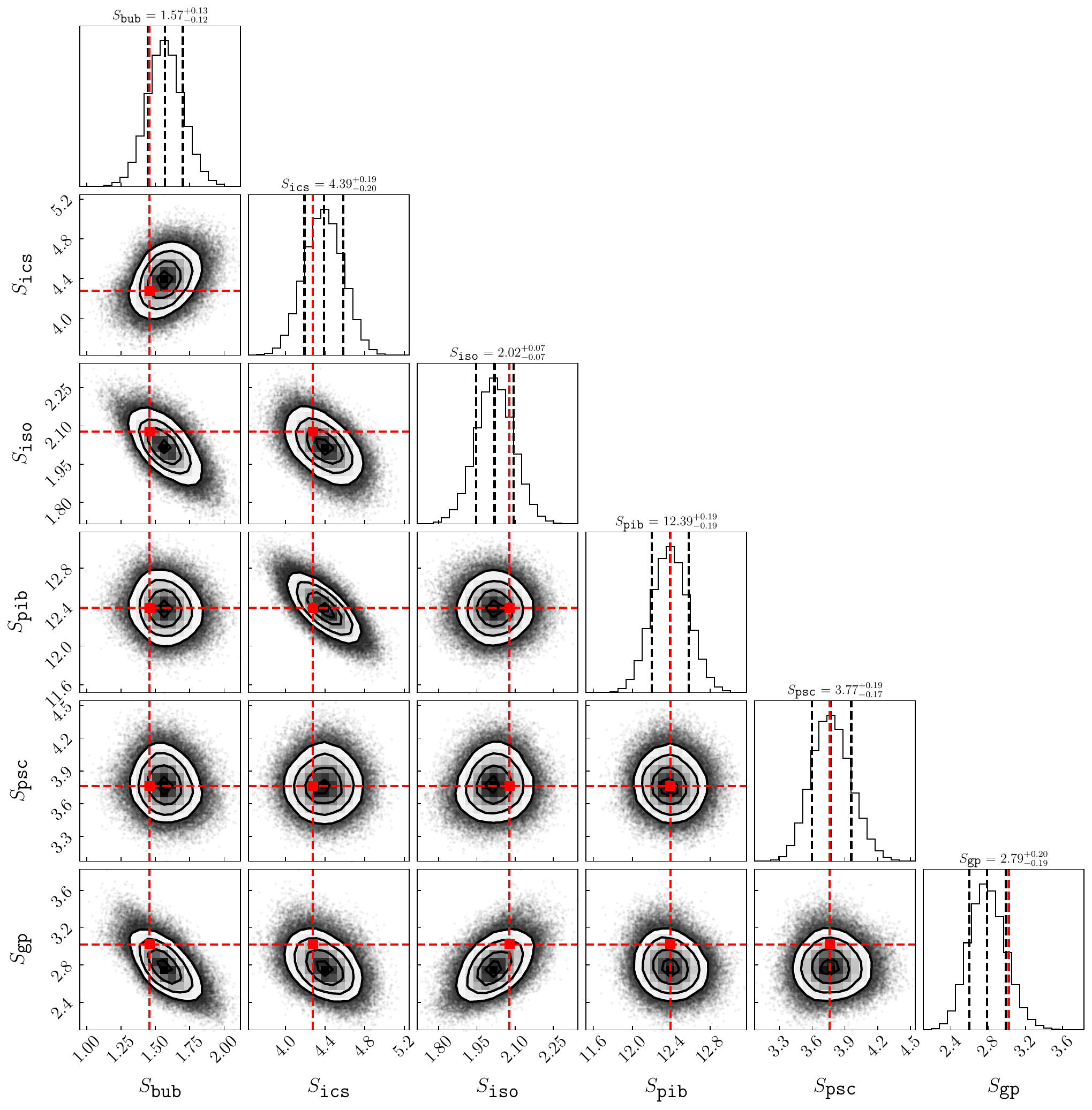} 
    \includegraphics[width = 0.8\columnwidth]{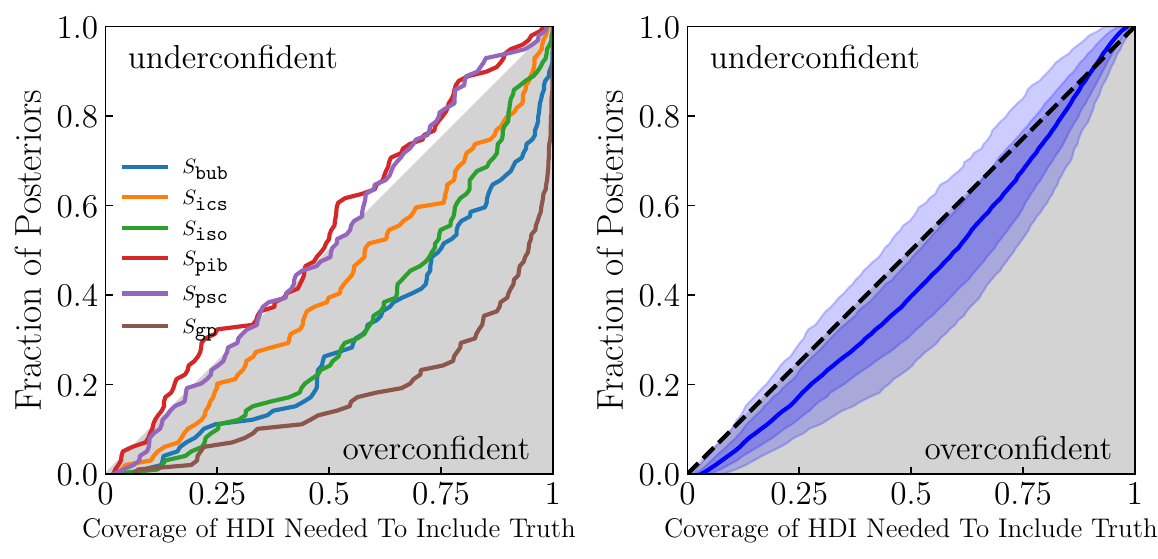}    
    \caption{Summary of fit results of fitting the GP model to 100 Poisson samples drawn from synthetic gamma-ray sources. \textit{(Upper Panel)} Average posterior of the gamma-ray source normalizations. True normalizations are denoted by the red dashed lines. \textit{(Lower Left)} Coverage plots of the gamma-ray source normalization posteriors. \textit{(Lower Right)} Coverage plot of 1,000 randomly-selected GP pixel rate posteriors (\textit{i.e.,} posteriors of $\exp[f(x_{i})]$).} \label{fig:poiss_summary}
\end{figure}

\subsubsection{Removing Poisson Noise} \label{sec:poiss}

 We expect our fit to be biased from the presence of Poisson noise in the measured counts, as we are fitting to only a single Poisson sample. Working with synthetic data, we can fit to many Poisson draws of the synthetic gamma-ray sources to reduce the effect of Poisson noise. We do this by performing identical fits to 100 Poisson samples and averaging the posteriors that we get from each fit. The results of the fits are shown in Fig.~\ref{fig:poiss_summary}. As we see in the left panel, the averaged posteriors of the various gamma-ray sources retain some bias. 

In the lower panels of Fig.~\ref{fig:poiss_summary}, we show two plots that quantify the confidence of the model's predictions. In the lower left panel of Fig.~\ref{fig:poiss_summary}, we plot the fraction of the 100 posteriors that contain the ground truth within a specific coverage of the \textit{Highest Density Interval} (HDI) -- generating a curve for each parameter. The coverage plots demonstrate again that the fainter templates and the GP are overconfident. Nevertheless, the individual pixel values of the GP are less overconfident than the overall normalization suggests, as we show by making coverage plots of 1,000 GP-values at randomly selected pixels in the lower right panel of Fig.~\ref{fig:poiss_summary}.

\begin{figure}[t!] 
    \centering
    \includegraphics[width = 0.5\columnwidth]{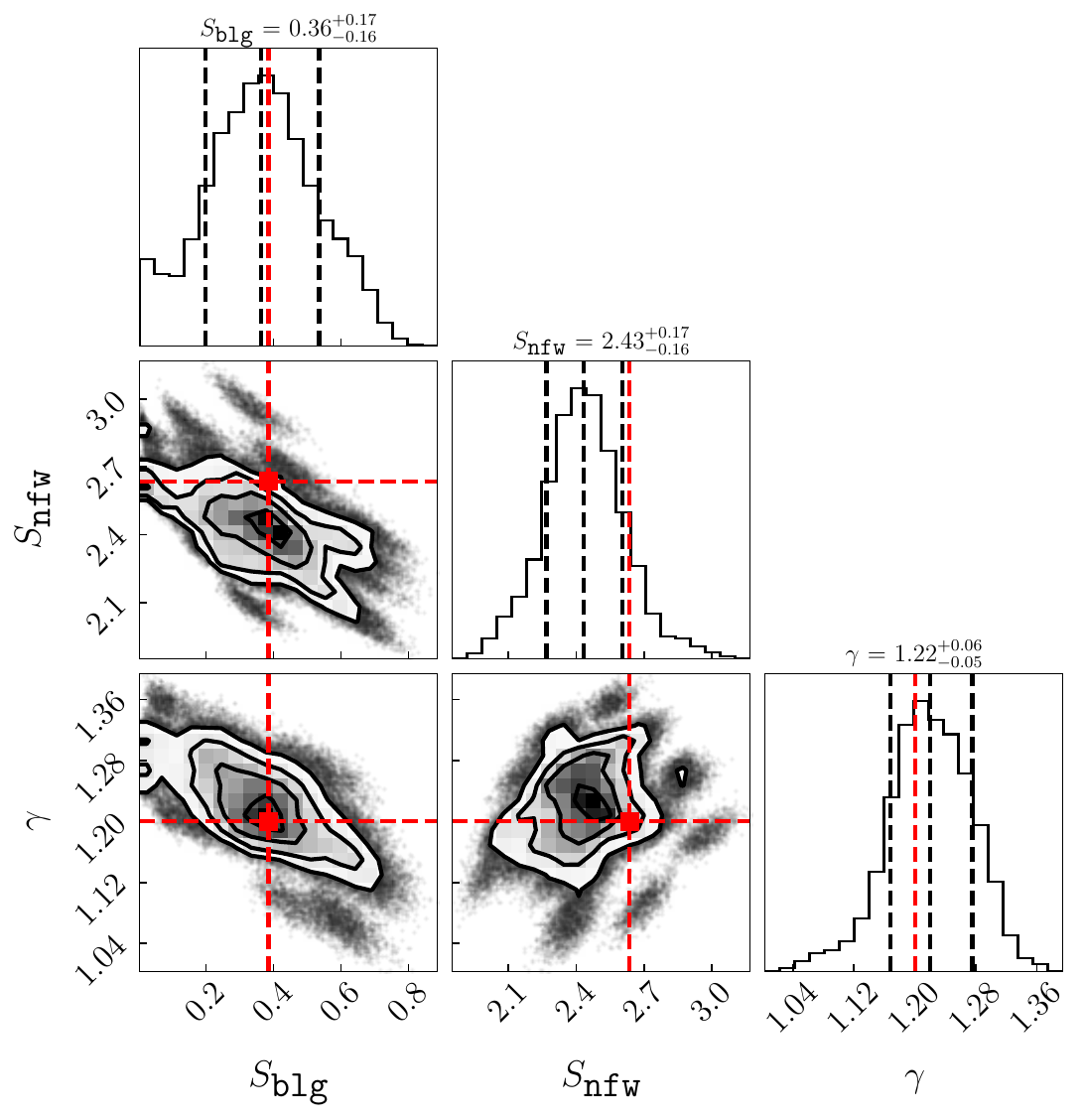}\includegraphics[width = 0.5\columnwidth]{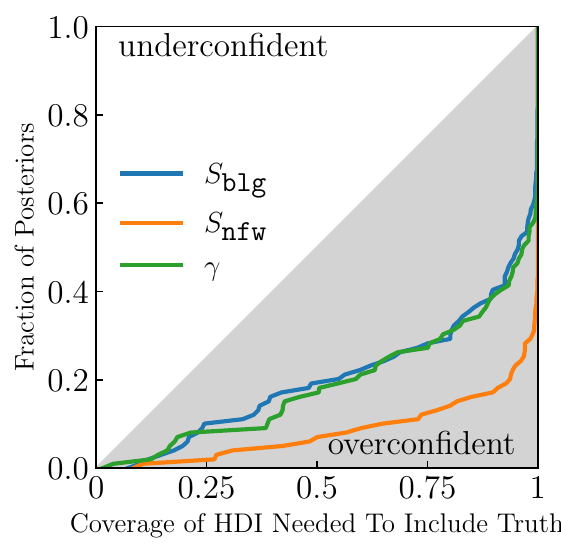}   
    \caption{Summary of results of fitting a template GCE model to 100 GP posteriors resulting from fits to different Poisson draws of the same synthetic gamma-ray sources. \textit{(Left)} Average posterior of the template parameters. True parameters are denoted by the red dashed lines. \textit{(Right)} Coverage plot of each GCE template parameter.} \label{fig:poiss_summary_gp}
\end{figure}

In addition to fitting 100 GP models, we also fit to the 100 GP posteriors using the GCE template model used to generate the data. The results are shown in Fig.~\ref{fig:poiss_summary_gp}. The templates extracted from each of our 100 fitted GP models are also centered near their true values, with roughly non-Gaussian shapes due to the overconfidence of each individual model (illustrated by the narrow posteriors such as Fig.~\ref{fig:temp_extract}). Therefore, when our template models are chosen correctly, the majority of uncertainty in our template fit to the GP arises from Poisson noise. The real data will have an additional systematic uncertainty corresponding to our choice of templates (see Appendix~\ref{app:mismodel}).

\subsection{Stability Tests} \label{app:stability}

In this section, we study to what extent the GP model fits change when we change either the GP model hyperparameters or the underlying GCE.

\subsubsection{GP Kernel Selection} \label{app:kernels}

We perform the same fit as we did in Section~\ref{sec:syndata}, using different kernels for the GP prior instead of the Matern32 kernel used in the main analysis (Eq.~\eqref{eq:matern}). We consider three additional kernels~\cite{rasmussen}:
\begin{align}
{\rm Matern52:}~\quad    k_{\theta}(x_1, x_2) & = \sigma^{2} (1 + \sqrt{5} r + 5r^{2} / 3) \exp(-\sqrt{5} r) \\
{\rm ExpSquared:}~\quad    k_{\theta}(x_1, x_2) & = \sigma^{2} \exp(-r^2 / 2) \\
{\rm RationalQuadratic:}~\quad    k_{\theta}(x_1, x_2) & = \sigma^{2} (1 + r^{2} / 2 \alpha)^{-\alpha},
\end{align}
where the distance $r$ is defined in Eq.~\eqref{eq:dist} with length-scale parameter $\rho$. The Matern52 kernel is similar to the Matern32 kernel, but is a distribution of second-order (rather than first-order) differentiable functions. The ExpSquared (radial basis function, RBF) kernel is the most widely used kernel and generates infinitely-differentiable functions. The RationalQuadratic kernel is equivalent to adding RBF kernels with different length scale parameters, as a result it varies smoothly across a range of length scales. Its functions are also infinitely-differentiable. The relative weighting of large- and small-scale variations is determined by $\alpha$, which we allow to float during the fit. 

\begin{figure}[t!] 
    \centering
    \includegraphics[width = 1\columnwidth]{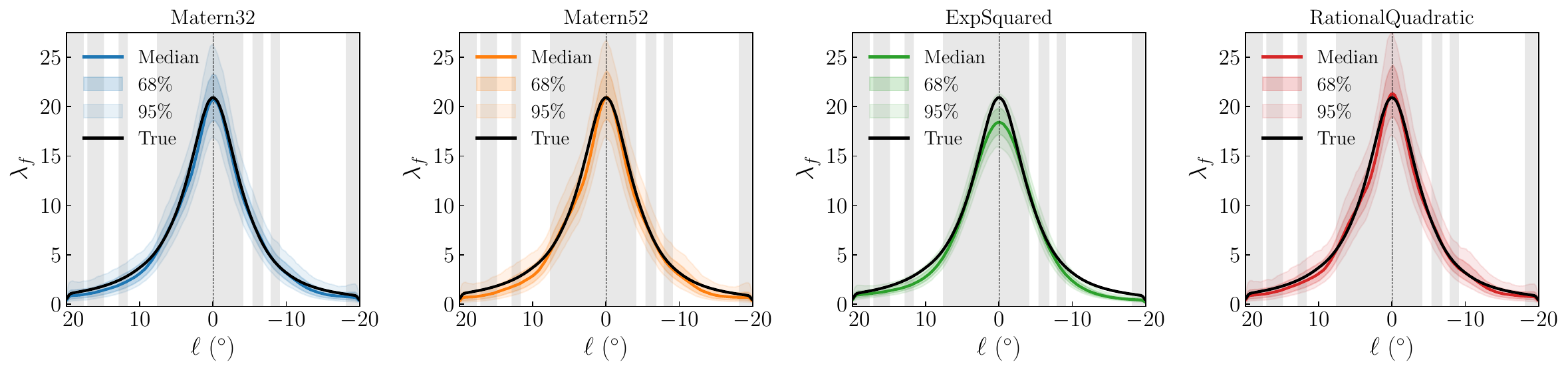}   
    \caption{Horizontal 1D posterior slices of $\lambda_{\rm gce}$ at $b = 3.375^{\circ}$ for fits with difference assumed GP kernel.} \label{fig:kernels}
\end{figure}

The result of the fits is shown in Fig.~\ref{fig:kernels}. The GCE posterior corresponding to the RBF kernel differs significantly from the others. This is due to the RBF's functions being unable to exhibit variations at scales smaller than the length-scale parameter $\rho$. Out of the remaining three kernels, we adopt the Matern32 kernel for our analysis since it is the most flexible (as the restrictions on the smoothness of its functions are more relaxed, being first-order differentiable as opposed to second-order or infinitely differentiable).

\subsubsection{Number of Inducing Points} \label{app:Nu}

Our fits are roughly independent of our choice for the number of inducing points $N_u$ (defined in Appendix~\ref{app:inducing}), as shown in Fig.~\ref{fig:nu_scan}, where we show normalized versions of the posteriors of each gamma-ray source corresponding to fits using different values for $N_u$. However, since the minimum kernel length scale is determined by the spacing between the inducing points, we choose a fairly large value of $N_u=300$ to allow for greater flexibility at the expense of computation time and memory. This choice corresponds to neighboring inducing points being separated by $\sim 1.15^{\circ}$. 

\begin{figure}[t!] 
    \centering
    \includegraphics[width = \columnwidth]{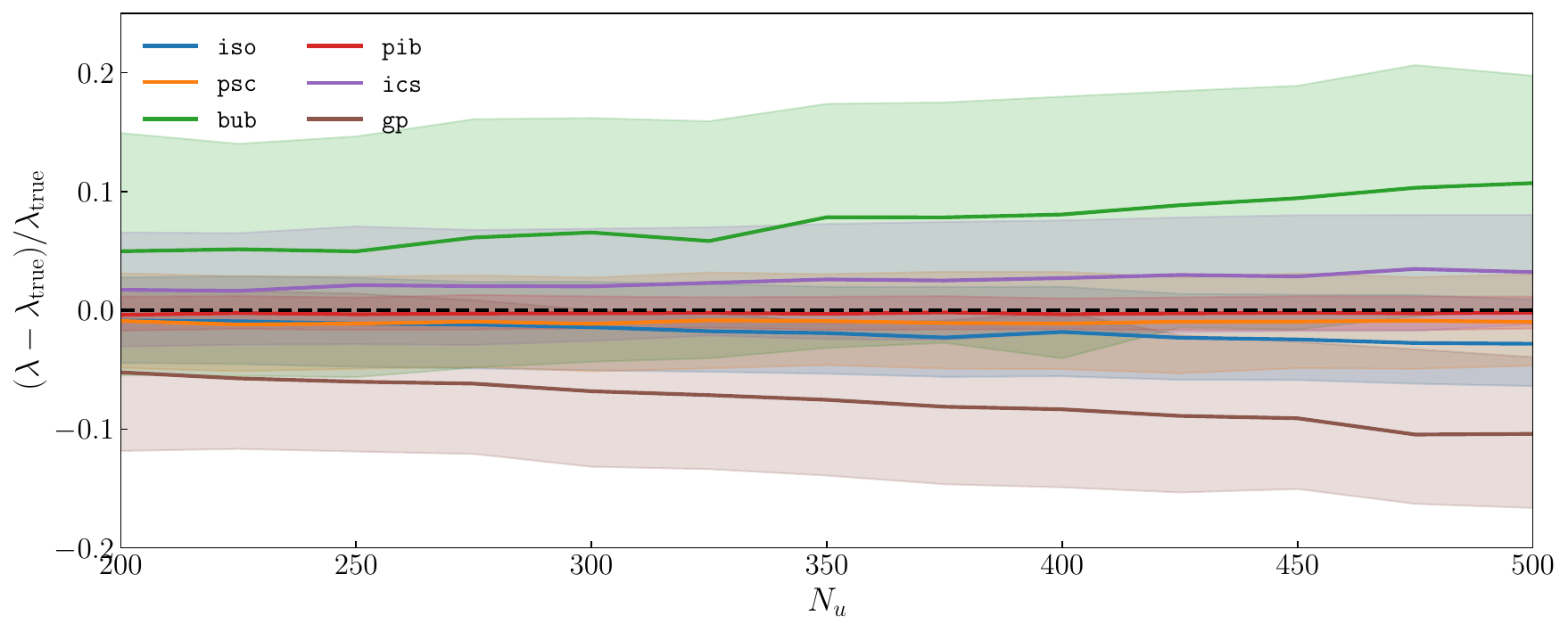}   
    \caption{Posteriors of gamma-ray components' total rate summed over the Inner ROI versus the $N_u$ of the GP model used to fit the GCE. The vertical axis is the relative difference $(\lambda - \lambda_{\rm true}) / \lambda_{\rm true}$ corresponding to each gamma-ray component. Each set of posteriors is the average of ten fits each using different Poisson draws of the synthetic rate as data to reduce the Poisson fluctuations of the result.} \label{fig:nu_scan}
\end{figure}

\subsubsection{Varying GCE Morphology} \label{app:blg_scan}

\begin{figure}[t!] 
    \centering
    \includegraphics[width = 1\columnwidth]{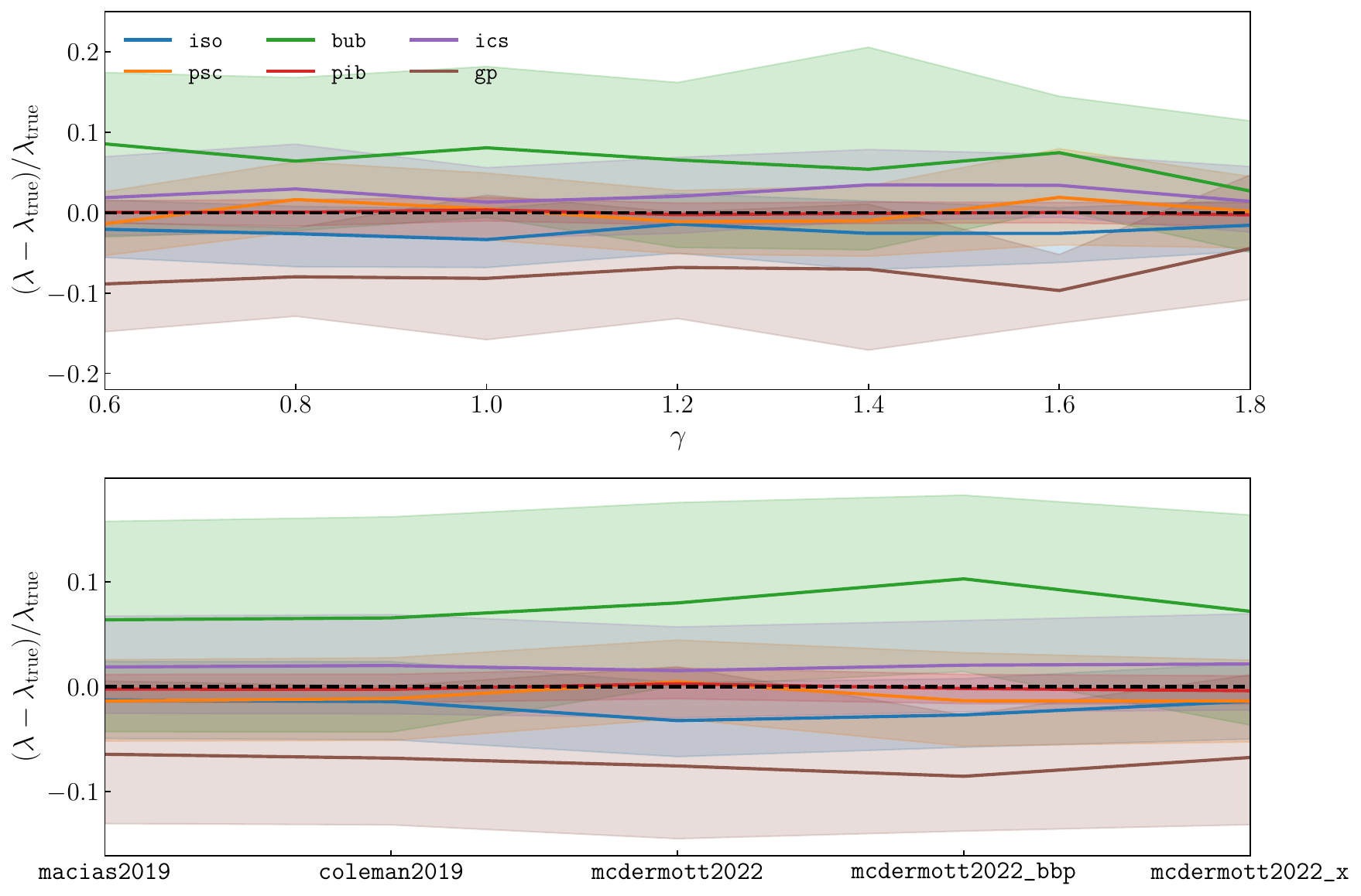}   
    \caption{Posteriors of gamma-ray components' total rate summed over the Inner ROI versus the spatial morphology of the synthetic GCE. The y-axis is the relative difference $(\lambda - \lambda_{\rm true}) / \lambda_{\rm true}$ corresponding to each gamma-ray component. Each set of posteriors is the average of ten fits each using different poisson draws of the synthetic rate as data to reduce the Poisson fluctuations of the result. \textit{(Left)} Posteriors versus NFW $\gamma$ parameter, where the bulge template is fixed to the \texttt{coleman2019} template. \textit{(Right)} Posteriors versus bulge template choice, where the NFW $\gamma$ parameter is fixed to 1.2.} \label{fig:syn_scan}
\end{figure}

We check how our fit is affected by the morphology of the underlying GCE, tested by varying the templates used to generate the synthetic data. We run our standard GP fit from Section~\ref{sec:syndata} on synthetic gamma-ray sources with the same normalizations as the fit in Section~\ref{sec:syndata}, varying the NFW $\gamma$ parameter and selecting different bulge templates to create the synthetic GCE. The results are shown in Fig.~\ref{fig:syn_scan}, which indicate that the quality of our fits are not affected significantly by our choice of GCE template.

\subsubsection{Removing the GCE}
\begin{figure}[t!] 
    \centering
    \includegraphics[width = 1\columnwidth]{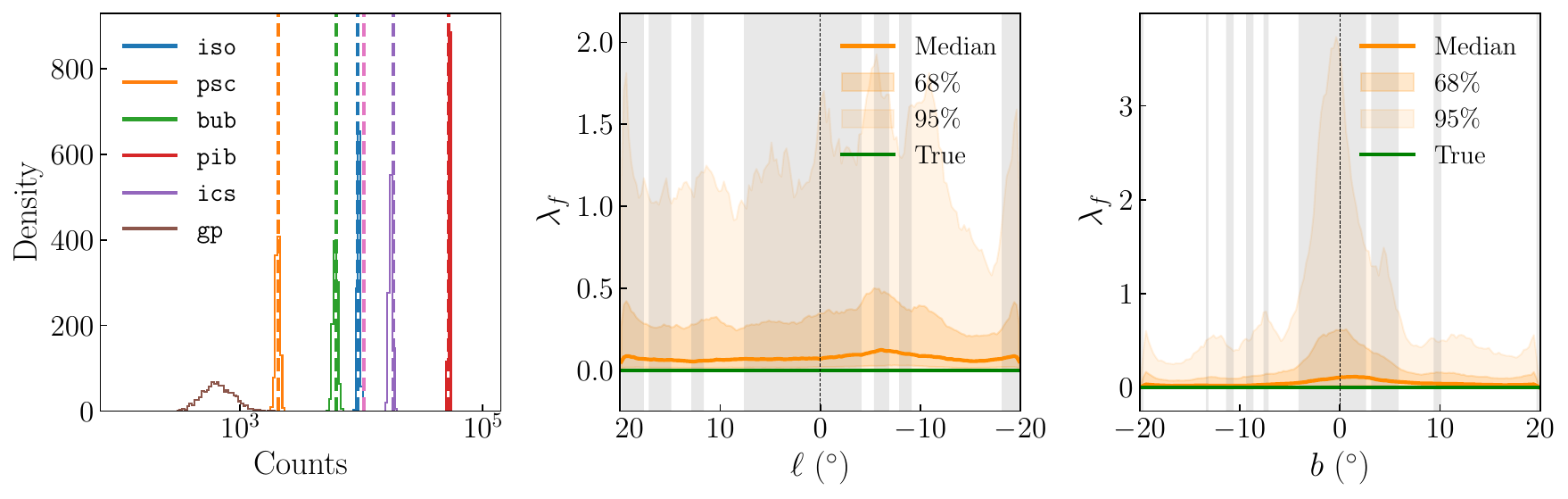} 
    \includegraphics[width = 0.66\columnwidth]{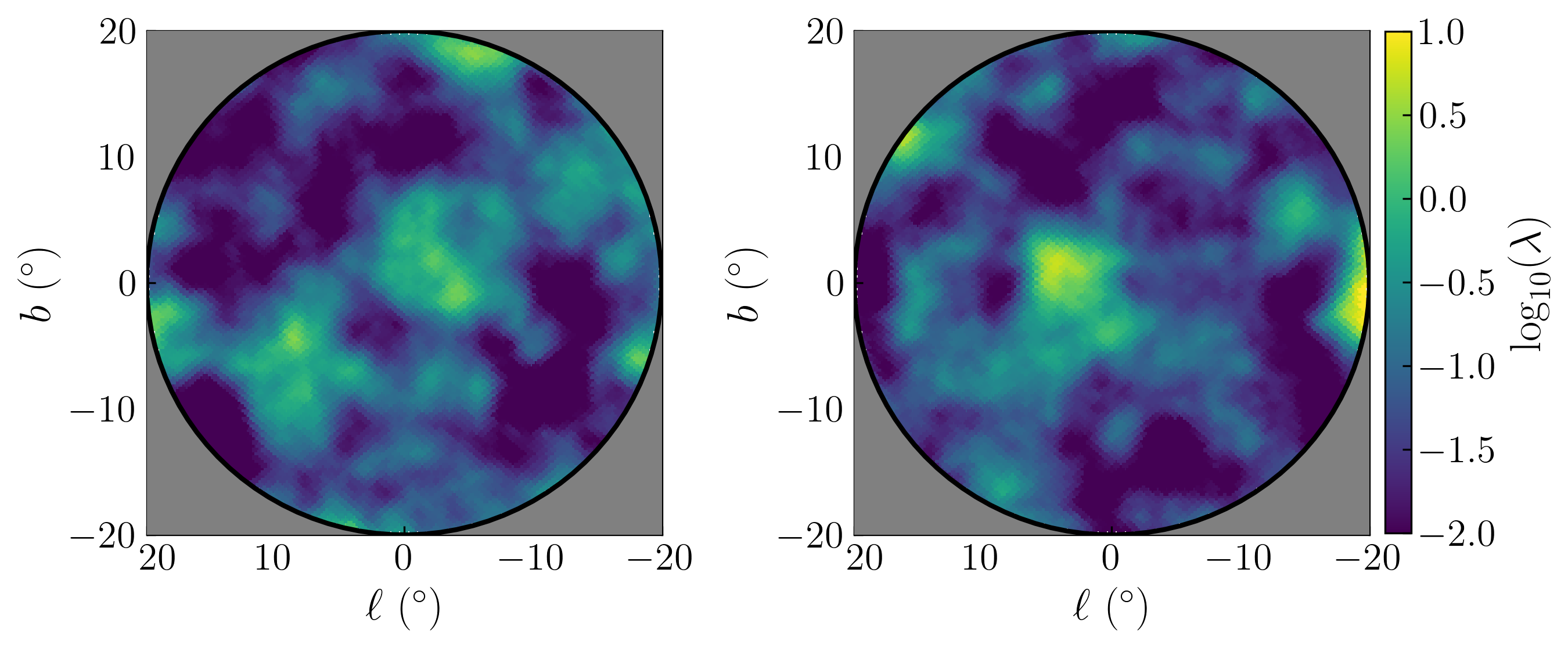}     
    \caption{Result of fit to synthetic data containing no GCE. \textit{(Upper Left)} Posterior samples of the total rate of all gamma-ray components, binned logarithmically. \textit{(Upper Center)} Horizontal 1D posterior slice of $\lambda_{\rm gce}$ at $b = 3.375^{\circ}$. \textit{(Upper Right)} Vertical 1D posterior slice of $\lambda_{\rm gce}$ at $\ell = -3.375^{\circ}$, and posteriors of the expected total number of counts in the ROI of every gamma-ray source. \textit{(Lower Row)} Posterior samples from the GP.} \label{fig:zero_gp}
\end{figure}

In this section, we verify that the GP model successfully fits the data when no GCE component is present. We generate the pseudodata using all the synthetic gamma-ray sources from Section~\ref{sec:syndata}, except the GCE. The results of this fit are summarized in Fig.~\ref{fig:zero_gp}. The top row shows that the GCE posterior is very faint, but nonzero. This is a consequence of the positivity of the GCE rate, stemming from our choice to model the GCE rate as the exponential of a function: $\lambda_{\rm gce}(\mathbf{x}) = \exp(f(\mathbf{x}))$. The bottom row shows that there is no clear preference for an excess signal in the Galactic Center.

\subsubsection{Different Point Source Masks} \label{app:ps}
\begin{figure}[t!] 
    \centering
    \includegraphics[width = 1\columnwidth]{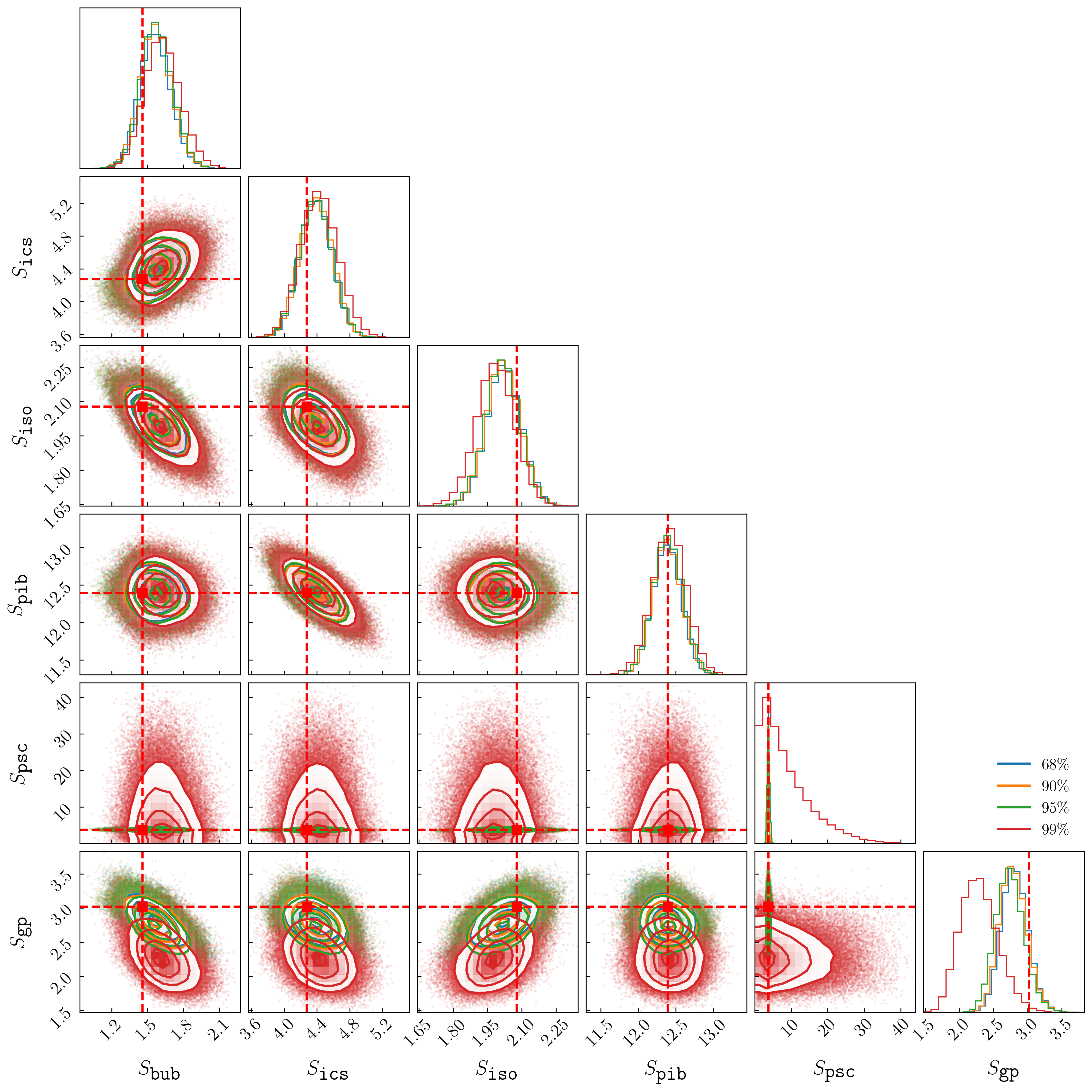}   
    \caption{Comparison of average posteriors corresponding to 100 fits for four different choices of point source masks. The true normalizations are denoted by the dashed red lines.} \label{fig:mask_res}
\end{figure}

Here, we show how the results of our fits change with respect to point source masking. We consider four point source masks with different containment fractions: 68\%, 90\%, 95\%, and 99\%. As in Appendix~\ref{sec:poiss}, we perform 100 identical fits to the synthetic data from Section~\ref{sec:syndata} for each choice of point source masks. The results are shown in Fig.~\ref{fig:mask_res}. The only fit that behaves differently is the one corresponding to masking at the 99\% containment level. The large masks make it difficult to constrain the point source catalog template. This introduces a degeneracy between the GP and the point source catalog templates, which leads to higher variance in both of their posteriors. There are regions in the sky that could be explained by the GP if the point sources were constrained, but since they are weakly constrained these regions can be explained by artificially increasing the intensity of the point sources' tails.

\subsection{Effects of Mismodeling Diffuse Emission Templates on GP Fit} \label{app:mismodel}

In the following plots, we quantify the extent to which mismodeling affects our fits. To do this, we generate three different sets of gamma-ray sources each assuming different diffuse emission models: Models O, A, or F for the \texttt{ics} and \texttt{pib} templates. The rest of the templates are the same as used in Section~\ref{sec:syndata}. The normalizations of all the synthetic templates are those from Section~\ref{sec:syndata}. The SVI fits run for 20,000 steps, but are otherwise identical to those presented in Section~\ref{sec:syndata}

\subsubsection{Mismodeling with a Single Diffuse Emission Model} \label{sssec:mismodel-single}

First, we compare fits using one of the three diffuse models to each of the synthetic datasets. That is, we consider fits to pseudodata generated using Model O with a model that assumes Model O, Model O data assuming Model A, 
and all other combinations.
To reduce the effect of Poisson noise, we take as our posterior the average of 10 posteriors each obtained by fitting to a different Poisson draw of the rate $\lambda(\textbf{x})$ generated by the synthetic gamma-ray sources.

\begin{figure}[t!] 
    \centering
    \includegraphics[width=0.8\columnwidth]{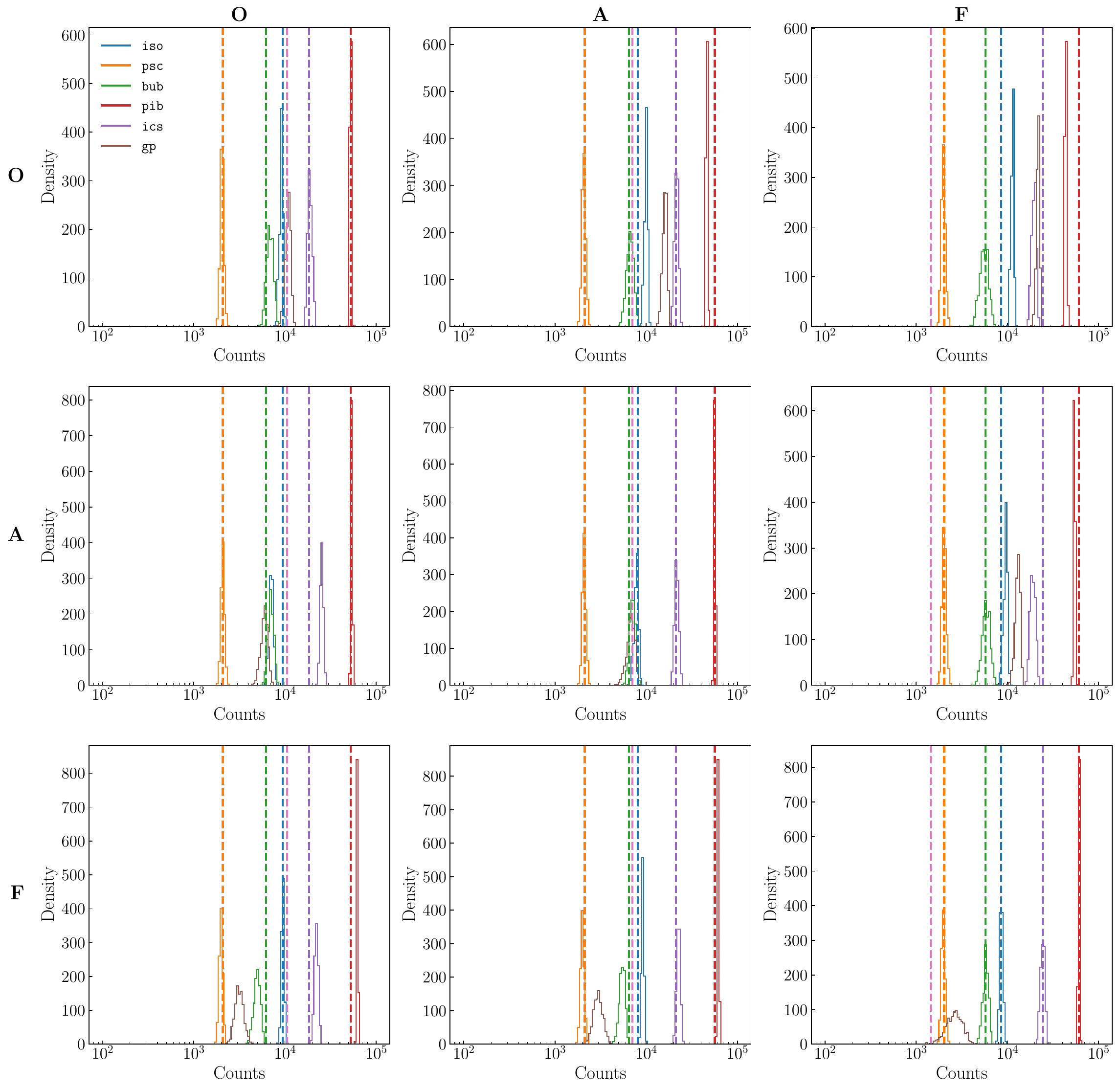}   
    \caption{Results from fits described in Section~\ref{sssec:mismodel-single}. Color-coded histograms showing the expected total number of counts of each gamma-ray source in the Inner ROI and are binned logarithmically. True normalizations are denoted by dashed lines. Each plot corresponds to different fits. Rows indicate the diffuse model assumed to fit the data. Columns indicate the synthetic dataset generated using one of the diffuse models. All posteriors are the average of 10 posteriors each obtained by fitting to a different Poisson draw of the synthetic gamma-ray sources.} \label{fig:dif_norms_1}
\end{figure}

In Fig.~\ref{fig:dif_norms_1}, we compare the posteriors of the expected total number of counts across the Inner ROI for all the gamma-ray sources. These results show that diffuse mismodeling affects both (1) our interpretation of the GCE and (2) our interpretation of the background templates (including the non-diffuse templates). The SVI training will determine the best configuration that explains the measured data, regardless of how well the modeled gamma-ray sources align with the true gamma-ray sources. 

\begin{figure}[t!] 
    \centering
    \includegraphics[width=\columnwidth]{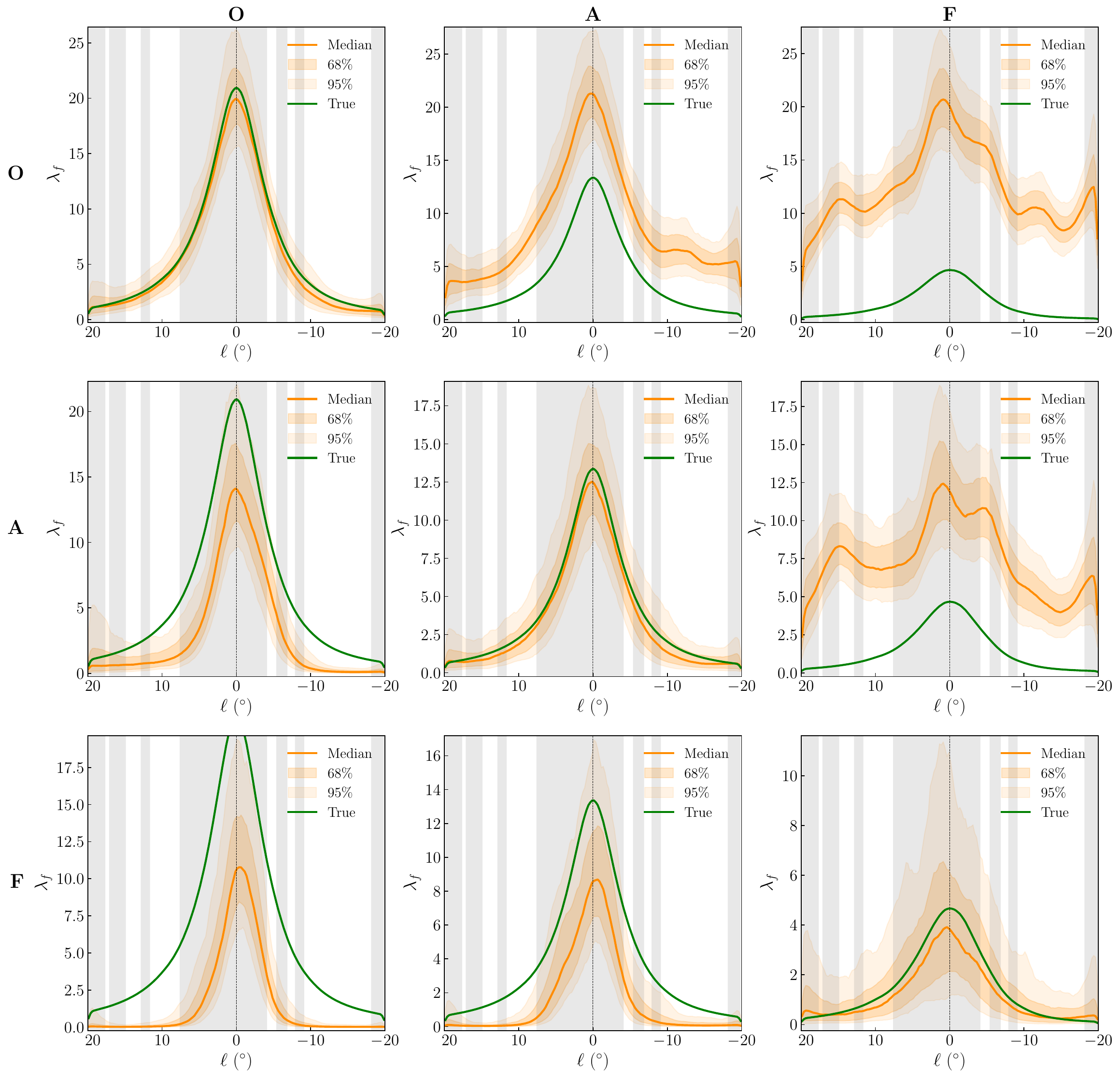}   
    \caption{Results from fits described in Section~\ref{sssec:mismodel-single}. Horizontal 1D slice of the GP posterior distribution of counts on Cartesian bins with lengths of $0.25^{\circ}$ for our fit to synthetic data. This slice is situated at $b = 3.375^{\circ}$. Each plot corresponds to different fits. Rows indicate the diffuse model assumed to fit the data. Columns indicate the synthetic dataset generated using one of the diffuse models. The posterior is the average of 10 posteriors each obtained by fitting to a different Poisson draw of the synthetic gamma-ray sources.} \label{fig:dif_hgp_1}
\end{figure}

\begin{figure}[t!] 
    \centering
    \includegraphics[width=\columnwidth]{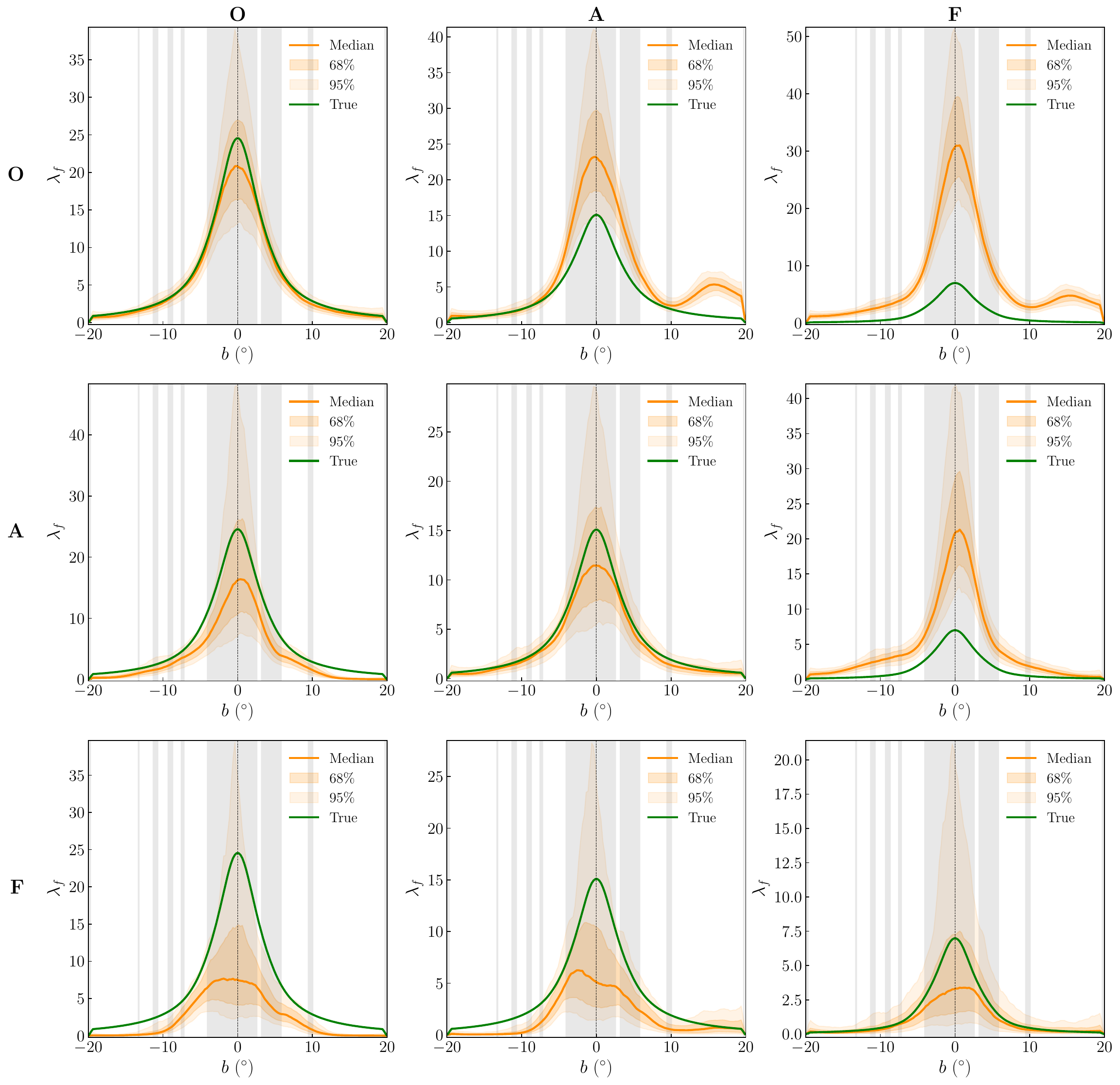}   
    \caption{Results from fits described in Section~\ref{sssec:mismodel-single}. Vertical 1D slice of the GP posterior distribution of counts on Cartesian bins with lengths of $0.25^{\circ}$ for our fit to synthetic data. This slice is situated at $\ell = -3.375^{\circ}$. Each plot corresponds to different fits. Rows indicate the diffuse model assumed to fit the data. Columns indicate the synthetic dataset generated using one of the diffuse models. The posterior is the average of 10 posteriors each obtained by fitting to a different Poisson draw of the synthetic gamma-ray sources.} \label{fig:dif_vgp_1}
\end{figure}

\begin{figure}[t!] 
    \centering
    \includegraphics[width=\columnwidth]{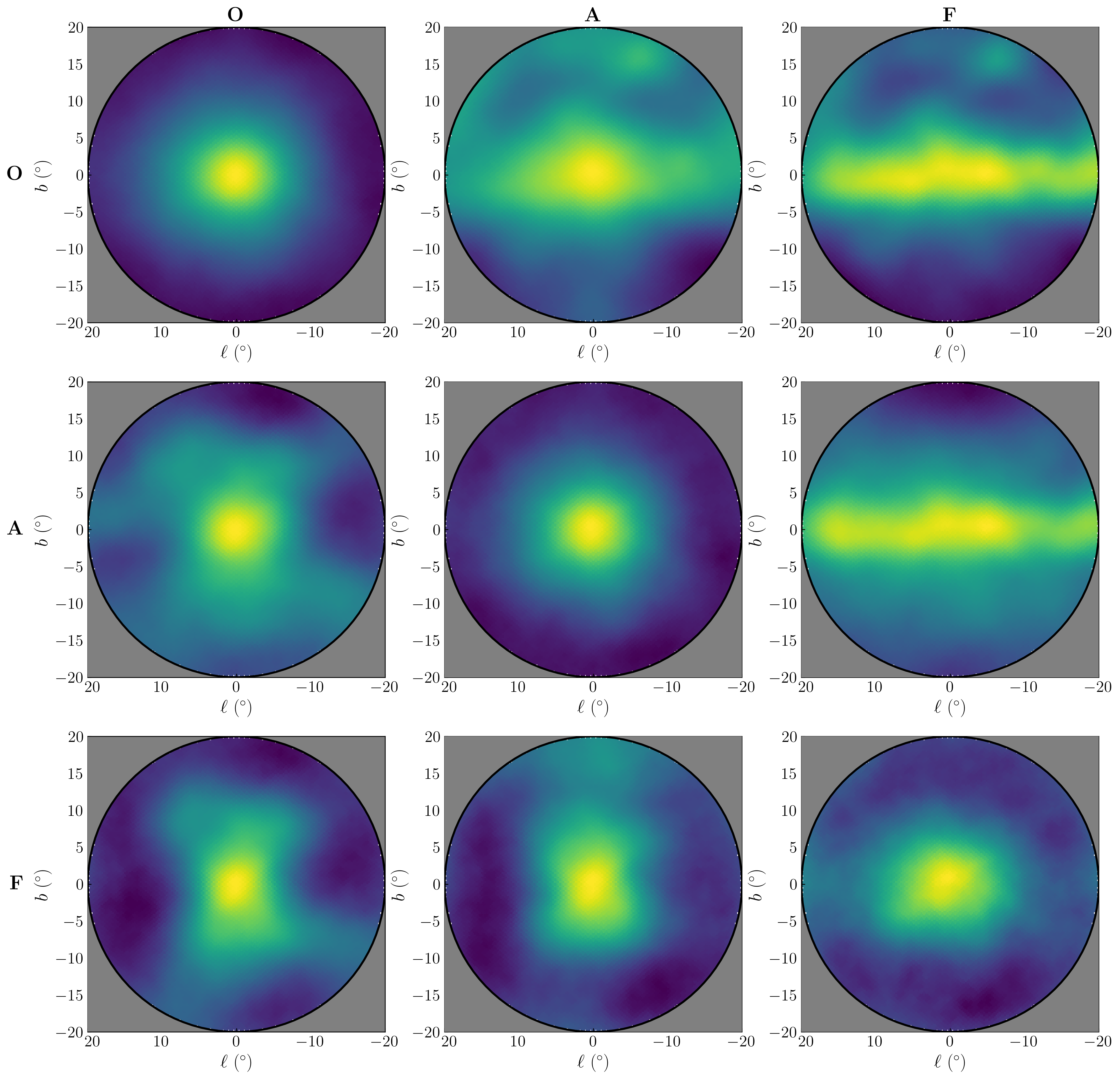}   
    \caption{Results from fits described in Section~\ref{sssec:mismodel-single}. Median GP posterior sample for different fits. Each plot corresponds to different fits. Rows indicate the diffuse model assumed to fit the data. Columns indicate the synthetic dataset generated using one of the diffuse models. The posterior is the average of 10 posteriors each obtained by fitting to a different Poisson draw of the synthetic gamma-ray sources.} \label{fig:dif_gp_1}
\end{figure}

In Fig.~\ref{fig:dif_hgp_1}, we compare the horizontal 1D slices of the GP posterior at $b = 3.375^{\circ}$. In Fig.~\ref{fig:dif_vgp_1}, we compare the vertical 1D slices of the GP posterior at $\ell = -3.375^{\circ}$. In Fig.~\ref{fig:dif_gp_1}, we compare the median GP posterior sample. These three results illustrate how the GCE interpretation changes when we assume the wrong diffuse model templates. Assuming the correct templates produces a GCE with a morphology that agrees with the true synthetic GCE. Failing to correctly model these sources causes the GP to correct for subsequent mismodeling from the background templates, sometimes causing it to inherit features characteristic of some of the background templates. For example, the first two plots in column F of Fig.~\ref{fig:dif_gp_1} show the GCE with a disk component, due to the inability for the \texttt{ics} template in Models O and A to model the synthetic \texttt{ics} generated from Model F. Similarly, the GCE inherits the localized source around $(\ell, b) = (-15^{\circ}, 5^{\circ})$ from the \texttt{pib} template in the second and third plots in the first row of Fig.~\ref{fig:dif_gp_1}. Lastly, the plot in the lower-left corner of Fig.~\ref{fig:dif_gp_1} suggests that the GCE may inherit features from the \texttt{bub} template. 

In Fig.~\ref{fig:dif_htot_1}, we compare the horizontal 1D slices of the total rate produced by all the gamma-ray sources at $b = 3.375^{\circ}$. In Fig.~\ref{fig:dif_vtot_1}, we compare the vertical 1D slices of the total rate produced by all the gamma-ray sources at $\ell = -3.375^{\circ}$. These plots illustrate how diffuse mismodeling affects the model's ability to describe the total rate $\lambda(\textbf{x})$.

In summary, these results show how the model leverages the GP's flexibility to correct for mismodeling arising from assuming inaccurate diffuse templates. This leads to inaccurate interpretations of the GCE morphology and the relative contributions of signal provided by the gamma-ray sources.

\subsubsection{Mismodeling with Multiple Diffuse Emission Models} \label{sssec:mismodel-multi}

Next, we consider models which fit using multiple templates that do not correspond to the templates used to generate the synthetic data. For example, we fit a model using a linear combination of the templates of Model O and A to pseudodata generated with Model F. We do the same for the remaining two combinations (${\rm OF} \rightarrow {\rm F}, {\rm AF} \rightarrow {\rm O}$).
As before, all posteriors are the average of ten posteriors obtained by fitting to different Poisson draws of the rate of the synthetic gamma-ray sources. During the fit, the relative normalizations of each diffuse model template is given a prior that has no preference for either mixed or unmixed models (see Appendix~\ref{app:multi-temp} for further discussion).

In the first row of Fig.~\ref{fig:dif_norms_2}, we compare the posteriors of the expected total number of counts across the Inner ROI for all the gamma-ray sources. In the second and third rows, we show the posteriors of the relative normalizations $\theta_{pib}$ and $\theta_{ics}$ corresponding to different models. The first row shows that the additional flexibility does not necessarily improve the alignment of the best-fit normalizations with the true normalizations. However, the fit in the second row (OF $\rightarrow$ A) shows close alignment with the true synthetic gamma-ray sources. The second and third rows show that (1) the \texttt{ics} template has a stronger preference for a single model template contributing to its signal and the \texttt{pib} model allows for a more evenly mixed linear combination, and (2) a particular model can be preferred for describing the \texttt{ics} template but not the \texttt{pib} template (\textit{e.g.}, see Model O in the second column). 

In Fig.~\ref{fig:dif_gp_2}, we compare the GP posteriors corresponding to each fit. As we saw in the previous set of figures, the fit OF $\rightarrow$ A gives an accurate description of the GCE. For the other two cases, we have morphologies similar to those encountered in Fig.~\ref{fig:dif_gp_1}, implying that the combination of models suffered from similar mismodeling effects when we assumed a single inaccurate diffuse model. 

\begin{figure}[t!] 
    \centering
    \includegraphics[width=\columnwidth]{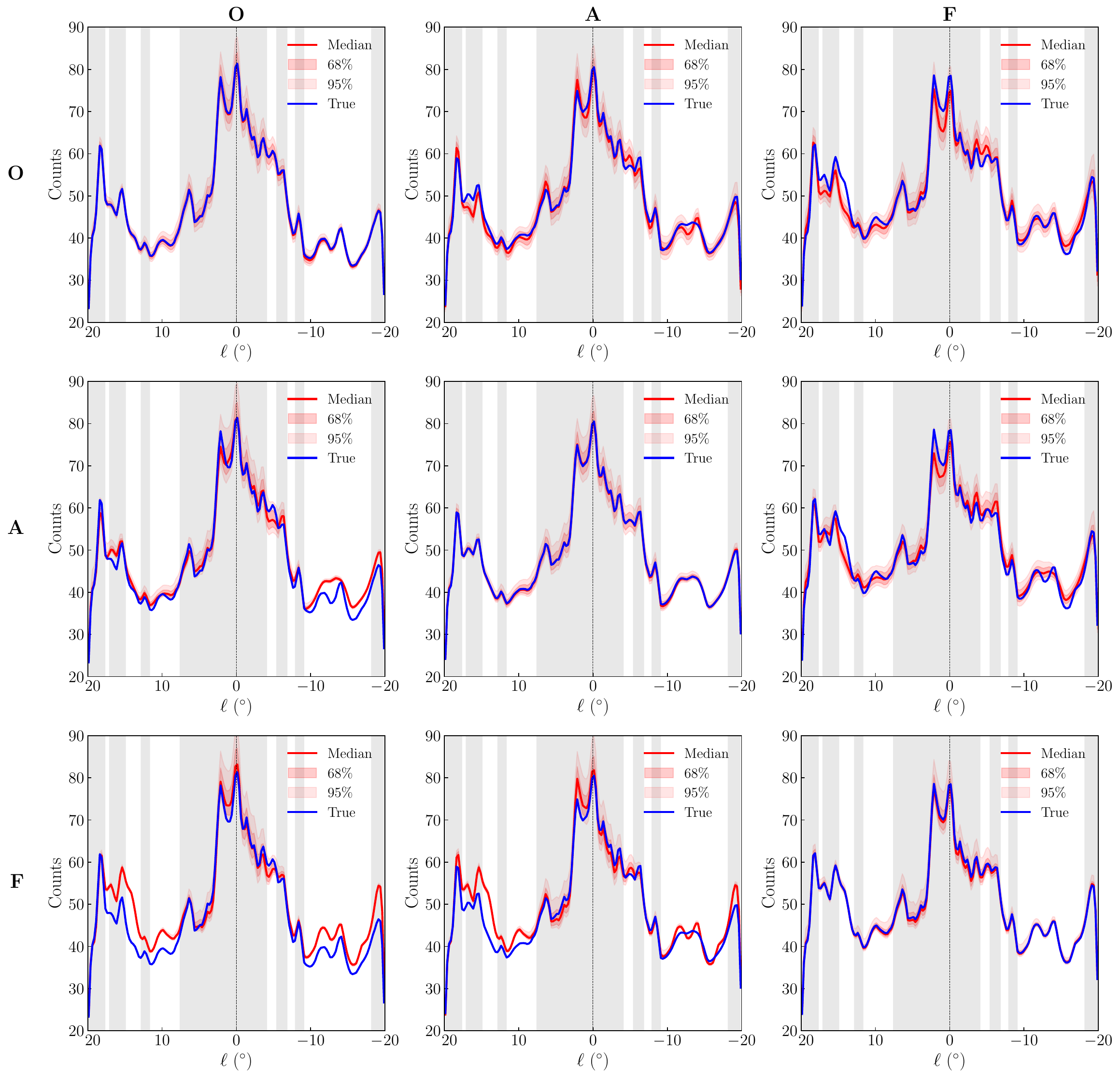}   
    \caption{Results from fits described in Section~\ref{sssec:mismodel-single}. Horizontal 1D slice of the posterior distribution of the total rate of emission from all gamma-ray sources on Cartesian bins with lengths of $0.25^{\circ}$ for our fit to synthetic data. This slice is situated at $b = 3.375^{\circ}$. Each plot corresponds to different fits. Rows indicate the diffuse model assumed to fit the data. Columns indicate the synthetic dataset generated using one of the diffuse models. The posterior is the average of 10 posteriors each obtained by fitting to a different Poisson draw of the synthetic gamma-ray sources.} \label{fig:dif_htot_1}
\end{figure}

\begin{figure}[t!] 
    \centering
    \includegraphics[width=\columnwidth]{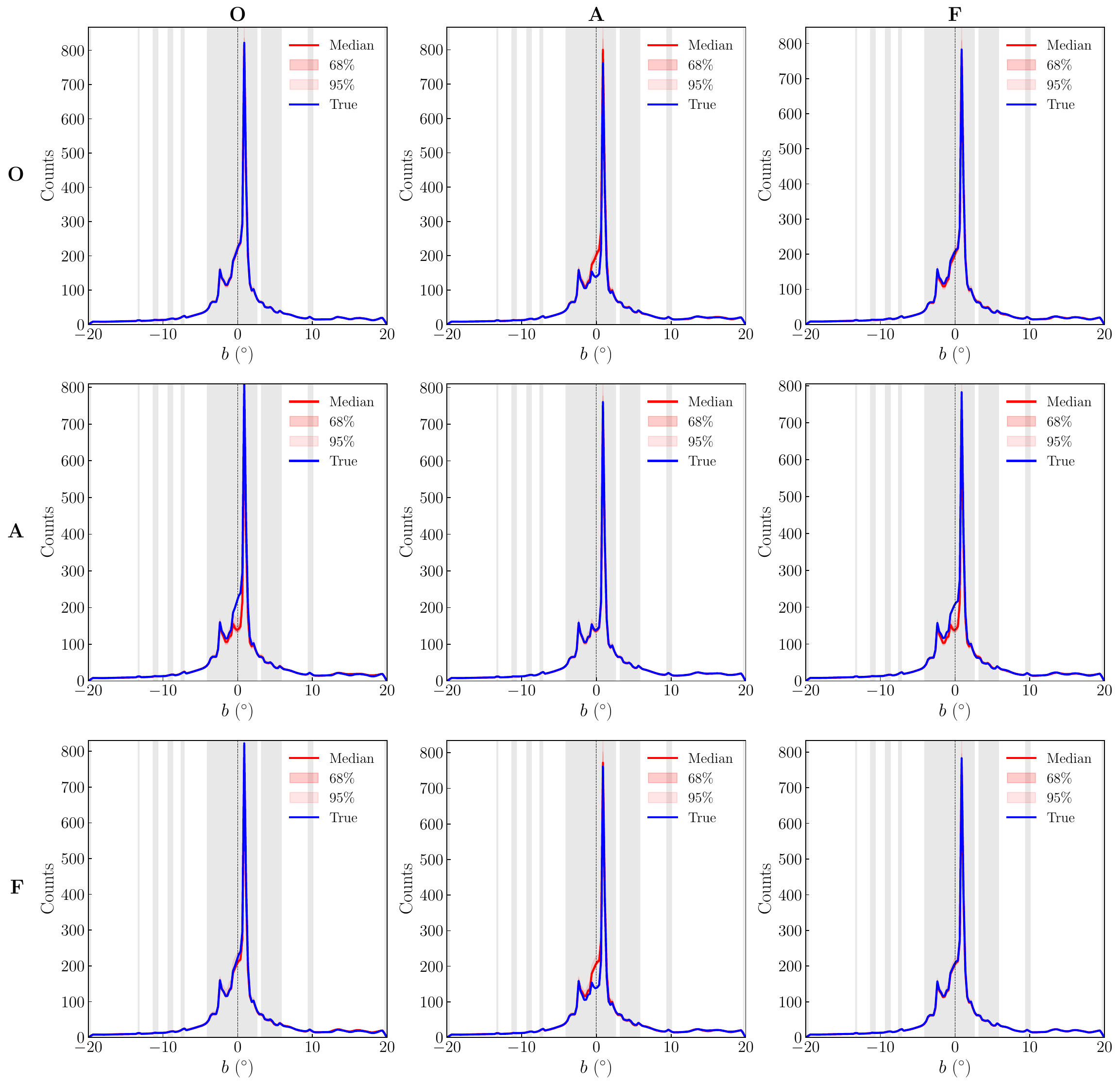}   
    \caption{Results from fits described in Section~\ref{sssec:mismodel-single}. Vertical 1D slice of the posterior distribution of the total rate of emission from all gamma-ray sources on Cartesian bins with lengths of $0.25^{\circ}$ for our fit to synthetic data. This slice is situated at $\ell = -3.375^{\circ}$. Each plot corresponds to different fits. Rows indicate the diffuse model assumed to fit the data. Columns indicate the synthetic dataset generated using one of the diffuse models. The posterior is the average of 10 posteriors each obtained by fitting to a different Poisson draw of the synthetic gamma-ray sources.} \label{fig:dif_vtot_1}
\end{figure}

\begin{figure}[t!] 
    \centering
    \includegraphics[width=\columnwidth]{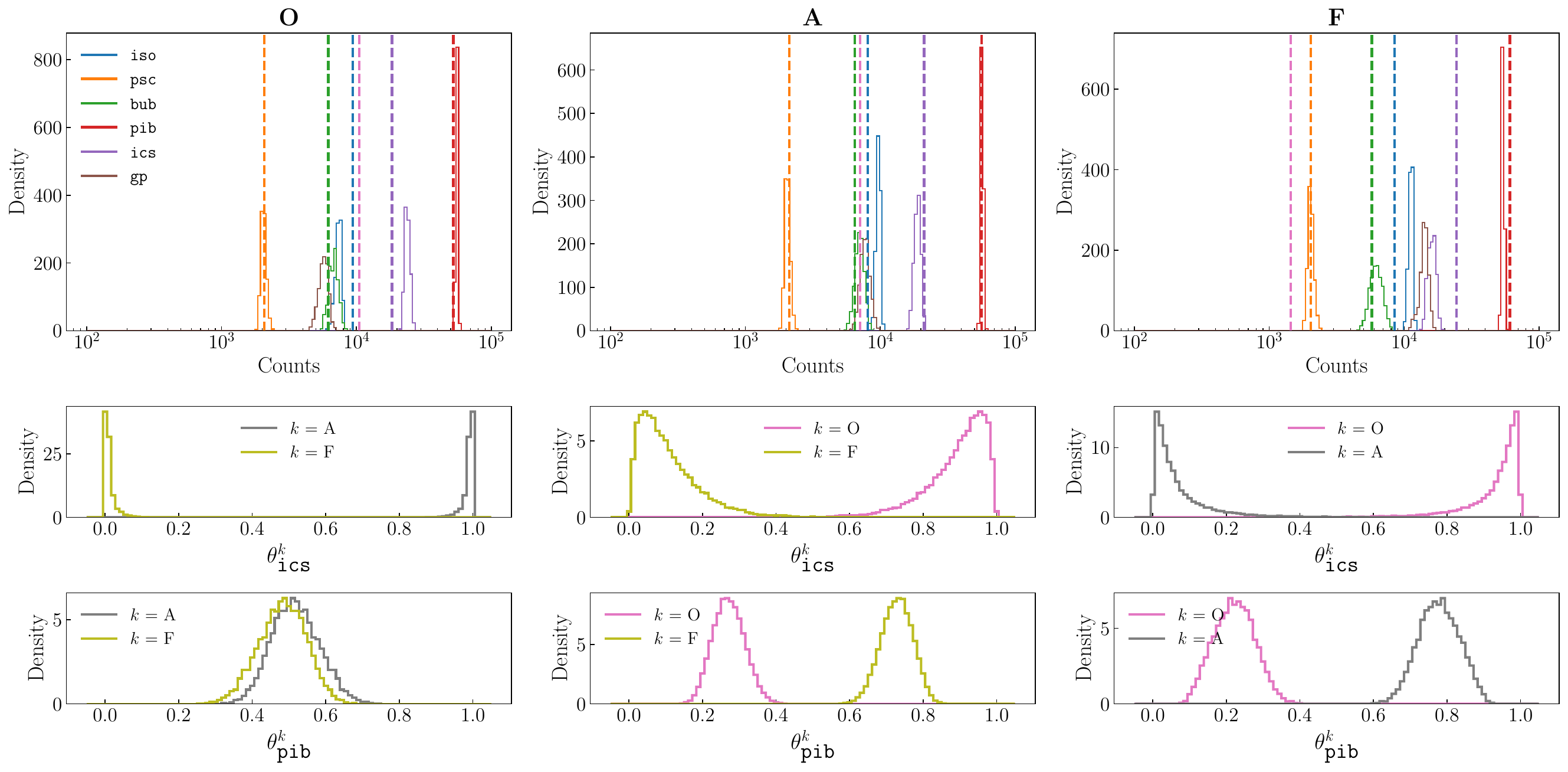}   
    \caption{Results from fits described in Section~\ref{sssec:mismodel-multi}. \textit{(Top)} Color-coded histograms showing the expected total number of counts of each gamma-ray source in the Inner ROI and are binned logarithmically. True normalizations are denoted by dashed lines. \textit{(Middle)} Posteriors of the relative normalizations of the \texttt{ics} templates. \textit{(Bottom)} Posteriors of the relative normalizations of the \texttt{pib} templates. Columns indicate the synthetic dataset generated using one of the diffuse models (and diffuse models used to fit the data). All posteriors are the average of 10 posteriors each obtained by fitting to a different Poisson draw of the synthetic gamma-ray sources.} \label{fig:dif_norms_2}
\end{figure}

\begin{figure}[t!] 
    \centering
    \includegraphics[width=\columnwidth]{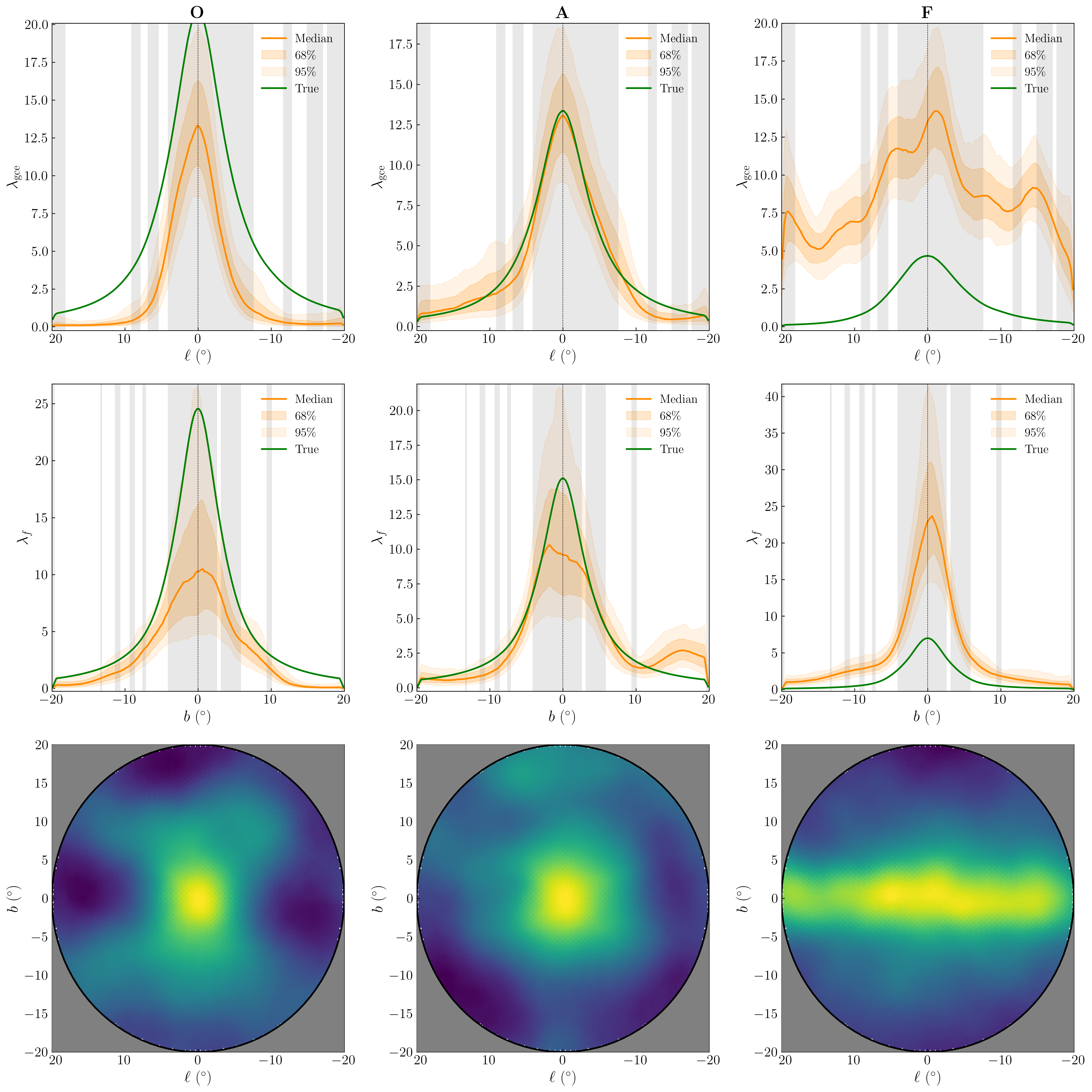}   
    \caption{Results from fits described in Section~\ref{sssec:mismodel-multi}. \textit{(Top)} Horizontal 1D slice of the GP posterior distribution of counts on Cartesian bins with lengths of $0.25^{\circ}$ for our fit to synthetic data. This slice is situated at $b = 3.375^{\circ}$. \textit{(Middle)} Vertical 1D slice of the GP posterior distribution of counts on Cartesian bins with lengths of $0.25^{\circ}$ for our fit to synthetic data. This slice is situated at $\ell = -3.375^{\circ}$.  \textit{(Bottom)} Median GP posterior sample for different fits. Columns indicate the synthetic dataset generated using one of the diffuse models (and diffuse models used to fit the data). All posteriors are the average of 10 posteriors each obtained by fitting to a different Poisson draw of the synthetic gamma-ray sources.} \label{fig:dif_gp_2}
\end{figure}

In Fig.~\ref{fig:dif_tot_2}, we compare the total rate of emission from all the gamma-ray sources along the horizontal line at $b = 3.375^{\circ}$ and the vertical line at $\ell = -3.375^{\circ}$. Compared to the single-model fits that do not assume the correct diffuse model (see Fig.~\ref{fig:dif_htot_1} and Fig.~\ref{fig:dif_vtot_1}), the model was able to select a configuration of the templates and the GP that better fit the total rate $\lambda(\textbf{x})$.

In summary, we found that modeling the data with a linear combination of (inaccurate) diffuse model templates can improve the model's ability to fit the data. However, the improved fit does not guarantee that each individual template will converge to a state that aligns with the corresponding true gamma-ray source. Moreover, the ability for the GP to align with the true GCE depends on how accurately the combination of diffuse models can describe the true diffuse gamma-ray sources.

\subsubsection{Fitting with Multiple Diffuse Emission Models Including True Template} \label{sssec:mismodel-all}

Lastly, we consider how models using linear combinations of diffuse models perform when the true model is included in the linear combination (${\rm OAF} \rightarrow {\rm O}$, ${\rm OAF} \rightarrow {\rm A}$, ${\rm OAF} \rightarrow {\rm F}$).
All posteriors are the average of ten posteriors obtained by fitting to different Poisson draws of the rate of the synthetic gamma-ray sources. The relative normalizations of each diffuse model template is given a prior that has no preference for either mixed or unmixed models. 

In the first row of Fig.~\ref{fig:dif_norms_3}, we compare the posteriors of the expected total number of counts across the Inner ROI for all the gamma-ray sources. In the second and third rows, we show the posteriors of the relative normalizations $\theta_{pib}$ and $\theta_{ics}$ corresponding to different models. The second and third columns indicate that the source normalizations align with the true normalizations when the model finds a preference for the true diffuse template model over the other models contributing to the signal. In the first column, the $\texttt{ics}$ templates are somewhat degenerate, leading to normalization posteriors that are not quite aligned with the truth. 

In Fig.~\ref{fig:dif_gp_3}, we compare the GP posteriors corresponding to each fit. Here, the GP has a morphology that aligns well with the true GCE template in all cases, though the posterior is somewhat biased towards a fainter GCE. 

\begin{figure}[t!] 
    \centering
    \includegraphics[width=\columnwidth]{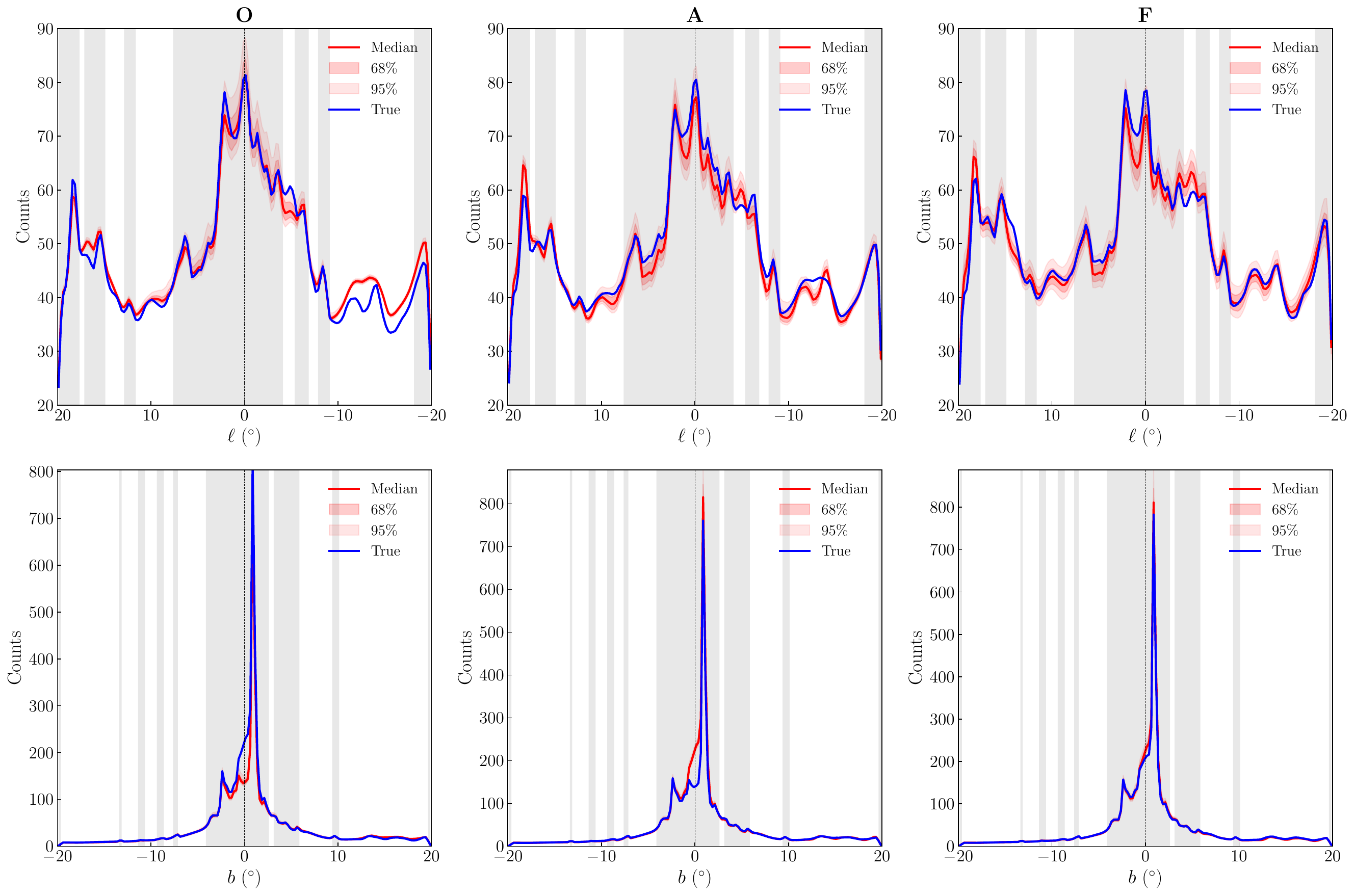}   
    \caption{Results from fits described in Section~\ref{sssec:mismodel-multi}. \textit{(Top)} Horizontal 1D slice of the posterior distribution of the total rate of emission from all gamma-ray sources on Cartesian bins with lengths of $0.25^{\circ}$ for our fit to synthetic data. This slice is situated at $b = 3.375^{\circ}$. \textit{(Bottom)} Vertical 1D slice of the posterior distribution of the total rate of emission from all gamma-ray sources on Cartesian bins with lengths of $0.25^{\circ}$ for our fit to synthetic data. This slice is situated at $\ell = -3.375^{\circ}$. Columns indicate the synthetic dataset generated using one of the diffuse models (and diffuse models used to fit the data). All posteriors are the average of 10 posteriors each obtained by fitting to a different Poisson draw of the synthetic gamma-ray sources.} \label{fig:dif_tot_2}
\end{figure}

\begin{figure}[t!] 
    \centering
    \includegraphics[width=\columnwidth]{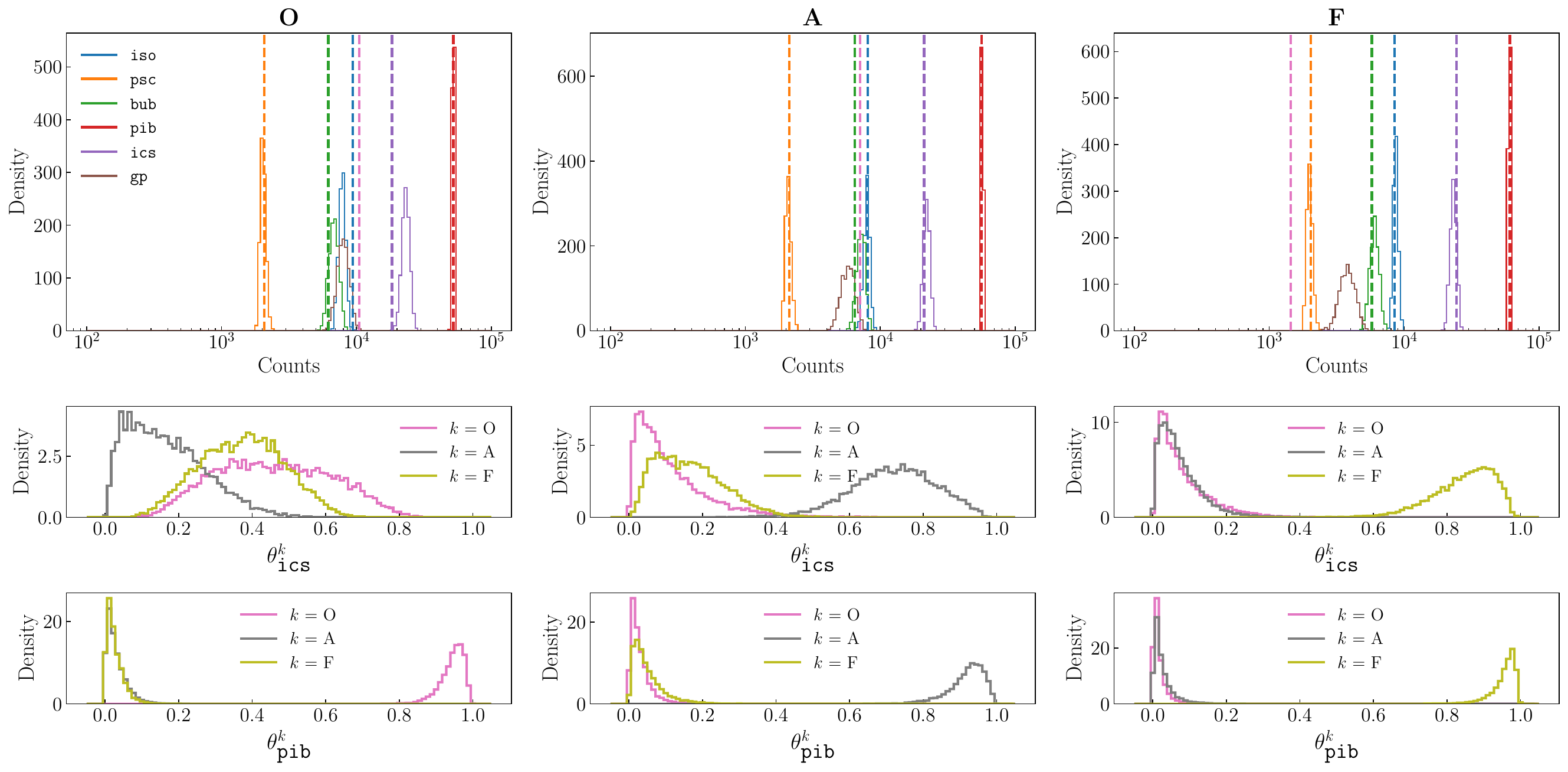}   
    \caption{Results from fits described in Section~\ref{sssec:mismodel-all}. \textit{(Top)} Color-coded histograms showing the expected total number of counts of each gamma-ray source in the Inner ROI and are binned logarithmically. True normalizations are denoted by dashed lines. \textit{(Middle)} Posteriors of the relative normalizations of the \texttt{ics} templates. \textit{(Bottom)} Posteriors of the relative normalizations of the \texttt{pib} templates. Columns indicate the synthetic dataset generated using one of the diffuse models. All posteriors are the average of 10 posteriors each obtained by fitting to a different Poisson draw of the synthetic gamma-ray sources.} \label{fig:dif_norms_3}
\end{figure}

\begin{figure}[t!] 
    \centering
    \includegraphics[width=\columnwidth]{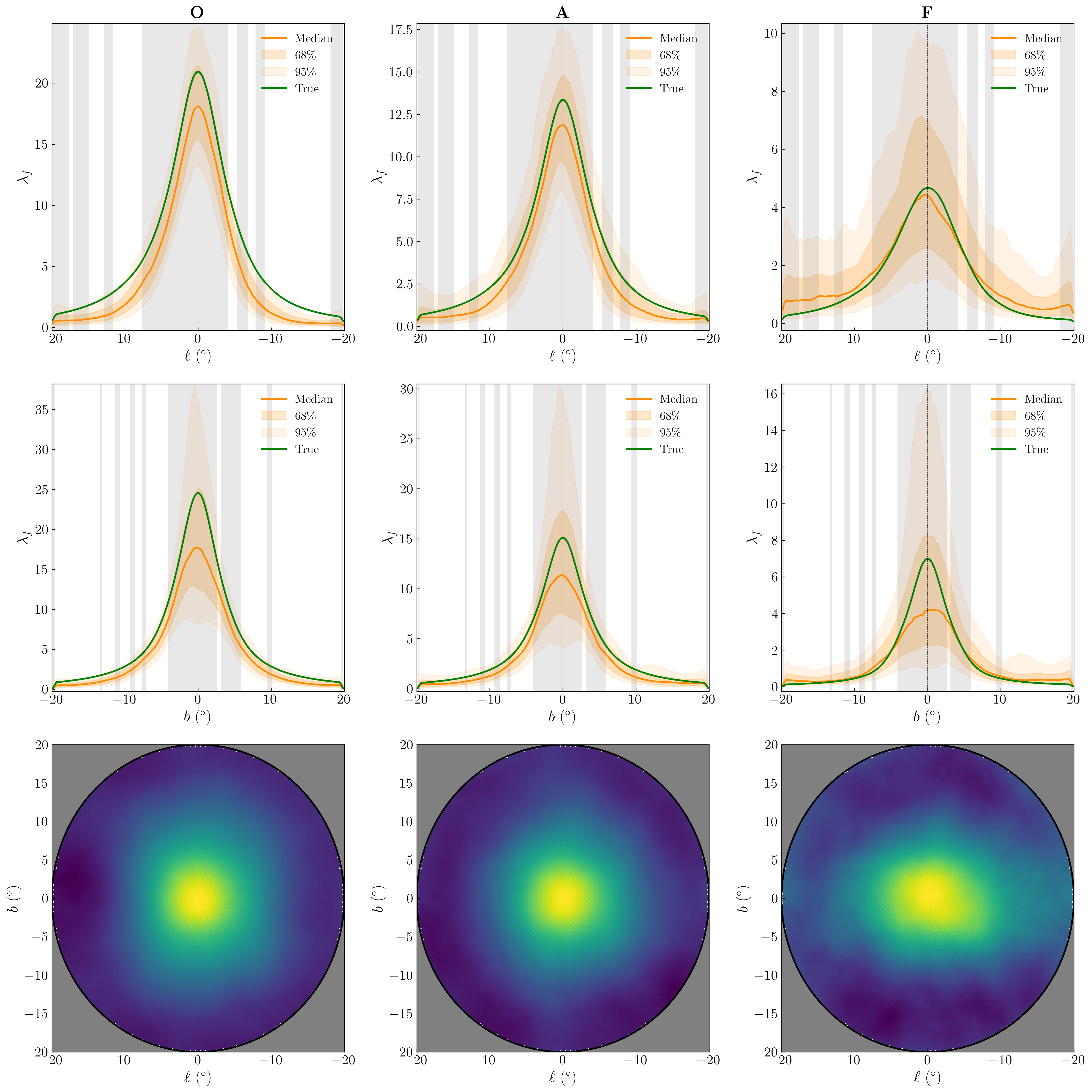}   
    \caption{Results from fits described in Section~\ref{sssec:mismodel-all}. \textit{(Top)} Horizontal 1D slice of the GP posterior distribution of counts on Cartesian bins with lengths of $0.25^{\circ}$ for our fit to synthetic data. This slice is situated at $b = 3.375^{\circ}$. \textit{(Middle)} Vertical 1D slice of the GP posterior distribution of counts on Cartesian bins with lengths of $0.25^{\circ}$ for our fit to synthetic data. This slice is situated at $\ell = -3.375^{\circ}$.  \textit{(Bottom)} Median GP posterior sample for different fits. Columns indicate the synthetic dataset generated using one of the diffuse models. All posteriors are the average of 10 posteriors each obtained by fitting to a different Poisson draw of the synthetic gamma-ray sources.} \label{fig:dif_gp_3}
\end{figure}

\begin{figure}[t!] 
    \centering
    \includegraphics[width=\columnwidth]{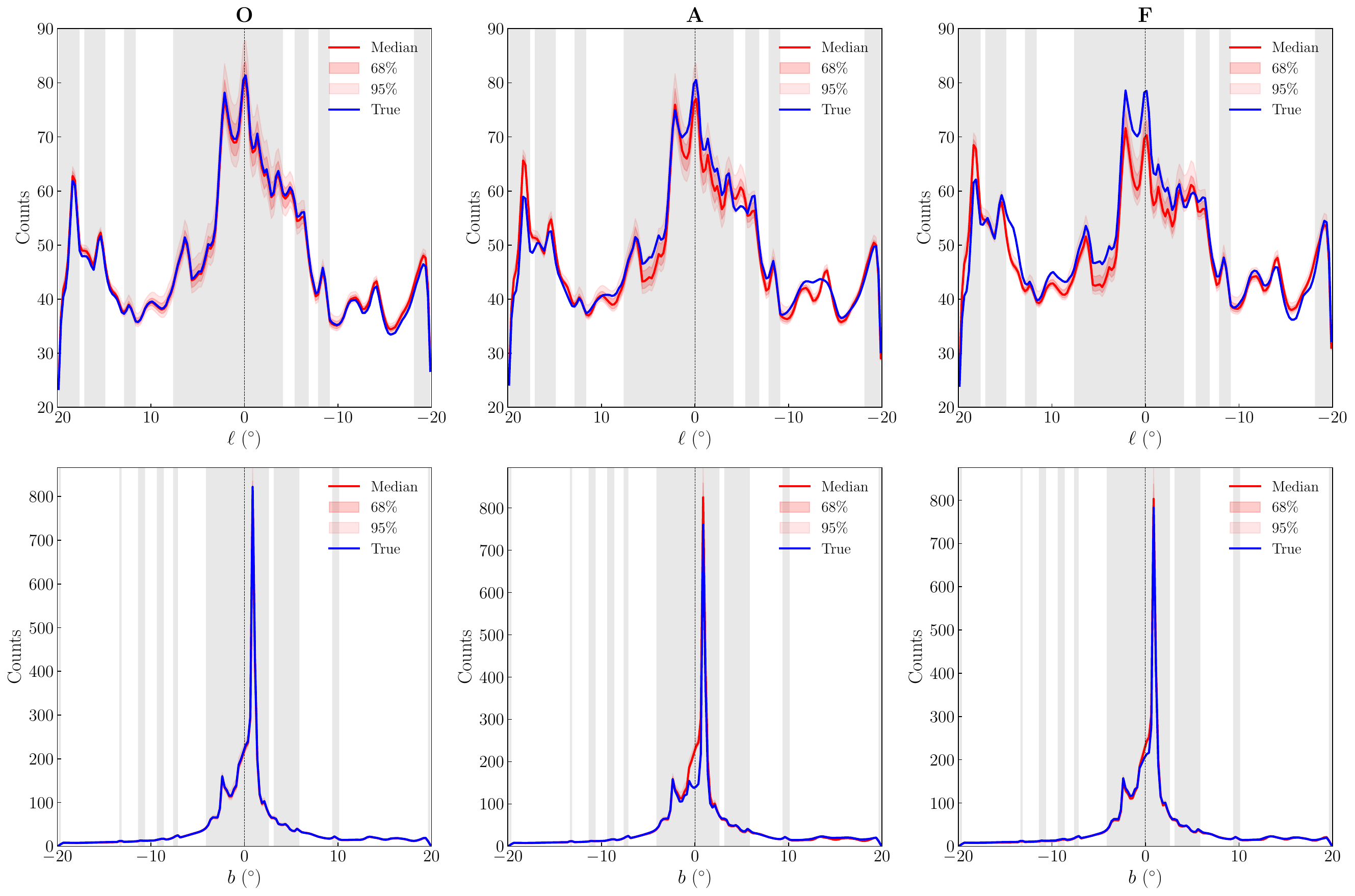}   
    \caption{Results from fits described in Section~\ref{sssec:mismodel-all}. \textit{(Top)} Horizontal 1D slice of the posterior distribution of the total rate of emission from all gamma-ray sources on Cartesian bins with lengths of $0.25^{\circ}$ for our fit to synthetic data. This slice is situated at $b = 3.375^{\circ}$. \textit{(Bottom)} Vertical 1D slice of the posterior distribution of the total rate of emission from all gamma-ray sources on Cartesian bins with lengths of $0.25^{\circ}$ for our fit to synthetic data. This slice is situated at $\ell = -3.375^{\circ}$. Columns indicate the synthetic dataset generated using one of the diffuse models. All posteriors are the average of 10 posteriors each obtained by fitting to a different Poisson draw of the synthetic gamma-ray sources.} \label{fig:dif_tot_3}
\end{figure}

In Fig.~\ref{fig:dif_tot_3}, we compare the total rate of emission from all the gamma-ray sources along the horizontal line at $b = 3.375^{\circ}$ and the vertical line at $\ell = -3.375^{\circ}$. In some cases, introducing the true template worsened the fit in some areas compared to the plots in Fig.~\ref{fig:dif_tot_2}, but a major improvement is seen in the first row (${\rm OAF} \rightarrow {\rm O}$). 

In summary, including the true diffuse template in a linear combination of templates modeling the same diffuse gamma-ray source may improve the fit to the overall rate $\lambda(\textbf{x})$ and the modeled gamma-ray source alignment with the true gamma-ray sources, as long as the templates in the linear combination are not very degenerate. This illustrates the extent at which our model can select a correct template among a linear combination of templates when fitting to data.
 
\end{document}